\documentclass[allclo]{FBSart}
\usepackage{epsfig,feynmf}
\usepackage{amssymb}

\usepackage{fancyheadings,epsfig}
\usepackage{feynmf}

\newcommand{\be}{\begin{equation}}
\newcommand{\ee}{\end{equation}}
\newcommand{\bea}{\begin{eqnarray}}
\newcommand{\eea}{\end{eqnarray}}
\newcommand{\SG}{\sigma}
\newcommand{\ep}{i\varepsilon}
\newcommand{\nn}{\nonumber}
\newcommand{\al}{\alpha}

\newcommand{\om}{\omega}

\title{ Implications of Analyticity to Solution of  Schwinger-Dyson Equations in Minkowski Space }

\author{Vladim\'{i}r \v{S}auli}
\institute{Dep. of Theor. Phys. NPI \v{R}e\v{z} near Prague, CAS, CZ}

\runningauthor{V. \v{S}auli }
\runningtitle{  Implications of analyticity to... }
\begin{document}

\maketitle
\begin{abstract}We review some recent developments in 
nonperturbative studies of quantum field theory (QFT) using the Schwinger-Dyson equations formulated directly in Minkowski space.    
We begin with the introduction of essential ideas of the integral representation in QFT 
and a discussion of renormalization in this approach.
The technique based on the integral representation of  Green's functions is exploited to solve Schwinger-Dyson equations in  several quantum field models, eg. in scalar models and in strong coupling $QED_{3+1}$ in the quenched and in the unquenched approximation. The phenomenon of dynamical chiral symmetry breaking in regularized theory is touched.
In QCD,  the analyticity of gluon  propagator on the complex momentum square plane is exploited to continue some recent lattice data to timelike momentum axis.
We find non-positive absorptive part contribution in the Landau gauge gluon propagator which is 
in agreement with some other new  recent analyzes. 
\end{abstract}

%keywords: Schwinger-Dyson equations, dispersion relation, Strong coupling QFT, spectral
%representation, confinement}

%preprint: hep-ph/0412188}

%%%%%%%%%%%%%%%%%%%%%%%%%%%%%%%%%%%%%%%%%%%%%%%%%%%%%%%%%%%%%%%%%%%%%%%%%%%%%%%%%%%%%%%%%%%%%%%%%%%%%%%%%%%%%%%%%%%%%%%%%%%%%%%%%%%%%%%%%%%%%%%%%%%%%%%%%%%%%%%%%%%%%%%%%%%%%%%%%%%%%%%%%%%%%%%%%%%%%%%%%%%%%%%%%%%%%%%%%%%%%%%%%%%%%%%%%%%%%%%%%%%

\section{Introduction}

This  work is devoted to studies of  quantum field models with
strong coupling. The {\em Schwinger-Dyson equations} (SDEs) in
momentum representation are solved and analyzed in Minkowski space.
In this introduction our motivations, broad contexts and main results
are briefly summarized.

These days the experimentally accessible particle physics is being
described in the framework of the (effective) quantum field theory
(QFT) by the Standard model of the strong and electroweak
interactions with the gauge group $SU_c(3)\otimes SU_L(2)\otimes
U_Y(1)$. When the perturbation theory  is applicable the
observables follow from the Lagrangian in relatively
straightforward way: the corresponding S-matrix elements are
generated from n-point {\em Green's functions} (GFs) which are
calculated and regularized/renormalized order by order according to
the standard rules. Order by order perturbation theory can be naturally generated from
expansion of SDEs: an infinite tower
of coupled equations 
connecting successively higher and higher points GFs. The GFs themselves are
not observable and they contain some redundant information: off-shell
behavior (connected e.g. with a possible field redefinition) and also
some other remnants of the particular calculation scheme employed,
e.g., gauge fixing parameters and renormalization scales. If the
program of the perturbation theory is carried out consistently up to the certain
order, the observables should be independent of all these unphysical
parameters.

Needless to say, many phenomena exist outside the reach of the perturbation theory
approach, especially when the strong sector of the Standard model --
{\em Quantum Chromodynamics} (QCD) -- or some extensions of the
Standard model are concerned: bound states, confinement, dynamical
mass generation etc. Going beyond perturbation theory requires development of
elaborated approaches and sophisticated tools, especially in the
strong coupling regime. Number of such approaches has been pursued:
lattice theories, Feynman-Schwinger representation, nonperturbative
treatment of SDEs, bag models \dots. These days neither of them
provides such a clear and unambiguous understanding of the physics as
we are accustomed to in the perturbative regime. The lattice approach
clearly stands by its own, since it does not rely on any
uncontrollable approximations: just the brute force discretization of
path integrals is employed. Lattice ``data'' should (in principle)
provide as good representation of the studied dynamics as real ones
(with additional merit of allowing to vary the dynamical input), but
converged unquenched results are in most cases just not available as
yet (however for some recent progress in QCD see \cite{lat2003},\cite{BHLPW2004},\cite{BHLPW2005}).

In this work we deal with the nonperturbative solutions of the SDEs.
If one could solve the full infinite set of SDEs, the complete
solution of the QFT will be available, coinciding with the results of
converged lattice calculations. In reality one has to truncate this
set of equations: the main weakness of SDEs phenomenology is the
necessity to employ some Ansatze for higher Greens functions.   The
reliability of these approximations can be estimated by comparison
with the known perturbation theory result in the regime of a soft coupling constant
and by comparison with the lattice data (and/or results of
alternative approaches like $\Phi$-derivable effective action, etc.) where available. The recent results (see e.g.
reviews
\cite{ROBERTS,ALKSMEK2001,MARROB2003,FISALK2003}
and references therein) provide some encouragement: they suggest that
SDEs  are viable tool for obtaining (eventually) the truly
nonperturbative answers for plethora of fundamental questions: {\it
dynamical chiral symmetry breaking}, {\it confinement} of colored
objects in QCD, the high temperature superconductivity in condensed
matter (modeled \cite{MAPA2003} by $QED_{2+1}$). In the QCD sector
there is also a decent agreement with some experimental data: for
meson decay constants like $F_{\pi}$, evolution of quark masses to
their phenomenologically known (constituent) values, bound states
properties obtained from Bethe-Salpeter or Fadeev equations (the
part of SDEs). 

The recent solutions of SDEs \cite{FISALK2003},\cite{BLOCH2003} nicely agree with the
lattice data \cite{BHLPW2004,BOBOLEWIZA2001,BCLM2002} 
(the explicit comparison is made in \cite{BLOCH2003,FISALK2003,FIES2004}). 
Thus, encouraging
connections between results of SDEs studies and fundamental theory
(QCD) and/or various phenomenological approaches (chiral perturbation
theory, constituent quark model, vector meson dominance \dots) are
emerging.

Most of these results were obtained by solution of SDEs in Euclidean
space. One of the main goal of this work is to develop an alternative
approach allowing to solve them directly in Minkowski space. In this
method (generalized) {\em spectral decompositions} of the GFs are
employed, based on their analytical properties. The main merit of this
approach is possibility to get solutions both for spacelike and
timelike momenta. Techniques of solving SDEs directly in Minkowski
space are much less developed than the corresponding Euclidean ones.
Therefore most of this paper deals with developing and testing such
solutions on some simple QFT models.

For large class of nonasymptotically free  theories
running couplings -- calculated from perturbation theory -- diverge at some finite but
usually large spacelike scale. If perturbation theory results are taken literally,
stretching them clearly beyond the range of applicability, also the
QCD coupling develops singularity for low momenta squared. In the
latter case the lattice  as well as SDEs results suggest that nonperturbative effects
alter behavior of running coupling  $\alpha_{QCD}(q^2)$ significantly, situation for
nonasymptotically free is less clear. This is why  we pay special attention to behavior
of our  solutions in the region of high momenta.

The paper is organized as follows. Section~2  serves as an
introduction: some well known general results and useful textbooks
 formulas are collected
and basics of the formalism employed later are discussed. 
The  integral
representation for GFs are reviewed and the calculation of dispersion relation
is exhibited at one and two loop example.
Section~3 is devoted to the simple scalar model $a\Phi^3+b\Phi^4$
for pedagogical reason. The momentum space SDE is converted to the 
Unitary Equation for propagator spectral density. The Unitary Equation is shown to be real
equation where  the singularity accompanying usual Minkowski space calculation 
are avoided. The relativistic problem of bounds states as an intrinsically
nonperturbative phenomena is briefly reviewed in this Section too.

 Section~4 represents  study of the SDEs in 3+1 QED. For this purpose we improve and reorganize the Unitary Equations already particularly  found in the papers \cite{SAULIJHEP,SAULIRUN} in a way that the numerical obstacles with singular integral kernels are avoided. 
Due to this the resulting numerical solutions of our Unitary Equations become more stable and accurate. 

One should mention at this point a possibility of getting the
solutions of the SDEs in Minkowski space with the help of analytical
continuation of the Euclidean ones.  Nevertheless in simple cases it
can be successful: we have recently done this for the gluon form
factor and we present our results in the Section devoted to the QCD.

Section~5 offers some further comments on spectral treatment of Schwinger-Dyson equations and possible further progress is discussed.
  
Section~6 presents the known attempt to solve the SDEs in QCD in Minkowski space. Furthermore,
$SU(3)$ Yang-Mills, i.e the pure gluodynamics is discussed there and analytical behavior of 
gluon propagator is deduced by the analytical continuation from Euclidean space to the analyticity domain boundary. To obtain numerically stable analytical continuation is not an easy task and
attempts to perform it are rare. As an input data we have  used  recently obtained lattice data \cite{lat2003}.     

%%%%%%%%%%%%%%%%%%%%%%%%%%%%%%%%%%%%%%%%%%%%%%%%%%%%%%%%%%%%%%%%%%%%%%%%%%%%%%%%%%%%%%%%%%%%%%%%%%%%%%%%%%%%%%%%%%%%%%%%%%%%%%%%%%%%%%%%%%%%%%%%%%%%%%%%%%%%%%%%%%%%%%%%%%%%%%%%%%%%%%%%%%%%%%%%%%%%%%%%%%%%%%%%%%%%%%%%%%%%%%%%%%%%%%%%%%%%%%%%%%%%%%%%%%%%%%%%%%%%%%%%%%%%%%%%%%%%%%%%%
\section{Integral Representations in Quantum Field Theory }

Solutions of the SDEs in this work 
are based on integral  representations of Green's functions. In this Section we briefly review
their derivation and then present two illustrative examples from the $\Phi^4$ model.

The development of spectral techniques can be stimulated by following consideration.
Let us assume that any GF can be written as an 
expansion of skeleton contributions, i.e.,  the diagrams where the propagators corresponding to 
the internal lines are exact. If we further assume the analyticity of these propagators
( in the  sense discussed in the next text), then one can apply the whole well established
knowledge  about the analytical structure of Feynman diagrams. The singularity structure of each Feynman graph 
allows a simple connection  between momentum dependence in spacelike and timelike regime, i.e.,  between Euclidean and Minkowski space. 
That the equations of motions which hold between Green's functions, i.e. SDEs , can be
translated into the corresponding relations between spectral functions was already known  
half century ago \cite{NAMBU1955,NAMBU1956}. The renormalization program has been carried
in the terms of spectral function as well.

This was also the basis of historical development  of 
the dispersion relation (for summary see \cite{DEWITT1960})
and of the Mandelstam representation  in the late 50's \cite{MAN1959}. Many of these efforts 
have been summarized in the standard  textbook of that time, see e.g. \cite{ELOP1966}.
This book  represents deductive approach to S-matrix theory, the subject rather popular in particle physics at that time. Fortunately, being mainly based on the analysis of Feynman graphs (and their infinite sum ), the   methods reviewed in this book ( including
also asymptotic high energy behavior, Regge poles,..) survive and 
have found their appropriate use, even after the time when  QCD was born and the trust 
to quantum field theory was renewed.

\subsection{The K\"allen--Lehmann  Representation}

In this Section we remind the derivation of the
Umezawa--Kamefuchi--K\"allen--Lehmann {\em spectral representation} (SR)
\cite{UMEKAM1951,KAL1952,KALL1958,LEH1954} for a two-point Green's
function.

Let us consider the vacuum expectation value of the product of two
real scalar fields
$
\label{soucin}
<0|\phi(x)\phi(y)|0>
$.
To arrive at the desired expression for the propagator 
\be
iG(x-y)=<0|T\phi(x)\phi(y)|0>
\ee
 one inserts the identity operator 
\be
\hat{1}=|0><0|+\sum_{\lambda}\int\frac{d^3{\bf p}}{(2\pi)^3\,
2\sqrt{{\bf p}^2+m^2}}|\lambda_{\bf p}><\lambda_{\bf p}|
\label{complet}\, ,
\ee
between the fields $\phi(x)$ and $\phi(y)$ in (\ref{soucin}):
\be
\label{krokjedna}
<0|\phi(x)\phi(y)|0>=\sum_{\lambda}\int\frac{d^3p}{(2\pi)^3\, 2
\sqrt{p^2+m^2}}<0|\phi(x)|\lambda_{\bf p}><\lambda_{\bf p}|\phi(y)|0> \, 
\ee
where the sum runs over complete set of $\lambda$-particles states $|\lambda_{\bf p}>$
with given momentum ${\bf p}$.
It is assumed that a spontaneous symmetry breaking does not take
place, i.e., that $<0|\phi(x)|0>=0$; if it does a space-time independent
constant appears on the r.h.s. Making use of the translational invariance,
transformation properties in respect to the Lorenz boost:
\bea
&&<0|\phi(x)|\lambda_{\bf p}>=<0|e^{i\hat{P}\cdot x}\phi(0)e^{-i\hat{P}\cdot x}|\lambda_{\bf p}>
\nn \\
&&=<0|\phi(0)|\lambda_{\bf p}>e^{-i\,p\cdot x}|_{p_o=E_p}=<0|\phi(0)|\lambda_0>
e^{-i\,p\cdot x}|_{p_o=E_p}
\eea
and of the integral representation of the step function, one gets for  $x_0>y_0$
\be
\label{repre}
<0|\phi(x)\phi(y)|0>_{x_0>y_0}= i\, \sum_{\lambda}\int\frac{d^4p}{(2\pi)^4}\,
\frac{\exp(-i\,p\cdot (x-y))}{p^2-m^2_{\lambda}+i\epsilon}
\, \left|<0|\phi(0)|\lambda_0>\right|^2 \, .
\ee
Taking (\ref{repre}) together with an analogous expression for an
opposite time ordering gives the Feynman propagator in the form
\bea
 G(x-y)&=&\int d\omega\, \bar{\sigma}(\omega)D(x-y,\omega) \, , \\
\bar{\sigma}(\omega)&=&\sum_\lambda \delta(\omega-m^2_{\lambda})
\left|<0|\phi(0)|\lambda_0>\right|^2 \, , \\
G(p)&=& \int d\omega\, \frac{\bar{\sigma}(\omega)}{p^2-\omega+i\epsilon} \, ,
\label{two_point}
\eea
where $\bar{\sigma}(\omega)$ is a positive  spectral density ({\it the Lehmann function}) and
$D$ stands for free-particle propagator with 'mass' $\sqrt{\omega}$,i.e.
\be
D(x-y,\omega)=\int\frac{d^4p}{(2\pi)^4}\,
\frac{\exp(-i\,p\cdot (x-y))}{p^2-\omega+i\epsilon}\,
\ee
and $D(p,\omega)$ is its Fourier transform.

To sum it up: the derivation of the spectral representation of the
two-point Green's function is based on three assumptions:
\begin{itemize}
\item the Poincare invariance of the theory,
\item existence of the physical states $|\lambda_{\bf p}>$,
\item the completeness relation (\ref{complet}).
\end{itemize}
The physical propagator usually contains the single particle pole and
the continuous part  stemming from the interaction. Therefore we
typically get:
\be
\label{propagator}
G(p)=\int d\omega \bar{\sigma}(\omega)D(p,\omega)
=r\, D(p,m^2)+\int d\omega \sigma(\omega)D(p,\omega) \, ,
\ee
where $D(p,m^2)$ is a free propagator with the physical mass, $r$ is
a residuum at $p^2= m^2$, and $\sigma$ is a nonsingular continuous
part, nonzero above certain kinematical threshold. For the free
scalar field $r=1$ and $\sigma=0$.

\subsection{Analyticity}

\begin{figure}[t]
\begin{center}
\includegraphics[height=6cm]{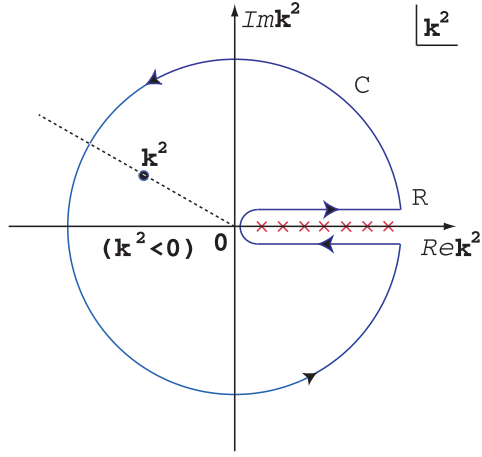}
\caption[99]{\label{planek}  The position of singularities on the complex momentum plane with the sketched integration contour $C$ used in the main text. The singularities can appear only for $Re\, k^2>0$ (crosses).}
\end{center}
\end{figure}

Suppose that the complex function $f(z)$ is analytical in the whole
complex plane of $z\equiv k^2$ except for the positive real axis $z > 0$.
Then we can choose a closed loop $C$ in the complex $k^2$ plane as in
Fig.~\ref{planek}, such that the complex function $f(z)$ is analytical inside
and on the loop $C$ . For a reference point $k^2$ inside $C$, the
Cauchy integral formula tells us that
\bea
  f(k^2) &=& {1 \over 2\pi i} \oint_{C} dz {f(z) \over z-k^2}
\\
  &=&    {1 \over 2\pi i} \int_{0}^{R} dz {f(z+i\epsilon)-f(z-i\epsilon) \over z-k^2}
+ {1 \over 2\pi i} \oint_{|z|=R} dz {f(z) \over z-k^2} \, ,
\label{Cif}
\eea
where we have separated the integral into two pieces: one is the
contribution from the paths above and below the positive real axis;
the other is the contribution to the integral over the circle of
radius $R$.

%%%%%%%%%%%%%%%%%%%%%%%%%%%%%%%%%%%%%%%%%%%%%%%%%%%%%%%%%%%%
%
%
%%%%%%%%%%%%%%%%%%%%%%%%%%%%%%%%%%%%%%%%%%%%%%%%%%%%%%%%%%%%

First, making use of the fact that $f(k^2)$ is real for $k^2<s_{min}$ for some nonnegative $s_{min}$
(by assumption, at least for $k^2<0$, i.e., for spacelike $k$), it
follows from  the Schwartz reflection principle that for $\Re\,
z>s_{min}$ it holds $f(z-i\epsilon)=f^*(z+i\epsilon)$ and hence
$f(z+i\epsilon)-f(z-i\epsilon)=f(z+i\epsilon)-f^{*}(z+i\epsilon) =
2i\, \Im\, f(z+i\epsilon)$. Hence the first term in r.h.s.\ of
(\ref{Cif}) picks up the discontinuity (the imaginary part) of the
function $f(z)$ along the positive real axis.

Next, the circle $|z|=R$ is parameterized as $z=Re^{i\theta}$:
\be
  f(k^2) = {1 \over \pi} \int_{0}^{R} dz\, {\Im\, f(z+i\epsilon) \over z-k^2}
+  \int_{0}^{2\pi} {d\theta \over 2\pi}\,
{Re^{i\theta} f(Re^{i\theta}) \over Re^{i\theta}-k^2} \, .
\ee
The second term on the r.h.s.\ of this equation has the bound
\be
  \Big| \int_{0}^{2\pi} {d\theta \over 2\pi}\,
  {Re^{i\theta} f(Re^{i\theta}) \over Re^{i\theta}-k^2}
  \Big| \le
  \int_{0}^{2\pi} {d\theta \over 2\pi}\,
  {R|f(Re^{i\theta})| \over |Re^{i\theta}- k^2|}
  \le   {\rm max}_{0<\theta<2\pi}|f(Re^{i\theta})|\,
{R \over R-|k^2|} \, .
\ee
Now we wish to take the radius $R$ of the circle to infinity. Thus, if
$|f(z)| \rightarrow 0$ as $|z| \rightarrow \infty$, then the second
term has no contribution as $R \rightarrow \infty$ and we arrive at the
desired Dispersion Relation (DR):

\be
  f(k^2) =  {1 \over \pi} \int_{0}^{\infty} dz\, {\Im\, f(z) \over z-k^2+i\epsilon}
  \, .
\label{dispersionrel0}
\ee
This can also be written as
\be
  f(k^2) =   \int_{0}^{\infty} ds\, {\rho(s) \over s-k^2+i\epsilon} \, ,
  \quad
  \rho(s) := {1 \over \pi} \Im\, f(s) \, ,
\label{dispersionrel}
\ee
 then the SR for propagator
(\ref{propagator})  exactly coincides with the DR
(\ref{dispersionrel}).

To solve the SDEs in Minkowski space we will need also integral representation for the inverse
of propagator. In the renormalizable theories  such spectral functions
 often do not vanish at infinity. This can be already viewed from the identity
\be
\Im \, f(z)^{-1}=-|f(z)|^{-1} \Im \,f(z)\, ,
\ee
where we identified the function $f$ with the propagator function G.  Then one makes the momentum space subtraction(s) and arrives at modified dispersion relations.  For instance, let us define a new function
\be
f_R(\zeta;z)=f(z)-f(\zeta)\, ,
\ee
and assume that $f_R(z)\rightarrow 0$. This gives the once-subtracted
DR:
\be \label{twice}
f_R(\zeta;z)= { (k^2-\zeta) \over \pi} \int_{0}^{\infty} dz\,
 {\Im\, f(z) \over (z-k^2+i\epsilon)(z-\zeta)} \, .
\ee
If the other divergence occurs, one can go on and subtract the
next term in the Taylor expansion of $f(z)$.

\subsection{Dispersion relations in scalar theory}

In this Section we derive the DRs first for the one-loop elastic two
particle scattering amplitude in the scalar $\Phi^4$ theory without
spontaneous symmetry breaking, i.e., in the QFT given by Lagrangian:
\bea
\label{HIGSATO-NESPONTATO}
{\cal L}&=&\frac{1}{2}\partial_{\mu}\Phi\, \partial^{\mu}\Phi -
{\cal V}(\Phi) \, , \\
\nn
{\cal V}(\Phi)&=&\frac{m_{0}^2}{2}\Phi^{2}+g_0\Phi^4
\, ,
\eea
where the global minimum condition $V(\Phi=0)=0$ entails
the zero value of the field vacuum expectation value.

Next, as a less trivial example,  we offer a detailed original 
proof of DR for a two-loop self-energy
function in the (off-shell)  momentum subtraction renormalization
scheme. The calculation is performed for the noninteger
dimension $D=4-\epsilon$, the limit $\epsilon \rightarrow 0$ is taken
after the appropriate subtractions.  However the 
results on imaginary part of the sunset diagram was already obtained several
years ago by the Cutkosky rules approach \cite{BBBB} (see also
\cite{BDR} for general case in  n-dimensions) the derivation presented here
follows an ideas of renormalization realized directly in momentum space. 

\subsection{Scattering amplitude at one skeleton loop}

The scattering matrix in $\Phi^4$ theory is composed from the
four-point Green's function. The one-loop skeleton contribution to
the irreducible GF reads
\begin{fmffile}{froz}
\bea
\Gamma_0^{(4)}(l^2)=
\parbox{2.0\unitlength}{%
\begin{fmfgraph}(2.0,2.0)
\fmfpen{thick}
\fmfleftn{i}{2}
\fmfrightn{o}{2}
\fmf{plain}{i1,v1}
\fmf{plain}{i2,v1}
\fmf{plain}{v2,o1}
\fmf{plain}{v2,o2}
\fmf{plain,left,tension=0.25}{v1,v2}
\fmf{plain,right,tension=0.25}{v1,v2}
\end{fmfgraph}}% end of parbox
=-i\, \frac{(4!g_0)^2}{2}\int \frac{d^dk}{(2\pi)^d}\,
G(k+l)G(k)\, ,
\eea
\end{fmffile}
where $g_0$ is a bare coupling and  $G$ represents fully dressed
propagator (\ref{propagator}). Substituting the Lehmann
representation for $G$ gives
\bea
\Gamma_0^{(4)}(l^2)&=&
 \frac{(4!g_0)^2}{2}\int d\al_1\, \bar{\sigma}(\al_1)\, \int d\al_2\, \bar{\sigma}(\al_2)\,
 I_0(l^2) \, ,
\nn \\
I_0(l^2)&=&\int\frac{d^dk}{(2\pi)^d}\,
D(k+l,\al_1)D(k,\al_2) \, .
\eea
Thus, the derivation of DR reduces to the familiar ``perturbative''
spectral representation of the scalar integral $I_0(l^2)$. The
unrenormalized vertex $\Gamma_0$ requires just one subtraction
 due to the presence of logarithmic UV divergence
\bea \label{jedenlup}
\Gamma^{(4)}(l^2,\zeta_g)&=&
\Gamma_0^{(4)}(l^2)-\Gamma_0^{(4)}(\zeta_g) \, ,
\nn \\
\Gamma^{(4)}(l^2,\zeta_g)&=&\frac{(4!g(\zeta_g))^2}{2}
\int d\al_1\, \bar{\sigma}(\al_1)\, \int d\al_2\, \bar{\sigma}(\al_2)\, I_R(l^2) \, ,
\nn \\
I_R(l^2)&=&\int\limits^{\infty}_{(\sqrt{\al_1}+\sqrt{\al_2})^2}
d\omega\, \frac{(l^2-\zeta_g)X(\al_1,\al_2,\omega)}{(\omega-\zeta_g)(l^2-\omega+\ep)}
\, ,
\eea
where the function $X$ is defined as
\bea   \label{xnula}
&&X(a,b;\omega)=\frac{\lambda^{1/2}(a,b,\omega)}{\omega}\,
\Theta(\omega-(\sqrt{a}+\sqrt{b})^2) \, ,
\nn \\
&&\lambda(x,y,z)=x^2+y^2+z^2-2xy-2yz-2xz \, .
\eea
 and $\zeta_g$ is the (square of) renormalization scale. In order to define the
renormalized coupling $g(\zeta_g)$ we choose the function
$\Gamma^{(4)}(l^2,\zeta_g)$ to be   renormalized  loop contribution of
quartic vertex and include the infinite constant $\Gamma_0(\zeta_g^2)$
into the counter-term part of $g_0$. The full scattering amplitude
$M(s,t,u)$ in the considered approximation is the sum of the infinite
chain of the irreducible contributions defined above:
\be
\label{amplitude}
M(s,t,u)=4!g(\zeta_g)+\sum_{x=s,t,u}\frac{\Gamma^{(4)}(x,\zeta_g)}
{1-\Gamma^{(4)}(x,\zeta_g)} \, ,
\ee
where one common renormalization scale $\zeta_g$ was used to
renormalize GF. $\Gamma$ in each channel characterized by the
Mandelstam variable $ s,t,u$ (labeling incoming (outgoing) momenta
$p_{1,2}$ ($p_{3,4}$), they are $s=(p_1+p_2)^2, t=(p_1-p_3)^2,
u=(p_1-p_4)^2 $). The first term on the l.h.s.\ of
Eq.~(\ref{amplitude}) represents the tree value of the amplitudes
$M$.

Several important remarks can be made about the obtained result.
\begin{itemize}
\item Integrating the single particle contribution we explicitly see that
the vertex becomes complex for the timelike $x>4m^2$ and its
absorptive part reads:
\bea
\label{ctyrato}
\Im\,  \Gamma(l^2)&&=\frac{g^2(\zeta_g)}{16\pi}
\left[r^2\sqrt{1-\frac{4m^2}{\omega}}\, \Theta(\omega-4m^2)\right.
+2r\int_{9m^2}^{\infty} d\al\, \sigma(\al)X(\al,m^2,\omega)
\nn \\
&&+\left.\int_{9m^2}^{\infty} d\al_2\, \sigma(\al_2)\,
\int_{9m^2}^{\infty} d\al_1\, \sigma(\al_1)
X(\al_1,\al_2,\omega)\right] \, ,
\eea
which is explicitly independent on the renormalization scale $\zeta_g$.
\item perturbation theory is trivially reproduced. Assuming $r\simeq 1$ the one-loop result in the perturbation theory  is given by the first term in
(\ref{ctyrato}).
\item The absorptive part does not vanish for large momenta, it is a constant.
For fixed momentum the  naive limit $\zeta_g\rightarrow \infty$
cannot be performed . Taking this explicitly then the original log. divergence 
reappear again. It is impossible to define infinite momentum frame renormalization scheme (the so called Weinberg scheme \cite{WEISCHW}).

\item
The summation
of the irreducible contributions leads to the Landau ghosts in the
amplitudes and the renormalized coupling can not be reasonably
defined without some regulators. This pathology is manifest already
at the one-loop level. This is because the $\Phi^4$ theory is the
(probably the best known) example of the nonasymptotically free theory.
\end{itemize}

\subsection{DR for sunset diagram in momentum subtraction renormalization
scheme}

In the relation (\ref{ctyrato}) knowledge of the Lehmann function
$\sigma$  is required. It can be taken from the perturbation theory or 
 some improved approximation of the full result can be used.  To
this end we continue with derivation of a similar relation for the
two-point Green function. The renormalized inverse propagator is
given by
\be G^{-1}(p^2)=p^2-m^2(\zeta)-\Pi_R(\zeta;p^2) \, ,
\ee
where $\Pi_R$ is the renormalized two-point proper self-energy function: 
\be
\label{subtrakce}\label{disperze1}
\Pi_R(\zeta;p^2)= \Pi(p^2)- \Pi(\zeta) -
\left. \frac{d \Pi(p^2)}{d p^2}\right|_{p^2=\zeta}(p^2-\zeta) \, .
\ee
since the  quadratically divergent self-energy is still logarithmically divergent
even after the ``mass subtraction''. This 'infinity'  proportional to $p^2$ is to be absorbed by the field-strength redefinition. The two loop proper selfenergy function to be renormalized is:
\begin{fmffile}{fsunset}
\bea \label{full}
\Pi_0(p^2)&=&
\parbox{2.0\unitlength}{%
\begin{fmfgraph}(2.0,1.5)
\fmfpen{thick}
\fmfleft{i}
\fmfright{o}
\fmf{plain}{i,v1}
%\fmfblob{0.1w}{v1}
\fmfv{d.sh=diamond,d.f=hatched,d.si=0.15w,label=$\Gamma_4$}{v1}
\fmf{plain}{v2,o}
\fmf{plain,left,tension=0.2}{v1,v2}
\fmf{plain,tension=0.2}{v1,v2}
\fmf{plain,right,tension=0.2}{v1,v2}
\end{fmfgraph}}=
\nonumber \\
&-&\frac{4!g}{6}\int
\frac{d^dk_1}{(2\pi)^d} \int \frac{d^dk_2}{(2\pi)^d}\, \Gamma^{[4]}\, G(k_1)
G(k_1+k_2) G(k_2+p) \, .
\eea
\end{fmffile}
Here, it is taken in the skeleton sunset
approximation: the corrections to the vertex are neglected and the
full quartic is replaced by its tree value $\Gamma^{[4]}_{tree}=4!g$.

The direct consequence of the subtraction procedure
(\ref{subtrakce}) is  the DR formula for the   renormalized
self-energy
\begin{equation}
\label{disperzato}\label{cojeco}
\Pi_R(\zeta;p^2)=\int d\omega\, \frac{(p^2-\zeta)^2\rho(\omega)}{(\omega-\zeta)^2
(p^2-\omega+\ep)} \, .
\end{equation}
Our task here is to find the weight function $\rho(\omega)$ for the
skeleton sunset approximation considered. As in the previous one-loop
case it is convenient to drop all prefactors and thus to perform
effectively the subtraction (\ref{subtrakce}) at the level of ``bare
propagators'':
\bea
\label{bare}
\Pi_0(p^2)&=&-\frac{(4!g)^2}{6}\int d\al_1\, \bar{\sigma}(\al_1) \int d\al_2\,
\bar{\sigma}(\al_2) \int d\al_3\, \bar{\sigma}(\al_3)\, J(p^2) \, ,
\nn \\
J(p^2)&=&\int \frac{d^dk_1}{(2\pi)^d}
\int \frac{d^dk_2}{(2\pi)^d}\,
D(k_1,\al_1)D(k_1+k_2,\al_2)D(k_2+p,\al_3) \, .
\nn \\
\eea
Our goal is to cast the function $J(p^2)$ into the integral over some
spectral variables, i.e., to manipulate integrals so that
integrations over momenta can be taken. While the one loop case (\ref{jedenlup}) proceed standard textbook derivation (note, the details of the derivation are provided in the Section 3, follow the relations (\ref{SELF}) to (\ref{PP}))
 the derivation the two-loop derivation is less trivial. We do it for this particular
example in detail in the Appendix A, here we present the result firstly published in 
\cite{BBBB},\cite{BDR}. The absorptive part of the integral $J(\omega)$ reads
\be
\rho_J(\omega)=\frac{1}{(4\pi)^4}
\int\limits_{(\sqrt{\al_1}+\sqrt{\al_2})^2}^{(\sqrt{\omega}-\sqrt{\al_3})}ds
\frac{ \sqrt{\lambda(s,\omega,\al_3)} \sqrt{\lambda(\al_2,s,\al_1)}}{s\rm\omega}
\, \Theta\left(\omega-(\sum_{i=1}^{3}\al_i^{1/2})^2\right) \, .
\ee
%}
%
which leads to the full expression for the spectral function in  DR (\ref{disperzato}) for sunset skeleton graph:
\bea \label{enda} \frac{\Im\,
\Pi_{R}(p^2)}{\pi}&=&\frac{(4!g)^2}{6(4\pi)^4} \left(\prod_{i=1}^{3}
\int d\al_i \bar{\sigma}(\al_i)\right)
\int^{(\sqrt{p^2}-\sqrt{\al_3})^2}_{(\sqrt{\al_1}+\sqrt{\al_2})^2}ds
\nn \\
&&\frac{\sqrt{\lambda(s,p^2,\al_3)\lambda(\al_2,s,\al_1)}}{s\,p^2}\,
\Theta\left(p^2-(\sum_{i=1}^{3}\al_i^{1/2})^2\right) \,.
\eea
In massive $\phi^4$ the spectral function is nonzero when $p^2>9m^2$, otherwise it is zero.

\subsection{Perturbation Theory Integral Representation in renormalizable theories}

The Perturbation Theory Integral Representation (PTIR) \cite{NAK1971}
is a natural extension of the Lehmann representation for the
propagator to  $n-$point Green's functions. The PTIR allows to
express the amplitudes as integrals over weight functions and a
momenta-dependent denominator with known singularity structure. Since
the Feynman parametric integral always exists for any Feynman diagram
defined by the perturbation theory, one can always define an integral representation
such that the number of independent integration parameters is equal
to that of invariant squares of external momenta. The general
expression for $n$-point amplitudes, which is considered to be a sum
of all Feynman diagrams with $n$ external legs, has been derived by
Nakanishi \cite{NAK1971}  for scalar theories with nonderivative
couplings. The PTIR is unique for unrenormalized amplitudes. The
general proof of the Uniqueness Theorem can be also found in the
Nakanishi's book \cite{NAK1971}. The renormalized 2,3,4-point
vertices can differ only by the presence of subtraction polynomials
as follows from renormalization procedure. The coefficients of the
subtraction polynomials then depend on the renormalization scheme
employed.

For theories with interactions containing field derivatives or
theories with particles of nonzero spin similar results to all order
of perturbation theory have never been proved, since tensor and spinor structure of
corresponding expressions becomes rather complicated with increasing
number of external legs of Feynman diagrams. We do not feel too
discouraged by this: the (generalized) PTIR can always be deduced for
a given Feynman diagram. Indeed, if we  assume the analyticity of the
lowest point Green's function -the propagators- then it is 
 sure that the PTIR is derivable for skeleton diagram under consideration. 

In what follows we briefly review examples of PTIR for scalar
amplitudes. We start with  PTIR  for the two-point correlation
function $\Pi(p^2)$, already discussed in detail above. Recall that
the expression for unrenormalized self-energy reads
\be \Pi_0(p^2)=\int dx\, \frac{\rho(x)}{p^2-x+i\epsilon} \, .
\ee
The renormalized $\Pi_R(p^2)$ 
turns out to be
\be  \label{cobylopom}
\Pi_R(p^2)=a+bp^2+\int dx \frac{\rho(x)(p^2-\zeta)^2}
{(p^2-x+i\epsilon)(x-\zeta)^2}  \, ,
\ee
where the finite coefficients $a, b$ are characteristic of a given
but arbitrary renormalization
scheme. Their size  indicates how a given scheme employed differs from the
 momentum subtraction renormalization
scheme where $a, b=0$, since the coefficients
\be
\int dx\, \frac{\rho(x)}{\zeta-x+i\epsilon}\quad;
\quad\frac{d}{dp^2}\left.\left[\int dx\,
\frac{\rho(x)}{p^2-x+i\epsilon}\right]\right|_{p^2=\zeta}
\ee
are fully absorbed into the Lagrangian counter-terms in this scheme.

For the  three-leg vertex function $\Gamma(p_1,p_2)$ analysis of
contributing Feynman diagrams leads to the PTIR derived by Nakanishi:
\be
\label{ptirtri}
\Gamma(p_1,p_2)=\int\limits_0^{\infty}d\alpha\prod_{i=1}^3\int\limits_0^1dz_i\,
\delta(1-\sum_{i=1}^3z_i)\,
\frac{\rho_3(\alpha,\vec{z})}{\alpha-\left(z_1p_1^2+z_2p_2^2+z_3p_3^2\right)-i\epsilon}
\, ,
\ee
where the momenta are conserved $p_1+p_2+p_3=0$, the invariant
squares are independent and the single weight function
$\rho_3(\alpha,\vec{z})$ is sufficient to describe the sum of all
relevant Feynman diagrams.

Further  analysis of  this vertex function gives various constrains
on the integration regions ${\alpha,\vec{z}}$, dependent on masses of
the particles. For instance,  for the $\Phi\Phi\phi$ interaction and
$m$ ($\mu$) being the mass of particle $\Phi$ ($\phi$),
Eq.~(\ref{ptirtri}) reduces to the following two-variable spectral
representation:
\be
\label{ptirtri2}
\Gamma(p,P)=\int\limits_0^{\infty}d\alpha\int\limits_{-1}^1
dz\frac{\rho(\alpha,z)}{\alpha-(p+z\frac{P}{2})^2-i\epsilon}
\ee
in the region $0<P^2<4m^2$, $P=p_1+p_2$, $p=p_1-p_2$. The spectral
function  $\rho(\alpha,z)$  is a positive regular function of the
spectral variables for the fixed value of $P^2$. Furthermore, if the
internal lines with momenta $p_1$ and $p_2$ correspond to the
particle $\Phi$ and the third external line corresponds to the
particle $\phi$, the following support of $\rho(\alpha,z)$ can be
derived:
\be
\rho(\alpha,z)=0, \quad \mbox{unless} \quad \alpha>
\left(m+\mu-(1-|z|)\frac{\sqrt{P^2}}{2}\right)^2 \quad.
\ee
This two variable representation was first proposed in an axiomatic
field theory \cite{DGR1959},\cite{IDA1960} and successfully used  to
solve the Bethe-Salpeter equation in various approximations
\cite{KUSIWI1997},\cite{SAUADA2003}. The paper \cite{FEL1960} solves
both the bound state equation and the scattering problem and applies
this two variable representation to QED even in the nonperturbative
regime.

The scattering matrix $M(p,q;P)$ describes the process
$\Phi_1\Phi_2\rightarrow \Phi_3\Phi_4$, where $p=q_1-q_2$ and
$q=q_3-q_4$ are the initial and final relative momenta, respectively,
and $P$ is the total four-momentum. It is given by the infinite
series of Feynman diagrams.  The full renormalized scattering matrix
can formally be written as \cite{NAK1971}:
\bea
\label{fullscat}
M(p,q;P)&=&\int\limits_0^{\infty} d \gamma\,
\int_{\Omega} d\vec{\xi}\, \left\{\frac{\rho_{st}(\gamma,\vec{\xi}\,
)}
{\gamma-\left[\sum_{i=1}^4\xi_iq_i^2+\xi_5s+\xi_6t\right]-i\epsilon}\right.
\nn \\
&+&\frac{\rho_{tu}(\gamma,\vec{\xi}\, )}
{\gamma-\left[\sum_{i=1}^4\xi_iq_i^2+\xi_5t+\xi_6u\right]-i\epsilon}
\nn \\
&+&\left.\frac{\rho_{us}(\gamma,\vec{\xi}\, )}
{\gamma-\left[\sum_{i=1}^4\xi_iq_i^2+\xi_5u+\xi_6s\right]-i\epsilon}\right\} \, ,
\eea
where $q_i^2$ is the square of the four-momentum carried by $\Phi_i$
and $s,t,$ and $u$ are the Mandelstam variables. Since only six of
these variables are independent (due to the relation
$q_1^2+q_2^2+q^3_2+q_4^2=s+t+u$), this is also the number of
independent $\xi-$parameters. Hence, one only has to introduce the
``mass'' parameter $\gamma$ and six dimensionless Feynman parameters
$\xi_i$ with one constraint. The symbol $\Omega$ denotes the
integration region of $\xi$ such that $\Omega=\{\xi|\, 0<\xi_i<1,
\sum \xi_i=1 (i=1,...,6)\}$.  The function $\rho_{ch}$ gives the
weight of the spectrum arising from the different channels which can
be denoted by $ch=\{st\},\{tu\},\{us\}$. Since any Feynman diagram
for the scattering matrix can be written in this form, it must be
also true for their sum. From the corresponding SDEs of considered
theory it simply follows that PTIR for $M$ must contain the weights
of lower point Green's functions. For instance, the simplest
contribution from $t$-channel is given by a free propagator
\be
M(p,q;P)=\frac{g^2}{m_{exch}^2-(p-q)^2-i\epsilon} \, ,
\ee
and the corresponding weight is simply a product of six delta
functions.

The first topologically less trivial diagrams are various boxes and
crossed-boxes.  Collecting various couplings to the constant prefactor $C$ then 
the two-variable
expression for crossed box diagram was derived some time ago by
Mandelstam \cite{MAN1959}:
\bea
I_{cross}&=&C\int_{4\mu^2}^{\infty}\int_{4m^2}^{\infty}
\frac{dt'}{t'-t-i\epsilon}\frac{du'}{u'-u-i\epsilon}
\frac{\theta(\kappa(u',t'))}{\sqrt{\kappa(u',t')}} \, ,
\nn \\
\kappa(u',t')&=&u't'\left[(t'-4\mu^2)(u'-4m^2)-4\mu^4\right] \, ,
\eea
where $\mu$ is mass of the exchanged particle $\phi$ (its propagators
cross inside the diagram), while $m$ is a mass of the scattered
particle (emitting and absorbing  $\phi$). Here, the function
\be
\sigma(u',t')= C\, \frac{\theta(\kappa(u',t'))}{\sqrt{\kappa(u',t')}} \, ,
\ee
is called the double spectral function. It can be easily converted to
the PTIR form:
\be
I_{cross}=\int\limits_{0}^{\infty}d\alpha\int\limits_0^1 dz\,
\frac{\rho_{ut}(\alpha,z)}{\alpha-zu-(1-z)t} \, ,
\ee
with the help of the Feynman identity.

%%%%%%%%%%%%%%%%%%%%%%%%%%%%%%%%%%%%%%%%%%%%%%%%%%%%%%%%%%%%%%%%%%%%%%%%%%%%%%%%%%%%%%%%%%%%%%%%%%%%%%%%%%%%%%%%%%%%%%%%%%%%%%%%%%%%%%%%%%%%%%%%%%%%%%%%%%%%%%%%%%%%%%%%%%%%%%%%%%%%%%%%%%%%%%%%%%%%%%%%%%%%%%%%%%%%%%%%%%%%%%%%%%%%%%%%%%%%%%%%%%%%%%%%%%%%%%%%%%%%%%%%%%%%%%%%%%%%%%%%%%%%%

\section{SDEs and Unitary Equations in the scalar toy model}
\label{chap5}

This Section should serve as a pedagogical introduction to the more general  strategy  
 of  Unitary Equations. As we will see, using the spectral decomposition of GFs in  SDEs allows to derive a real nonsingular integral equation for a real spectral function $\sigma$.
For this purpose we consider a simple quantum model describing the self-interaction of one real spinless field $\phi$. The Lagrangian density for our toy model  reads
\be \label{lagr}
{\cal L}=\frac{1}{2}\partial_{\mu}\phi_0(x)\partial^{\mu}\phi_0(x)-
\frac{1}{2}m_0^{2}\phi_0^{2}(x)-g_0\phi_0^{3}(x)+ \lambda_0\phi_0^{4}(x)+... \, ,
\ee
where the subscript $0$ indicates the unrenormalized  quantities and the dots mean the neglected interactions with neglected field content.

The main philosophy of the derivation does not depend on the details of a given model.
The method can be directly applied to all theories  where the dispersion relation approach is accessible. We also  explain and review the spectral technique used in the calculations of relativistic bound states in the subsequent Section.
For this purpose we follow the main ideas of the paper \cite{SAUADA2003} where the (massive and gauged) Wick-Cutkosky model and its response on electromagnetic interaction was considered. Furthermore, we not only exhibit the appropriate derivation of Unitary Equations but we also present the numerical solution and we offer the comparison with the results obtained via usual Euclidean formalism.

Using the diagrammatic representation for the selfenergy the SDE for scalar propagator $G$ can be depicted as:
\be
\label{USDE}
G_0^{-1}(p^2)=p^2-m_0^2-\Pi_0(p^2) \, ,
\ee
\begin{fmffile}{fskalar}
\be  
\Pi_0(q^2)=
\parbox{2.0\unitlength}{%
\begin{fmfgraph}(2.0,2.0)
\fmfpen{thick}
\fmfleft{i}
\fmfright{o}
\fmf{plain}{v,v}
\fmf{plain}{i,v,v,o}
%\fmf{plain,left=90}{v,v}
%\fmfdot{v}
\end{fmfgraph}}
\, \, + \, \,  
\parbox{2.0\unitlength}{%
\begin{fmfgraph}(2.0,2.0)
\fmfpen{thick}
\fmfleft{i}
\fmfright{o}
\fmf{plain}{i,v1}
\fmfblob{0.1w}{v1}
\fmf{plain}{v2,o}
\fmf{plain,left,tension=0.25}{v1,v2}
\fmf{plain,right,tension=0.25}{v1,v2}
\end{fmfgraph}}% end of parbox
\, \, +\, \, 
\parbox{2.5\unitlength}{%
\begin{fmfgraph}(2.5,2.0)
\fmfpen{thick}
\fmfleft{i}
\fmfright{o}
\fmf{plain}{i,v1}
%\fmfblob{0.1w}{v1}
\fmfv{d.sh=diamond,d.f=hatched,d.si=0.15w,l=$\Gamma_4$}{v1}
\fmf{plain}{v2,o}
\fmf{plain,left,tension=0.2}{v1,v2}
\fmf{plain,tension=0.2}{v1,v2}
\fmf{plain,right,tension=0.2}{v1,v2}
\end{fmfgraph}}
\, \,  +  \, \,  
\parbox{2.5\unitlength}{%
\begin{fmfgraph}(2.5,2.0)
\fmfpen{thick}
\fmfleft{i}
\fmfright{o}
\fmf{plain}{i,v1}
\fmfblob{0.1w}{v2}
\fmfblob{0.1w}{v3}
\fmf{plain}{v3,o}
\fmf{plain,left,tension=0.2}{v1,v3}
\fmf{plain,left,tension=0.1}{v1,v2}
\fmf{plain,right,tension=0.1}{v1,v2}
\fmf{plain,right,tension=0.2}{v2,v3}
\end{fmfgraph}}% end of parbox
\nonumber \, ,
\ee
\end{fmffile}
where the full irreducible vertex
functions $\Gamma_0^{[3]}$ ($\Gamma_0^{[4]}$) are represented by blobs 
(diamond) and they satisfy their own SDEs (for the derivation of SDEs form the effective action see some standard textbook,e.g. \cite{ITZYKSON}, but also \cite{RIVERS}).  All the internal lines stand for $G$ ,i.e. for fully dressed propagators. For the  explanatory simplicity 
we neglect the two loops diagram,  how to  take into account the last two diagrams will be explained  at the end of this Section. In our simple approximation  the selfenergy to be considered is given by the first two diagrams above. Their contribution is: 
\be
\Pi_0(p^2)=i\lambda\int\frac{d^4q}{(2\pi)^4}G_0(q)+i3g_0 \int\frac{d^4q}{(2\pi)^4}\Gamma_0^{[3]}(p-q,q)G_0(p-q)G_0(q) \, .
\ee
Furthermore, we approximate the trilinear and quartic vertices  by their tree value: $\Gamma^{[3]}=3!g$ and $\Gamma^{[4]}=4!\lambda$ (note, contrary to the usual convention used in QED, we very naturally include the coupling constants into the definition of proper GFs). In such approximation the quartic interaction $\lambda\Phi^4$ as well as the field  do not need renormalization since we have neglected higher diagrams. Thus the divergent integral defining $\Pi_0$ requires only  the mass renormalization:
\be \label{jednasubstrakce}
\Pi_R(\zeta;p^2)=\Pi_0(p^2)-\Pi_0(\zeta) \, ,
\ee
where the   divergence (quadratic  here)  is absorbed to the counterterm part of the Lagrangian:
\be
m_0^2=m^2(\zeta)+\delta m^2 \quad , \quad \delta m^2=\Pi_0(\zeta).
\ee

The renormalized equation for the propagator then  reads 
\bea \label{jetonut}
G^{-1}(p^2)&=&p^2-m(\zeta)^2-\Pi_R(\zeta;p^2) \, ,
\nn \\
 \Pi_R(\zeta;p^2)&=&\Pi(p^2)-\Pi(\zeta)\, ,
\label{yanez} \\
\Pi(p^2)&=&i3g \int\frac{d^4q}{(2\pi)^4}\Gamma^{[3]}G(p-q)G(q) 
\label{SDE2}\, ,
\eea
where we have omitted the seagull diagram contribution in expression for $\Pi(p^2)$ since it exactly vanishes after the subtraction and where we simply take $G_0(p^2)=G(p^2)$ since  
the field strength renormalization constant $Z=1$ here (for  more complete discussion of the multiplicative renormalization in scalar models see the paper \cite{SAULIPHD}).

The key point of our spectral technique is the derivation of the Unitary Equations.
The Unitary Equations relate the imaginary part of the propagator -the spectral function $\sigma$- with the imaginary and the  real part of  proper GF (this is the reason why we call them the Unitary Equations). 
Dealing with explicitly  massive model here,  the generic spectral
decomposition of the renormalized propagator (\ref{propagator}) is used, where
the spectral function of the stable  particle is
\be
\label{propa2}
\bar{\sigma}(\alpha)=r\delta(m^2-\alpha)+\sigma(\alpha) \, .
\ee
When propagator SR is used in the equation for $\Pi$ Eq.~ (\ref{SDE2}) it  leads to the DR
for the renormalized selfenergy  $\Pi_R(\zeta;p^2)$. 
Its form  is  well known from the perturbation theory treatment and is given by the DR already discussed in the  Section 3.  However here, for purpose of convenience of the reader and in order to make our explanation  self-contained, we also give  the 
appropriate derivation here.

Substituting   SR for propagators  into (\ref{jetonut}) the
selfenergy  $\Pi$ reads
\bea
\label{SELF}
\Pi(p^2)&=&18g^{2}
\int d\alpha d\beta
\bar{\sigma}(\alpha)\bar{\sigma}(\beta)
I(p^2)\, ,
\\ 
I(p^2)&=&i\int \frac{d^4l}{(2\pi)^4}
\frac{1}{((p+l)^{2}-\alpha+i\epsilon)(l^{2}-\beta+i\epsilon)}\, .
\nn
\eea
Clearly $\Pi_R(\zeta;p^2)$ is given by the integral over the analogically defined  
$I_R(\zeta;p^2)=I(p^2)-I(\zeta)$. Using the Feynman parameterization $I_R$ can be written as
%
%{ \small
\bea
I_R(\zeta;p^2)&&=\int_0^1dx\int \frac{d^4l}{(2\pi)^4}\left\{
\frac{i}{\left[l^2+p^2x(1-x)-\al(1-x)-\beta x+\ep\right]^2}
-(p^2\rightarrow \zeta)\right\}
\nn \\
&&=\int \frac{d^4l}{(2\pi)^4}\int\limits_0^1
\frac{-2i\,dx\,dz\,(p^2-\zeta)x(1-x)}{\left[l^2+(p^2-\zeta)x(1-x)z
+\zeta x(1-x)-\al(1-x)-\beta x+\ep\right]^3}
\nn \\
&&=\int_0^1dx\frac{dz}{z}\frac{(4\pi)^{-2}(p^2-\zeta)}{p^2-\Omega+\ep}\, ,
\eea
%}
%
where we have labeled $\Omega=\zeta+\frac{-\zeta+\al/x+\beta/(1-x)}{z}$.
Making a substitution $z\rightarrow\omega$ such that $\om=\Omega$, it yields
\be
I_R(\zeta;p^2)=\frac{1}{(4\pi)^2}\int_0^1dx\int_{\frac{\al}{x}+\frac{\beta}{1-x}}^{\infty}
\frac{p^2-\zeta}{(p^2-\omega+\ep)(\omega-\zeta)} \, .
\ee
Changing the ordering of the integrations (note, the variables $\alpha,\beta$ are positive) and putting $I$ into $\Pi$ we obtain the desired DR:
\be   \label{labuan}
\Pi_R(\zeta;p^2)= \int_{0}^{\infty}d \om 
\frac{\rho(\om)\left[\frac{p^2-\zeta}{\om-\zeta}\right]^n}{p^{2}-\om+i\epsilon}\, ,
\ee
where $n=1$ here (while $n=2$ when the field renormalization enters the calculation and the second subtraction is required, here we use $n$ for general purpose) and where the absorptive part
$\pi\rho$ is given as
\bea \label{PP}
\rho(\omega)&=&C_g
\int d\alpha d\beta \bar{\sigma}(\alpha)\bar{\sigma}(\beta)
\frac{\lambda^{1/2}(\alpha,\beta,\omega)}{\omega}
\Theta(\omega-(\sqrt{\alpha}+\sqrt{\beta})^2)
\nn \\
&=&C_g r^2\sqrt{1-\frac{4m^2}{\omega}}\Theta(\omega-4m^2)
+\, .\, .\, .\, ,
\eea
where $\lambda$ is the Khallen triangle function defined in (\ref{xnula}) and
the dots means the terms with one or two $\sigma$'s.
(Thorough this paper we always use the letter $\rho$ to denote the weight function that appears in selfenergy  in order to carefully  distinguish from the spectral function 
$\sigma$ , i.e. from $\Im G(p^2)/\pi$).  $C_g$ is used to label  the coupling strength defined as $C_g=\frac{18g^2}{(4\pi)^2}$.

The derivation of Unitary Equations is achieved by using  the functional identity 
\be \label{functional}
\frac{1}{x'-x+\ep}=P. \frac{1}{x'-x}-i\pi\delta(x'-x)\, ,
\ee
where  $P \cdot$ denotes the principal value integration.
$GG^{-1}=1$ can be written as:
\be \label{stuin}
\left[\frac{r}{p^2-m^2}+\int_{4m^2}^{\infty} d \alpha\,
\frac{\sigma(\alpha)}{p^2- \alpha- i \epsilon}
\nn \right]\left[p^2-m^2(\zeta)-\int d \om 
\frac{\rho(\om)\left[\frac{p^2-\zeta}{\om-\zeta}\right]^n}{p^{2}-\om+i\epsilon}\right]=1\, ,
\ee
we arrive at the equations between the spectral functions $\sigma$ and $\rho$ and their principal values integral. We eliminate the unpleasant (since unknown) principal value integral over the function $\sigma$ from the equation for the real part
\bea \label{realp}
P.\int d\omega \frac{\sigma(\om)}{p^2-\om}=\frac{1-\pi^2\rho(p^2)\sigma(p^2)}{b(p^2)}-
\frac{r}{p^2-m^2}\, ,
\\ \nn
b(p^2)\equiv \Re G^{-1}(p^2)= p^2-m^2(\zeta)-P.\int d \om 
\frac{\rho(\om)\left[\frac{p^2-\zeta}{\om-\zeta}\right]^n}{p^{2}-\om}\, .
\eea
The equation for imaginary part of Eq. (\ref{stuin}) can be written as
\be \label{imp}
\sigma(p^2)b(p^2)-
\rho(p^2)\left(\frac{r}{p^2-m^2}+
P.\int d\omega \frac{\sigma(\om)}{p^2-\om}\right)=0\, .
\ee
Substituting (\ref{realp}) into the Eq.~(\ref{imp}) we arrive at
the desired Unitary Equation:
\be
\label{symb}
\sigma(p^2)=\frac{\rho(p^2)}{b^2(p^2)+\pi^2\rho^2(p^2)}\, ,
\ee
 as follows from the rel. (\ref{PP}), the continuous spectral function  $\sigma$
starts to be smoothly nonzero from the perturbative 
threshold $4m^2$.
Note that the function $b$ still contain the principal value integral
this can evaluated numerically \cite{SAULIPHD} igeneral, 
but in simple one loop case  here it is known analytically.
The form of the Eq. (\ref{symb}) is based only on  the analytic property of the propagator, it does not depend on a details of  interaction. 
All the dynamical information is contained in the 
expression for $\rho$ which follows from the SDEs solution here, or from the usual Feynman diagram in the case of  perturbation theory  treatment.   
At the end we should mention that the derivation of Unitary Equations has been performed for momentum subtraction scheme only  for simplicity. 
It should be stressed here, that the extension to any other renormalization  scheme is very straightforward. Remind the changes in  DR which arise  due to  finite change in the  field renormalization constant $Z$. It would be  reflected by the presence of nontrivial subtracting polynom before the DR~(\ref{cobylopom}). This can be easily incorporated into the Unitary Equations.

If the pole is assumed in the propagator then  the knowledge of the pole  mass $m$
is required in the presented spectral treatment. From its definition $G^{-1}(m^2)=0$ we can obtain 
the following equation
\be  \label{klokan}
m^2=m^2(\zeta)+\int d\omega\frac{\rho(\omega)}{m^2-\omega}
\left[\frac{m^2-\zeta}{\omega-\zeta}\right]^n \, .
\ee
The propagator residuum can be obtained by the on-shell differentiation:    
\be \label{rezpi}
r=\lim_{p^2\rightarrow m^2}
\frac{p^2-m^2}{p^2-m^2(\zeta)-\Pi(\zeta;p^2)}
=\frac{1}{1+\int d\om\frac{\rho(\om)}{(\om-m^2)^2}}.
\ee

\subsection{Comparison with the Euclidean solution}

It is instructive to compare our Minkowski-space result with the one obtained in  Euclidean approach. For this purpose we consider the propagator SDE truncated and renormalized as before.

After the Wick rotation and angular $\beta$, $\phi$ integrations
of the four-dimensional sphere
\be
\int d\Omega_4=\frac{1}{2\pi^2}
\int_0^{\pi} d\theta \sin^2(\theta)\int_0^{\pi}
d\beta \sin(\beta)\int_0^{2\pi}d\phi \, ,
\ee
the renormalized SDE for propagator becomes
\be
\label{euklidak}
\Pi(x)=m^2(0)-\frac{2C_g}{\pi}\int_0^{\infty}dy\int_0^{\pi}
d\theta\, \frac{\sin^2(\theta)y}{y+\Pi(y)}\,
\left[\frac{1}{z+\Pi(z)}-\frac{1}{y+\Pi(y)}\right]\, ,
\ee
where we made subtraction at zero renormalization scale
$\zeta=0$. For purpose of brevity  we have defined $\Pi(y)=\Pi_R(y=p^2_E;\zeta=0)+
m^2(0)$,  variables $x,y,z$ are squares of Euclidean  momenta such that
$z=x+y-2\sqrt{xy}\cos \theta$

In our numerical treatment we put $m^2(0)=1$ in arbitrary units.
The coupling strength $C_g$ is then scaled through the  dimension-full coupling constant  $g$ which is taken in the units of $m(0)$. 

\begin{figure}
\centerline{\epsfig{figure=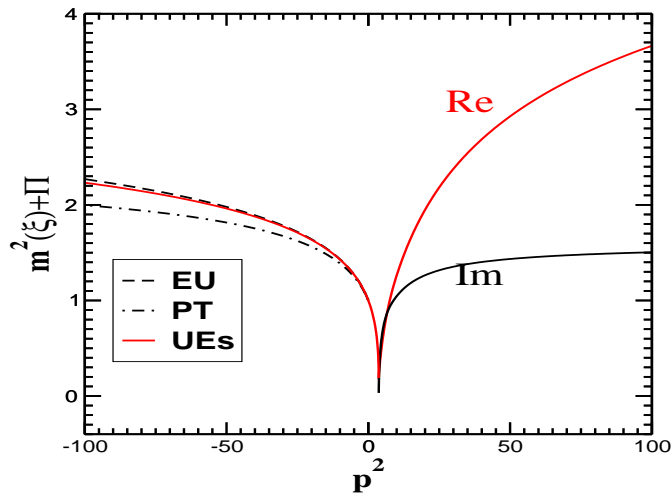,width=8.0truecm,height=14.0truecm,angle=270} }
\caption[99]
{\label{sandokan} The renormalized mass plus  selfenergy for the coupling strength $C_g=\frac{18g^2}{(4\pi)^2}=
0.5m^2(0)$. Minkowski result is labeled by Unitary Equations, EU stands for  the solution of Euclidean SDE, PT means  one loop perturbation theory. }
\end{figure}

Within  the same choice of renormalization scale $\zeta$ and the same choice of renormalized mass
the Minkowski problem has been solved to. The resulting 'total selfenergy' function 
$\Pi_R(p^2;0)+m^2(0)$ is plotted in Fig.~\ref{sandokan} for the coupling strength
$C_g=0.5m^2(0)$.  The Unitary Equations as well as the Eq. 
(\ref{euklidak}) has been solved by the iterations without some peculiar troubles.
We use Gaussian quadrature method for the integration. 
Note only that in order to calculate $\Pi(z_i)$ in (\ref{euklidak}) we need to interpolate (extrapolate) between the 'fitted' points $\Pi(x_i)$. To achieve good numerical accuracy in the 
Minkowski approach we have found it is necessary to use relatively large number of mesh points. This is a consequence of the presence of Heaviside step functions in the kernel and we take 800 of mesh points in the case of figure's calculation. The value of obtained physical mass is $m^2=0.92m^2(0)$ for $C_g=0.5m^2(0)$. Furthermore, we compare with the one loop perturbation  theory. These results are calculated with the physical pole mass as it has been determined from Unitary Equations. Note that determining pole mass from Eq.~(\ref{klokan}) in a framework of perturbation theory would lead to the small negligible shift of pole mass and threshold as well.

\subsection{The relativistic bound-states}
%CCCCCCCCCCCCCCCCCCCCCCCCCCCCCCCCCCCCCCCCCCCCCCCCCCCCCCCCCCCCCCCCCCCCCCCCc
%%%%%%%%%%%%%%%%%%%%%%%%%%%%%%%%%%%%%%%%%%%%%%%%%%%%%%%%%%%%%%%%%%%%%%%%%%%%%%%%%%%%%

\begin{figure}[t]
\centerline{  \mbox{\psfig{figure=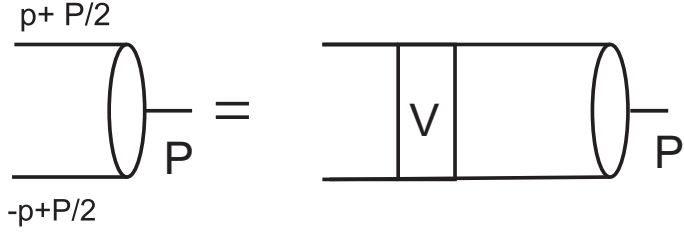,height=3.0truecm,angle=0}} }
\caption[99]{\label{figBSE} Diagrammatic representation of the Bethe-Salpeter equation for the bound state vertex function.}
\end{figure}

In  quantum field theory the two body bound state is described by 
the  bound state vertex function or, equivalently, by 
Bethe-Salpeter  amplitudes, both of  them are solutions of the 
corresponding (see Fig.~\ref{figBSE}) covariant four-dimensional 
Bethe-Salpeter equation \cite{BETSAP1951}.  In the 
so-called ladder approximation the  scattering matrix is given  by the 
sum of the generated ladders.   
When the more realistic models are considered  (hadronic physics for example)   
 one  is forced to use dressed correlation functions instead of their 
bare approximations. 

Here we are considering simple super-renormalizable model with cubic interaction
 $\phi_i^2\phi_3\,;i=1,2$
of three massive fields ( thus this model can be called massive Wick-Cutkosky model),
 where in the approximation that allows us  to exploit the results of previous Section. 
Having the propagators calculated
within  the bare vertex truncation of the SDEs we combine them with the
dressed ladder approximation for the scalar s-wave bound state 
amplitudes. To do this explicitly we follow the treatment described in \cite{SAUADA2003}
where the  spectral technique was used to obtain the accurate results 
 directly in Minkowski space.
In the mentioned paper the   analytic formula has been derived 
for the resulting equation which significantly simplify  the numerical 
treatment. For the extensive but certainly not exhaustive review of up to date 
known or used methods in scalar models
see also \cite{SAUADA2003}.

For our purpose we utilize the PTIR already reviewed in the  Section 3.
Very similar to the SDEs treatment reviewed in the  previous Sections,
 here the  Bethe-Salpeter equation written in momentum space is converted into a 
real integral equation for a real weight function.  
 This then allows us to treat the ladder Bethe-Salpeter equation in which all 
propagators (of constituents and of the exchanged  particle) are fully 
dressed. This is achieved by the implementation of  SR of the propagator
which  is determined by the appropriate set of SDEs. Having solved equations for  spectral functions, one can easily determine 
the Bethe-Salpeter  amplitudes in an arbitrary reference frame.

\subsection{Solution of Bethe-Salpeter equation in Minkowski space}

The Bethe-Salpeter  amplitude for bound state $(\phi_1,\phi_2)$ in momentum space is 
defined through the Fourier transform of 
\be
\langle 0|T\phi_1(x_1)\phi_2(x_2)|P\rangle
= e^{-iP\cdot X}
\int{d^4p\over (2\pi)^4} e^{-ip\cdot x} \Phi(p,P)\ ,
\ee
where $X\equiv \eta_1x_1+\eta_2x_2$ and $x\equiv x_1-x_2$, so that 
$x_1=X+\eta_2x$, $x_2=X-\eta_1x$. Here $p_{1,2}$ are the  four-momenta 
of particles corresponding to the fields  $\phi_{1,2}$ that constitute 
the bound state  $(\phi_1,\phi_2)$.  The total and relative momenta are 
then given as $P= p_1 +p_2 $ and $ p=(\eta_2p_1-\eta_1p_2)$, 
respectively, and $P^2=M^2$, where $M$ is the  mass of the bound state. 
Finally, $P\cdot X + p\cdot x = p_1\cdot x_1 + p_2\cdot x_2$. From now 
on we will put $\eta_1= \eta_2= 1/2$ for simplicity. 
 
Introducing the scalar Bethe-Salpeter  vertex function $\Gamma=iG_1^{-1}G_2^{-1} \Phi $, 
the homogeneous  Bethe-Salpeter equation for a s-wave bound state reads 
\be  \label{bse}
\Gamma(p,P) = i\int{d^4k\over(2\pi)^4}\,
V(p,k;P) G_1(k+P/2)G_2(-k+P/2) \Gamma(k,P)\, . 
\ee
%
%Fig 1

The bound states appear as  poles  of the scattering matrix at total momenta $0\leq P^2 < (m_1+ m_2)^2$ from which the constrain on the normalization condition follows (for the derivation see for instance \cite{SAULIPHD}). 

\begin{figure}[t]
\centerline{  \mbox{\psfig{figure=WCrho.eps,
height=6.5truecm,width=7.5truecm,angle=0}}}

\centerline{\mbox{\psfig{figure=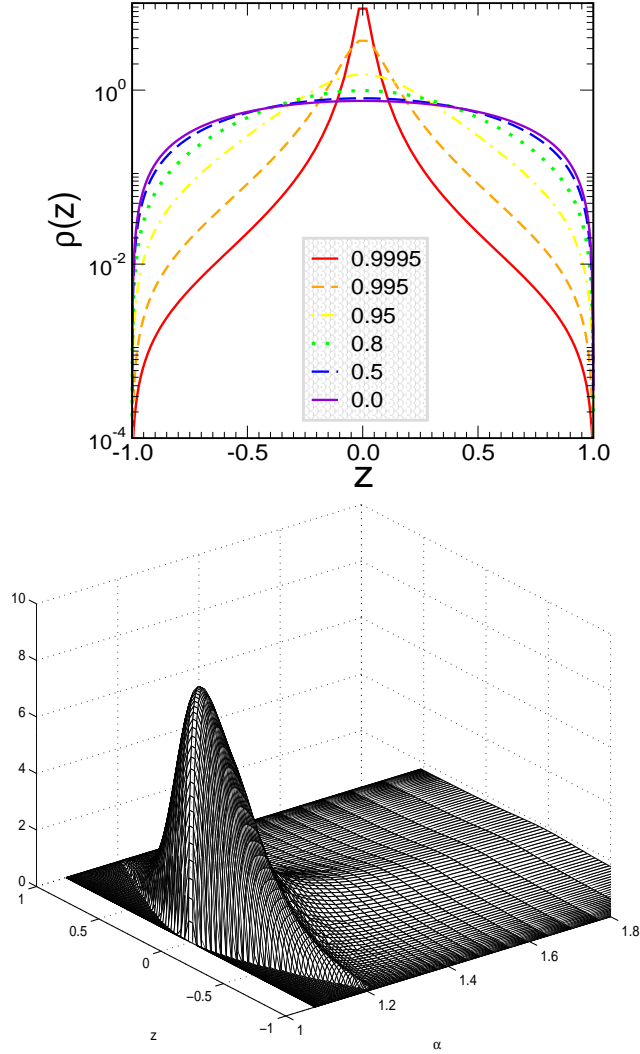,
height=7.5truecm,width=8.5truecm,angle=0}}}
\caption[99]{\label{fweighty} 
Top figure displays the weight function $\rho(z)$ in the original Wick-Cutkosky model
 $(m_3=0)$ for several values of $\eta=\sqrt{P^2}/2m$.  The Bottom figure displays rescaled weight function in the massive Wick-Cutkosky model for $m_3=0.1m$ and $\eta=0.95$. }
\end{figure}

In the original  Wick-Cutkosky model  \cite{WICK}  the exchanged boson is massless 
in the kernel $V$ and no radiative corrections are considered. This model is particularly 
interesting because it is the only example  which 
is solvable exactly  \cite{WICK}. For this model the s-wave bound state 
the PTIR reduces to the one dimensional expression:  
\be  \label{wcm}
\Gamma(p,P)=\int\limits_{-1}^1 dz \,
\frac{\rho(z)}{m^2-(p^2+zp\cdot P+P^2/4)- i\epsilon} \, \, ,
\ee  
where the function $\rho(z)$ satisfies the equation
\bea \label{origWCM}
\rho (z') &=& \frac{g^2}{(4\pi)^2} \int\limits_{-1}^1 d z V^{[1]}(z',z)\, \rho (z) \, , \\
V^{[1]}(z',z) &=& \sum_{s=\pm}\, \frac{\Theta(s(z-z'))\, T_s}{2(m^2-S)} \, , \nn
\\ 
S&=& \frac{1- z'^2}{4}P^2\quad ; \quad T_\pm= \frac{1 \pm z'}{1 \pm z}.
\nn
\eea
The resulting weight functions  $\rho(z)$ for various fraction of binding  $\eta=\sqrt{P^2}/2m$ are displayed in Fig. ~\ref{fweighty}. 

When the mass of exchanged particle is nonzero or even when the other diagram is considered in the kernel of Bethe-Salpeter equation then the appropriate  PTIR for the Bethe-Salpeter  vertex function is two dimensional \cite{NAK1971} (see also(\ref{ptirtri2})) 
\be  \label{nula}
\Gamma(p,P)= \int\limits_{-1}^1 dz \, 
\int\limits_{\alpha_{min}(z)}^{\infty}d\alpha\, \,
  \frac{\rho^{[n]}(\alpha,z)}{[\alpha-(p^2+zp\cdot P+P^2/4)-i\epsilon]^n} \, \, .
\ee
Thus the equation for  original Wick-Cutkosky model is regarded   as only the special case of (\ref{nula})
where the weight factorizes and is singular in $\alpha$.
However here, all the particles are massive thus the positive integer $n$ represents a dummy parameter constrained only by convergence of the solution. The choice $n=2$ was found to be a reasonable one. 

In this case the Bethe-Salpeter equation can be converted to the following real integral equation for the 
real spectral function: 
\be   \label{majn}
\rho^{[2]}(\alpha',z')=\frac{g^2}{(4\pi)^2} \int\limits_{-1}^{1}dz
\int\limits_{\alpha_{min}(z)}^{\infty} 
d\alpha\, \, V^{[2]}(\alpha',z';\alpha,z)\, \rho^{[2]}(\alpha,z)  \, ,
\ee 
The  interaction kernel is $V(p,k;P)=g^2G_3(p-k)$ in the dressed ladder approximation, where 
all the propagators are understood to be fully dressed.
For this case the kernel in spectral Eq. (\ref{majn}) $V^{[2]}(\alpha',z';\alpha,z)$ has been derived in \cite{SAUADA2003}. The explicit  formula  reads
\bea \label{jadro}
V^{[2]}(\alpha',z',\alpha,z)&=&\sum_{T} 
\prod_{j=1}^{3}\int d\al_j\bar{\sigma}(\al_j)
 \sum_{i=\pm} \frac{\theta(x_i)\theta(1-x_i)\theta(D)}{2J(\alpha,z)^2}\, \nonumber\\
&& \hspace*{-2.5truecm}
\left\{\frac{T J(\alpha,z)}{(1-x_i)|E(x_i,S,\alpha')| }
-{\mbox{sgn}}\left(E(x_i,S,\alpha')\right)\ln(1-x_i)\right\}_{x_i=x_i(T)}
  \, ,
\label{lacolie}  
\eea 
where  the functions entering the above relation are 
\bea
x_{\pm}(T)&=& \frac{\alpha'-\alpha_3-R_T \pm \sqrt{D}}{2(\alpha'_S)} \, ,
\nn \\
D&=& (R_T-\alpha'+\alpha_3)^2- 4\alpha_3(R_T-S) \, ,
\nn \\
E(x_\pm,S,\alpha')&=& \alpha'- \frac{\alpha_3}{x_\pm^2} -S\, , 
\nn \\
R_T&=& J(\alpha,z)T+ 
  \frac{\alpha_1+\alpha_2}{2}+\frac{\alpha_1-\alpha_2}{2}\, z' \, ,
\nn \\
J(\alpha,z)&=&  
\alpha-\frac{\alpha_1+\alpha_2}{2}-\frac{\alpha_1-\alpha_2}{2}\, z \, ,
\eea
and where the symbol $\sum_T$ should be understood as
\be
\sum_T f(T)= f(0)- \theta(z-z')f(T_+)- \theta(z'-z)f(T_-) \, ,
\ee
where $f$ is an arbitrary function. 

The spectral  function $\bar{\sigma}_i$ in Eq.~ (\ref{lacolie}) corresponds with the SR of i-particle propagator of i-field in the considered and relevant part of the Lagrangian   
\bea \label{chargevcml}
{\cal L}_{(strong)}= - g \phi_{1}^+ \phi_{1} \phi_{3}
 - { g \over 2}\, \phi_{2}^2 \phi_{3} \,  .
\eea
These were obtained from the Unitary Equations. Note  that  $n=2$ in DRs (\ref{labuan})  
 because of two subtractions were made, i.e. 
\be  \label{of} 
\Pi_{iR}(p^2)=\Pi_i(p^2)- \Pi_i(\zeta)-
 (p^2-\zeta_i)\, \frac{d\Pi_i(p^2)}{dp^2}|_{p^2=\zeta_i}\, \, , 
\ee 
since the on-shell renormalization prescription $\zeta_i=m_i^2$ with unit residuum was chosen.

The derivation of DRs for all three $\Pi_{iR}$ is very straightforward and follows the lines of previous Section (for some details see \cite{SAUADA2003}).
 The absorptive part of $\Pi$ is given by $\rho_i/\pi$, these are 
\bea \label{vysl}
\rho_{i}(\omega)&=& \frac{g^2}{(4\pi)^2}\,
\int d\alpha\, d\beta\, 
X(\alpha,\beta;\omega)\bar{\sigma}_3(\alpha)
\bar{\sigma}_i(\beta) \, ,\quad \quad i=1,2 \, ,
\nn \\
\rho_{3}(\omega)&=& \frac{g^2}{(4\pi)^2} \, 
\sum_{i=1,2}\int d\alpha\, d\beta\,
X(\alpha,\beta;\omega)\bar{\sigma}_i(\alpha) \bar{\sigma}_i(\beta) 
\, ,
\eea
where $X$ is defined by Eq. (\ref{xnula}).

In the ladder approximation the obtained spectrum  agree with the one obtained by other techniques \cite{NIETJO1996,KUSIWI1997,AHLALK1999,KUWI1995}. In dressed ladder approximation there is no other reliable result published in literature to be compared with.
 However the qualitative agreement with the paper
\cite{AHLALK1999}  was found: The critical value of coupling constant $g_c$ (defined by the constraint for the renormalization constant: $Z(g_c)=0$) as it has read from SDEs gives the physical domain of applicability of  Bethe-Salpeter equation. 

It is interesting to note here that  within the   use of 
Feynman-Schwinger representation method there was found in  \cite{SAGRTJ2004}
that the crossed boxes are dominates in scalar models studied here. For a 
comparison of different technique, including Bethe-Salpeter see the result in this paper. 
The observed dumping of bound state energy is  due to the bose statistic of the matter field in the model considered here (while there is an expected  cancellation between higher order skeletons 
when one consider the bound states of fermions). 

At this place we should mention the so called 'ghost' solution of Bethe-Salpeter equation.  These, probably unphysical solutions (see \cite{AHLALK1999} and references therein), have been found only for rather strong coupling in ladder approximation.  However, the couplings below the critical one allow only solutions for relatively weakly bound states and neither 'ghost' solution of the Bethe-Salpeter equation was found. It suggests the domain of validity of SDEs restricts the solutions of the Bethe-Salpeter equation in such way that only the normal solutions exist.

%%%%%%%%%%%%%%%%%%%%%%%%%%%%%%%%%%%%%%%%%%%%%%%%%%%%%%%%%%%%%%%%%%%%%%%%%%%%%%%%%%%%%%%%%%%%%%%%%%%%%%%%%%%%%%%%%%%%%%%%%%%%%%%%%%%%%%%%%%%%%%%%%%%%%%%%%%%%%%%%%%%%%%%%%%%%%%%%%%%%%%%%%%%%%%%%%%%%%%%%%%%%%%%%%%%%%%%%%%%%%%%%%%%%%%%%%%%%%%%%%%%%%%%%%%%%%%%%%%%%%%%%%%%%%%%%%%%%%%%%%%%%%%%%%%%%%%%%

\subsection{Elastic Electromagnetic Form Factor}

 Although the elastic form factor represents a simple dynamical 
observable, its Minkowski calculation represents  a nontrivial task. For 
this purpose we consider the massive Wick-Cutkosky model given by a  
Lagrangian  gauged as follows: 
\bea \label{gaugedwcm}
{\cal L}&=& (D^{\mu}\phi_1)^+ D_{\mu}\phi_1
+\frac{1}{2}\partial_{\mu}\phi_2\partial^{\mu}\phi_2
+\frac{1}{2}\partial_{\mu}\phi_3\partial^{\mu}\phi_3-
\frac{1}{4}F_{\mu\nu}F^{\mu\nu} -V(\phi_i) \, ,
\nn \\
V(\phi_i)&=& \left(m_1^2+g\phi_3\right)\phi_1^+\phi_1 +
\left( \frac{m_2^2}{2}+
{g \over 2}\, \phi_3\right)\phi^2_2+\frac{1}{2}m_3^2\phi_3^2 \, ,  
\eea
where the covariant derivative is $D_{\mu}=\partial_{\mu}-ieA_{\mu}$. 

The electromagnetic form factors parametrize the response of bound 
systems to external electromagnetic field. The calculation of these 
observables within the Bethe-Salpeter  framework proceeds along the Mandelstam's 
formalism \cite{MAN1955}. In 
the form factor calculations the effects of scalar dressing were not 
taken into account.

The current conservation implies  the 
parametrization of the current matrix element $G^{\mu}$ in terms of 
the single real form factor $G(Q^2)$ 
\be  \label{formf}
G^{\mu}(P_f,P_i)=G(Q^2)(P_i+P_f)^{\mu} \, ,
\ee
where $Q^2=-q^2$, so that  $Q^2$ is positive for  elastic 
kinematics. 

\begin{figure}[t]
\centerline{  \mbox{\psfig{figure=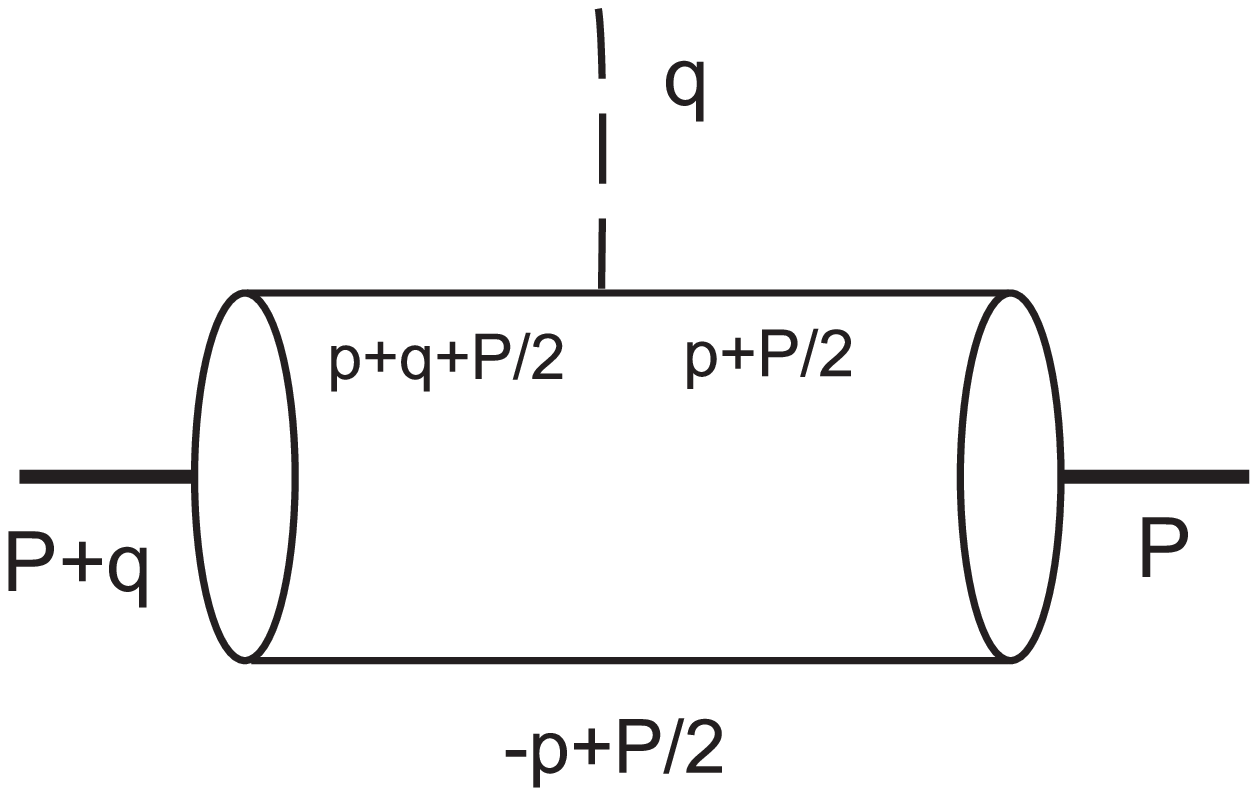,height=4.0truecm,angle=0}} }
\caption[99]{\label{figGmu} Diagrammatic representation of the electromagnetic current bound state matrix element.}
\end{figure}

The matrix element of the current in relativistic impulse approximation is depicted in Fig.~\ref{figGmu}.  The matrix element is  given in terms of the Bethe-Salpeter  vertex functions as
\bea  
&&G^{\mu}(P+q,P)=i\, \int\frac{d^4p}{(2\pi)^4}\,  
\bar\Gamma(p+\frac{q}{2},P+q) \nonumber\\ 
&& \left[ D(p_f;m_1^2) j_1^{\mu}(p_f,p_i) D(p_i;m_1^2) \,
D(-p+P/2;m_2^2) \right] \Gamma(p,P) \, ,
\label{chargeff}
\eea
where we denote $P=P_i$ and $j_1^{\mu}$ represents one-body current for 
particle $\phi_1$, which for the bare particle reads  
$j_1^{\mu}(p_f,p_i)= p_f^{\mu}+p_i^{\mu}$, where $p_i,p_f$ is initial 
and final momentum of charged particle inside the loop in 
Fig.~\ref{figGmu}, i.e., $p_i= p+ P/2, p_f= q+ p+ P/2$. 

The main result, as far as charge form factor is concerned, is the  rewriting of the rhs of Eq.~(\ref{chargeff}) directly in terms of the spectral weights of the bound state vertex function.  It allows the evaluation of the form factor by calculating the dispersion relation:
\be \label{drforf}
G(Q^2)=\int d \omega \frac{\rho_G(\omega)}{Q^2+\omega-i\epsilon}
\ee
without having to reconstruct the vertex 
functions $\Gamma(p,P)$ from their spectral representation. Unfortunately 
the derived expression for $\rho_G$ in  the DR (\ref{drforf}) contains
some additional integrations which the authors were not able to remove.
Due to this and in order to avoid sizable  numerical errors the results have been calculated only for positive $Q^2$. However a dramatic and exciting changes can be expected at timelike momentum regime. This remained unexplored until now.  
 
%%%%%%%%%%%%%%%%%%%%%%%%%%%%%%%%%%%%%%%%%%%%%%%%%%%%%%%%%%%%%%%%%%%%%%%%%%%%%%%%%%%%%%%%%%%%%%%%%%%%%%%%%%%%%%%%%%%%%%%%%%%%%%%%%%%%%%%%%%%%%%%%%%%%%%%%%%%%%%%%%%%%%%%%%%%%%%%%%%%%%%%%%%%%%%%%%%%%%%%%%%%%%%%%%%%%%%%%%%%%%%%%%%%%%%%%%%%%%%%%%%%%%%%%%%%%%%%%%%%%%%%%%%%%%%%%

\section{Strong coupling QED}

In this Section we  investigate   SDEs in 3+1 dimensional QED. 
We follow the prospects presented in the papers \cite{SAULIJHEP,SAULIRUN}, 
in which the solution for this theory in Euclidean and Minkowski space 
were compared for the first time. 

The first part of this Section is devoted to the solution for the
electron propagator in  ladder approximation, in the second
part an extension to the unquenched case (with the photon polarization
included)  is made. In the latter case the running coupling is
considered  self-consistently: 
the fermion mass and photon polarization function have been
calculated by solving the coupled SDEs for electron and photon
propagators in Landau gauge. In this paper we do not use the  vertex ansatze as in 
the paper \cite{SAULIRUN} but we  consider the first term of skeleton expansion
(i.e. bare $\gamma^{\mu}$ vertex is used). Having small quantitative effect, because of Landau gauge, this not only drastically simplified the calculation effort but allows us to perform many integrations analytically. Likewise in the Section where we discuss a scalar model we write down  the Unitary equations  where  the principal value integrations are excluded.
 This is a clear advantage when compared with earlier version of Unitary Equations \cite{SAULIJHEP,SAULIRUN} 
where this possibility was overlooked. 

In both approximations we are looking  for solutions with zero
and nonzero electron mass in Lagrangian. In the first case the
chirally symmetric solution (with massless electrons) always exists.
In the Euclidean formalism we obtained agreement with many previous
studies \cite{HSW1995,HSW1997,KONNAK1992}, for sufficiently large coupling $\alpha$ also nontrivial solution
for the mass function. On the other hand, no such solution was
found in our spectral Minkowski approach. This is a strong indication
that the dynamical fermion mass function has in this case a
complicated (complex) structure, which is not reflected by
assumptions of our spectral Ansatze.

We will be dealing mostly with the strong coupling regime, which
is far from  ``real-life QED'', for which  $\alpha_{QED}$ (in
experimentally accessible energy region) is small and use of perturbation theory  is
fully justified. Of course, all our solutions agree in weak
coupling limit with the perturbation theory. 

As it is already stated in the introduction the basic motivation 
is to decode the timelike structure of nonperturbatively obtained 
Green's functions. For this purpose the models considered here are an ideal playground.
However, in this Section the reader can find at least two other reasons to consider  strong-coupling QED dynamics.

The strong-coupling Abelian dynamics was considered as one of the
candidates for explanation of  electroweak symmetry breaking
\cite{PAGELS}. Such or a similar models  can  approximate
the dynamics in the strong coupling sector in various
Technicolors model, e.g., Slowly Walking Technicolors
\cite{WEI1979,HOL1985,CHIV2001,APP2003}, for which ladder approximation of QED-like
theory seems to provide a  test background for modeling  of the  (Techni)lepton
propagators (for a possible indirect  experimental evidence of
Technicolor-like model see \cite{CHRSHR2005}). Moreover, the
strong-coupling Abelian model is not only a suitable playground for
studies of supercritical phenomena like dynamical mass generation,
but it is suitable also for investigations of the analytical
structure of the fermion propagator. It
was 30 years ago when Fukuda and Kugo \cite{FUKKUG1976} observed the
disappearance of the  real pole of the fermion propagator in the
ladder QED and it was argued that this is a signal for confinement of
the fermion. Again, the main motivation is the simplicity of the
model, in alternative more complicated models the nonperturbative
phenomena are even more difficult to study quantitatively. We believe
that it is worth reviewing the recent progress in strong coupling
QED.

Our attention is mainly focused on the SDE for the fermion propagator, because
no nontrivial nonperturbative effects (i.e. effects not known from perturbation theory) 
were found in the photon propagator. However, let us remind the solution of perturbation theory renormalization group equations
 for the running coupling \cite{GOHOLIRA1998},
which in two loop approximations exhibits Landau singularity at the scale 
\bea
\Lambda_{L}=m_{R} \exp\left[
\frac{1}{\beta_{1}e_{R}^{2}}
\left( \frac{\beta_{2} e_{R}^{2}}{\beta_{1}+\beta_{2}e_{R}^{2}}\right)^{\beta_{2}/\beta_{1}^{2}}
\right] \, .
\eea
where $\beta_i$ are the coefficients of QED beta function
\be \beta=\frac{e_R^3}{12\pi^2}+\frac{e_R^5}{64\pi^4}+...
=\beta_{1}e_R^3+ \beta_{2}e_R^5+ \dots
\ee
giving us numerically $\Lambda_{L} \simeq 10^{227}$GeV if only the electrons are considered. For the particle spectrum of the standard model one gets $\Lambda_{L} \simeq 10^{34}$GeV which is many orders above the Planck scale and hence probably without any physical relevance. In the minimal Supersymmetric Standard Model with four Higgs, which offers a solution to the strong CP problem, the Landau pole moves down to $\Lambda_{L} \simeq 10^{17}$GeV. Thus the Landau pole is by no means academic and to resolve this problem one has to consider the nonperturbative formulation of QED. For this purpose we will consider coupled SDEs in unquenched QED but here rather in the strong coupling regime where the Landau singularity is expected at a much lower scale.

%%%%%%%%%%%%%%%%%%%%%%%%%%%%%%%%%%%%%%%%%%%%%%%%%%%%%%%%%%%%%%%%%%%%%%%%%%%%%%%%%%%%%%%%%%%%%%%%%%%%%%%%%%%%%%%%%%%%%%%%%%%%%%%%%%%%%%%%%%%%%%%%%%%%%%%%%%%%%%%%%%%%%%%%%%%%%%%%%%%%%%%%%%%%%%%%%%%%%%%%%%%%%%%%%%%%%%%%%%%%%%%%%%%%%%%%%%%%%%%%%%%%%%%%%%%%%%%%%%%%%%%%%%%%%%%

\subsection{QED SDEs}

The  SDEs for one particle irreducible Green functions can be derived by 
applying the functional differentiation
on  effective action $\Gamma[\psi_c,\bar{\psi}_c,A_c^{\mu}]$, where $\psi_c(A_c^{\mu})$ is  semiclassical vacuum expectation values of fermion (photon) field in the presence of the fictitious sources. For instance, the inverse of
photon propagator  is obtained as:
\be
(G^{-1})_{\mu\nu}(x,y) =  -\frac{\delta^2
\Gamma[\psi_c,\bar{\psi}_c,A_c^{\mu}]} {\delta A_{c\mu}(x)\delta A_{c\mu}(y)}
\, ,
\ee
where all effective fields are set to zero at the end of the derivation.
The details of derivation of SDEs   can be found
in the in standard textbooks,  (see e.g.
pp.~289-294 of Ref.~\cite{ITZYKSON}, but also \cite{ROBERTS,RIVERS}). 
Here we review the SDEs for propagators in momentum space for completeness 
here. The SDEs for the inverse of  photon propagator $G^{\mu\nu}$ and for the 
inverse of electron propagator $S$  read
\begin{fmffile}{fqed}
\bea
&&\left[G^{-1}\right]^{\mu\nu}\!(q) = - q^2 \left [g^{\mu\nu} +
\left(\frac{1}{\xi_A}-1\right)\frac{q^\mu q^\nu}{q^2} \right] +
\Pi^{\mu\nu}(q) \, , \label{eq2.18}
 \\
&&\Pi^{\mu\nu}(q)=\parbox{2.0\unitlength}{%
\begin{fmfgraph}(2.0,1.5)
\fmfpen{thick}
\fmfleft{i}
\fmfright{o}
\fmf{photon}{i,v1}
\fmfblob{0.2w}{v1}
\fmf{photon}{v2,o}
\fmf{plain,left,tension=0.25}{v1,v2}
\fmf{plain,right,tension=0.25}{v1,v2}
\end{fmfgraph}}% end of parbox
= ie^2 Tr \int \frac{d^4k}{(2\pi)^4}\,  
\gamma^\mu  S(k)  \Gamma^\nu(k,k-q)  S(k-q)  \, ,
\nn \\
\label{polar}
\eea
\end{fmffile}
\begin{fmffile}{fqed2}
\bea
&&S^{-1}(p) = p_\mu \gamma^\mu - m_0 - \Sigma(p) \, , \label{eq44}\\
&&\Sigma(p)= 
\parbox{2.0\unitlength}{%
\begin{fmfgraph}(2.0,1.5)
\fmfpen{thick}
\fmfleft{i}
\fmfright{o}
\fmfblob{0.2w}{v}
\fmf{plain}{i,v}
\fmf{photon,left,tension=0.25}{v,v2}
\fmf{plain,tension=0.25}{v,v2}
\fmf{plain}{v2,o}
\end{fmfgraph}}% end of parbox
=i e^2\int \frac{d^4k}{(2\pi)^4} \,
\gamma^\mu \, S(k) \, \Gamma^\nu(k,p) \, G_{\nu\mu}(k-p) \, ,
\label{elec-self}
\eea
\end{fmffile}
where $\Pi^{\mu\nu}(q)$ is  the vacuum
polarization tensor,  $\Sigma(p)$  is the fermion self-energy  and 
the blobs in the diagrams 
represent photon-fermion vertex function $\Gamma^\mu$  satisfying its own SDE
which can be sketched as in the Fig.~\ref{setdse}.
Further, the parameter $\xi_A$ in Eq. (\ref{eq2.18}) follows from the common choice
of covariant gauge fixing term $-(1/2\xi_A)\,(\partial_\mu A^\mu)^2$ of 
the QED action \cite{FAPO1967}.

From the conservation of the gauge current the following  Ward-Takahashi identity  for the photon-fermion vertex follows:
\be
\label{WTI} (p-l)_{\mu}\, \Gamma^{\mu}(p,l) = S^{-1}(p)-S^{-1}(l) \, ,
\ee
while the same implies transversality of the polarization tensor:
\be \label{ducati}
q_\nu\, \Pi^{\mu\nu}(q) = 0 \, ,
\ee
noting here that Eq.~(\ref{ducati}) also follows from Eq.~(\ref{polar})
using Eq.~(\ref{WTI}) provided the divergent integrals in (\ref{polar}) are regularized by a translation invariant manner.

Therefore, in this case, the vacuum polarization tensor can be parameterized in
terms of single scalar function $\Pi(q^2)$:
\be
\Pi^{\mu\nu}(q) = q^2\left[g^{\mu\nu} - \frac{q^\mu
q^\nu}{q^2}\right] \Pi(q^2) \, .
\label{eq29}
\ee
After the inversion, it yields for the photon propagator 
\be
G_{\mu\nu}(q) = - \frac{1}{q^2}
\left[ H(q^2)\left(g_{\mu\nu} - \frac{q_\mu q_\nu}{q^2}\right)
+ \xi_A \frac{q_\mu q_\nu}{q^2} \right] \, ,
\label{photprop}
\ee
where we define the photon renormalization function $H(q^2)$ as:
\be
H(q^2) \equiv \frac{1}{1-\Pi(q^2)} \, .
\label{photfunc}
\ee
From (\ref{photprop}) one finds that the  bare photon propagator is given by:
\be
G_{\mu\nu}^0(q) = - \frac{1}{q^2}
\left[ \left(g_{\mu\nu} - \frac{q_\mu q_\nu}{q^2}\right)
+ \xi_A \frac{q_\mu q_\nu}{q^2} \right] \, .
\ee

The general form for fermion propagator  (\ref{eq44}) is
\be
S^{-1}(p) = A(p^2) \, p_\mu
\gamma^\mu + B(p^2) \, ,
\ee
or alternatively:
\be
S(p) = \frac{F(p^2)}{\not \!p - M(p^2)} \, ,
\label{defAB}
\ee
where $F(p^2)=A^{-1}(p^2)$ is called the  fermion wavefunction
renormalization and $M(p^2)=B/A$ is the  fermion
mass function. Clearly the fermion propagator for a free fermion
field or  bare fermion propagator is given by $\Sigma=0$ in Eq. (\ref{eq44}).

\begin{figure}
\begin{fmffile}{fvertex}
\begin{equation} 
\Gamma^{\mu}=\,
\parbox{2.0\unitlength}{%
\begin{fmfgraph}(2.0,1.5)
\fmfpen{thick}
\fmfleft{i}
\fmfrightn{o}{2}
\fmfblob{0.2w}{v}
\fmf{photon}{i,v}
\fmf{plain}{v,o1}
\fmf{plain}{v,o2} 
\end{fmfgraph}}% end of parbox
\,\,\, = \,\,\,
\parbox{2.0\unitlength}{%
\begin{fmfgraph}(2.0,1.5)
\fmfpen{thick}
\fmfleft{i}
\fmfrightn{o}{2}
\fmf{photon}{i,v}
\fmf{plain}{v,o1}
\fmf{plain}{v,o2} 
\end{fmfgraph}}% end of parbox
\, \, \, + \, \, \,
\parbox{2.0\unitlength}{%
\begin{fmfgraph}(2.0,1.5)
\fmfpen{thick}
\fmfleft{i}
\fmfrightn{o}{2}
\fmfblob{0.2w}{v}
\fmf{photon}{i,v}
\fmf{plain,left,tension=0.25}{v,v2}
\fmf{plain,right,tension=0.25}{v,v2}
\fmfv{d.sh=square,d.f=hatched,d.si=0.25w,l=$\Gamma_4$}{v2}
\fmf{plain}{v2,o2}
\fmf{plain}{v2,o1}
\end{fmfgraph}}% end of parbox
\ee
\end{fmffile}
\caption[caption]{\label{setdse}Diagrammatical representation of the Schwinger-Dyson
equations for vertex function. Like in the diagrams for selfenergy the  blobs (box) represent the
exact vertex and the internal lines represent the exact propagators--
wavy line stands for the photon and the straight line labels fermion propagator.}
\end{figure}

\subsection{Physical constraints on the solution}

The model discussed in Section \ref{chap5} was too simple and thus does not render 
some very important issues encountered when one deals with more 
realistic models, e.g. the models for which we use a  gauge theory concept as a starting point.
Therefore we discus here the physical criteria which make the solution of SDEs (and any other 
nonperturbatively obtained solution of Green's functions) physically meaningful.    

In Quantum Field Theory an experimental observable is given by the transition probability associated with the appropriate part of the S-matrix.  When dealing with a renormalizable gauge theory the S-matrix element describing a given process should be completely independent on the renormalization scale (it is renormalization group invariant) and on the particular choice of gauge-fixing parameters (it is gauge-fixing independent). Furthermore -- as shown above -- the GFs should satisfy the Ward identities.  For a sufficiently small coupling constant the usual perturbation theory in conventional gauges offers a safe way to satisfy all the required invariance. It is well known that the gauge variant (unphysical) parts usually involved in the original GFs cancel out in a given S-matrix element.  To this end, there is recent progress in the so-called {\it Pinch Technique} \cite{BINPAP2002,BINPAP2004} which offers gauge-fixing independent Green's functions to all orders of perturbation theory both for the Abelian and nonAbelian case.  It is also remarkable that similar to the {\it Background-Field Method} the Pinch Technique GFs automatically satisfy the naive Ward identities rather than more complicated Slavnov-Taylor identities.

The problem of gauge invariance in the nonperturbative SDEs formalism is not yet fully solved.
Adopted techniques should maximally respect the gauge identities such as Ward identities, Nielsen identities \cite{NIEL} and Landau-Khalatnikov transformation \cite{LKT1,LKT2} when class of linear covariant gauges in QED is used. Clearly, the best we could have from this point of view is the systematic method of building the renormalized blocks of GFs which are already free of unphysical
information. Very promising but rather complicated method is to circumvent the gauge fixing problem by the use of stochastic quantization. This unconventional  idea is based on the introduction of a fifth stochastic-time coordinate, while the usual observables are obtained in the equilibrium. (The {\it stochastic quantization} and its relation to the usual methods is up to date  exhaustively reviewed in \cite{PROGRES}).  Since contrary to the effective action approach here, this method does not require gauge fixing at all and hence it offers  also the gauge fixing invariant GFs. Further, it is important to note that within the certain approximations the stochastic quantization yields the same Yang-Mills SDEs  as conventional approach in Landau gauge \cite{ZWAN2004}. The second promising method could be above mentioned Pinch Technique which was originally developed  in the context of SDEs \cite{COR1982,CORPAP1989} in 1980's. However, in this strategy the requirement of  multiplicative renormalizability has been ignored and to the best of author's knowledge from that time no progress has been made in utilization of Pinch technique in the SDEs formalism.

The gauge technique \cite{SALDEL1964,DELZHA1984}, being based on the spectral representation
offers GFs satisfying Ward identity in the whole Minkowski space. Within the context of gauge technique R.~Delbourgo in Ref.~\cite{DELB1999} highlights some of the related problems hindering the studies of SDEs. Among these, the first one is how well the Green's function solutions obey the Landau-Khalatnikov transformation and the second is how closely do the nonperturbative answers coincide with the perturbation series if we expand them in the power of couplings.

Making a simple Ansatz for a gauge vertices one can ensure Ward identity irrespective of analytical
properties of GF's. In QED the simplest solution is the Ball-Chiu vertex \cite{BALCHI1980}, while  any improved Ansatz can differ at most in the transversal part of vertex function. For an exhaustive  list of the modeling vertices in QED see the papers \cite{BASRAY,MONTANO}. However
stressed here that the recent numerical study \cite{BAPERA2006} of the gauge dependence of the chiral condensate leads us to the conclusion that the Ward identity alone is not sufficient to ensure the gauge independence of the physical observable. Further, it is argued that it is essential to apply full Landau-Khalatnikov transformation to the dynamically generated mass function as also advocated in \cite{BARA2005}. This step seems to be necessary in order to obtain a reliable results in other gauges then Landau gauge.

With the exception of the gauge technique  to solve the system of SDEs within the help of spectral technique means to calculate  complicated formula which should lead to the evaluation of  dispersion relation for proper GFs at the end. Therefore we instead considering some Ansatz for the vertex function we rather use the (two simplest) skeleton expansion of self-energies and our calculation is performed in  conventional Landau gauge where $\xi_A=0$. 
Such skeleton expansion can be  systematically improved by considering further and further terms in the multiloop expansion. In agreement with the issues made in  the paper \cite{ARI} we assume that the usage of Landau gauge should lead to the approximately gauge invariant GFs and  gauge-fixing independent observables.

The renormalizability means that  divergences may be removed (order-by order in perturbation theory) upon the field redefinitions
\bea
\psi\rightarrow\sqrt{Z_{2}}\psi, & &
A_{\mu}\rightarrow\sqrt{Z_{3}}A_{\mu},
\eea
and vertex renormalization
\be \gamma^{\mu}\rightarrow Z_{1} \gamma^{\mu}.
\ee
If
the renormalization scheme respects the Ward-Takahashi identity, i.e.
\be \label{Z_1=Z_2}
Z_1=Z_2
\ee
in the exact theory we say that QED is {\it
multiplicatively renormalizable}.

To summarize, we impose the physical
criteria, which will allow one to determine the solution of SDEs to
be physically meaningful (in general renormalizable  gauge theory). 
\begin{itemize}
\item The theory should be  multiplicatively renormalizable,
which implies:
\\ -the number of possible subtractions for a given GF (in momentum subtraction scheme).
\\ -the relation between various renormalization constants (and hence the coefficient of subtracting polynomials).
\\-renormalized GFs  satisfy WTIs.
\\-various possible renormalization schemes are physically equivalent
\\-the position of the pole in the propagators is scheme-invariant.
\end{itemize}

In addition, for the theory to be tractable in our Minkowski approach the following is assumed or required 
\begin{itemize}
\item The propagators are analytical functions in the whole complex plane of the momenta 
except the real positive axis, which implies:
\\- the vertices satisfy integral representation known from perturbation theory (here it is assured by skeleton expansion of GFs)
\\- the proper Green's functions satisfy dispersion relations
\end{itemize}
However, note here: neither the existence of a pole nor the positivity of the SR is  needed.

\subsection{Ladder approximation of electron SDE with nonzero bare mass}

In ladder approximation the vacuum polarization is neglected:
the photon propagator is taken in its bare form and the
photon-fermion-fermion vertex is approximated by the pure
$\gamma^{\mu}$ matrix. The relative technical simplicity  of
calculations  in the ladder approximation for electron SDE is an attractive feature
rendering very interesting study without some large
computational effort. It allows transparent comparison with the
Euclidean and the perturbation theory results as well.

Using  bare vertex and bare photon propagator leads to the
following ladder fermion SDE:
\bea
\label{vombat}
S^{-1}(p)&=&Z_2 [\not\!p -m_0]-\Sigma(p) \, ,
\nn \\
\Sigma(p)&=&ie^2\int\frac{d^4k}{(2\pi)^4}\,
G^{\mu\nu}_{free}(k)\gamma_{\mu} S(p-k)\gamma_{\nu} \, .
\eea

Further we introduce  the mass renormalization constant $Z_{m}$
and/or  the mass counterterm $\delta_m$
\be
 m_0=Z_{m}\, m(\zeta)\, ; \quad \quad \delta_{m}=m_0-m(\zeta) \, ,
\ee
relating the renormalized mass $m(\zeta)$ at the renormalization scale
$\zeta$ to the bare mass $m_0$ and recall that $Z_1=Z_2=1$ in ladder approximation in Landau gauge. 

\subsection{Euclidean formulation}

First of all, in order to make a careful and constructive comparison with the standard
Euclidean results  we review the basics of the standard
Euclidean formulation \cite{FUKKUG1976,HSW1995,HSW1997}
results presented in the literature.  
Taking the traces $Tr$ and $Tr\not\! p$ in the  first Eq.  (\ref{vombat}) the SDE is 
transformed into one equation for the mass $M_E=B_E$ : 
\bea
A_E(x)&=&1 \, ,
\nn \\
B_E(x)&=&m_0+\frac{3\alpha}{4\pi}\int\limits_0^{\infty} 
dy K(x,y) \frac{B_E(y)}{y+B_E^2(y)} \, ,
\label{lzice} \\
K(x,y)&=&\frac{2y}{x+y+\sqrt{(x-y)^2}}\, ,
\nn 
\eea
where Wick rotation  and angle integration have been done and  where we have used  $Z_2=1$, $x\equiv p^2_E=-p^2$, $y\equiv k^2_E=-k^2$. The renormalized Eq.~(\ref{lzice}) then reads
\bea     \label{euclid}
B_E(\zeta,x)&=&m(\zeta)+\frac{3\alpha}{4\pi}\int\limits_0^{\infty} dy V(\zeta,x,y) \frac{B_E(\zeta,y)}{y+B_E^2(\zeta,y)}\, ,
 \\
V(\zeta,x,y)&=&K(x,y)-K(\zeta,y)\, .
\nn
\eea

Recall here that Eq. (\ref{lzice}) with zero bare mass ($m_0=0$) 
provides nontrivial solution  only for $\alpha>\alpha_c$ 
(the so called supercritical QED, $\alpha_c$ is critical coupling)
while $B(p^2)=0$ for $\alpha<\alpha_c$ (chiral symmetric phase). 
Recall also here that the supercritical solution requires the implementation of a UV regulator $\Lambda_H$ owinig to Miransky scaling \cite{MIRANSKY,VOLODA}:
\be   \label{miransky}
B(0)\simeq\Lambda_H\sqrt{1-\frac{\alpha_c}{\alpha}}\, ,
\ee
where $\Lambda_H$ is  the naive (translation invariance violating) integral momentum cutoff.

\begin{figure}[t]
\centerline{\epsfig{figure=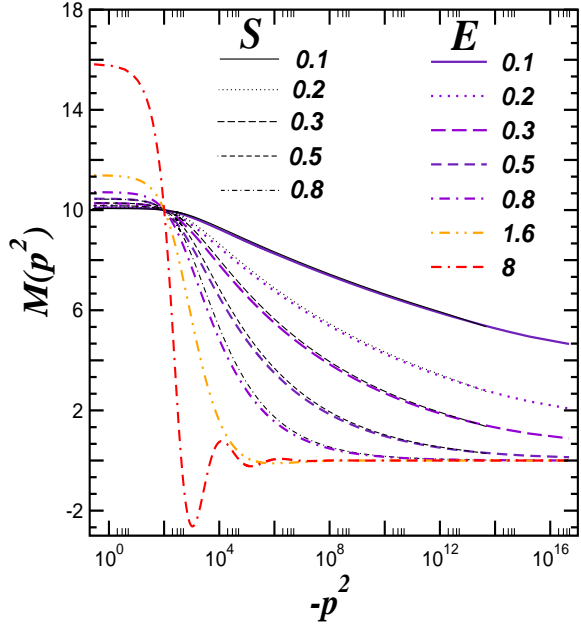,width=10truecm,height=12truecm,angle=270}}
\caption[caption]{Spacelike solutions for mass function of the electron propagator
as have been obtained within spectral (S)  and  Euclidean (E) formalism. \label{Fspace.eps}}
\end{figure}

In this paper we solved the SDE (\ref{euclid}) with the renormalization choice 
$M(\zeta=-10^2)=10$. The results are displayed in Fig.~\ref{Fspace.eps}.  It is not surprising  that for $\alpha>\alpha_c$ the obtained results do not agree with perturbation theory at all. In agreement with the studies \cite{FUKKUG1976,HSW1997} the expected damping of the  mass function to its negative values is observed in  supercritical phase of QED.  

%%%%%%%%%%%%%%%%%%%%%%%%%%%%%%%%%%%%%%%%%%%%%%%%%%%%%%%%
\subsection{Minkowski formulation}
%%%%%%%%%%%%%%%%%%%%%%%%%%%%%%%%%%%%%%%%%%%%%%%%%%%%%%%%%%%%%%%%%%

As in the scalar theory discussed earlier, we use  
the SR for the propagator and arrive at DRs for selfenergy.
Although the results of this Section are presented for ladder approximation QED 
the renormalization wave function $A$ will be formally kept during the calculation
for a possible more general future  studies. 
There are two independent spectral functions
in SR for QED fermion propagator:
\be \label{hobitka}
S(p)= \int ds\, \frac{\not \! p 
\bar{\SG}_v(s)+\bar{\SG}_s(s)}{p^2-s+\ep}=
\frac{r}{\not\!p-m}+\int ds \frac{\not\!p \SG_v(s)+\SG_s(s)}{p^2-s+\ep}\, ,
\ee
where $m$ is the physical electron mass. As usual the propagator is split into a real pole part and a part that is generated by the interactions.  To avoid confusion from the beginning, it is stressed here that the assumption (\ref{hobitka}) does not imply (but does not exclude) real or complex pure pole singularity structure of the electron propagator. Thanks to the masslessness of the  photon, the branch point corresponds to the physical mass of the electron. The  gauge-dependent complex singularity appears after the limit  $p^2\rightarrow m^2_+$, while under the threshold $p^2\rightarrow m^2_-$ the propagator is assumed to be a real function in any case. This known fact is reflected by the threshold behavior of the continuous  spectral function $\SG_{v,s}$ (in the previous scalar case the spectral function was zero at the threshold, however here $\SG_{v,s}(m^2_-)\neq 0$ in general).   

We split the unrenormalized self-energy (\ref{vombat}) into its Dirac vector and Dirac
scalar part
\be
\Sigma(p)=\not\!p \, a(p^2)+b(p^2) \,.
\ee
The self-energy in Eq.~(\ref{vombat}) is only logarithmically divergent and one subtraction:
\be \label{acko}
a_R(\zeta;p^2)=a(p^2)- a(\zeta)=\int d\omega
\frac{\rho_v(\omega)(p^2-\zeta)}{(p^2-\omega+\ep)(\omega-\zeta)} \, ,
\ee
\be \label{becko} 
b_R(\zeta;p^2)= b(p^2)- b(\zeta)=\int d\omega
\frac{\rho_s(\omega)(p^2-\zeta)}{(p^2-\omega+\ep)(\omega-\zeta)}\, .
\ee 
is sufficient to renormalize the self-energy: $\Sigma_R(\zeta,p)=\not\!\! p \, a_R(\zeta;p^2)+b_R(\zeta;p^2)$, where two scalar functions $a,b$ satisfy the DRs written also above.

Relatively straightforward calculations \cite{SAULIJHEP,SAULIPHD} lead
to the results for the imaginary parts of the functions (\ref{acko}),(\ref{becko})
with the result $\rho_v=0$ while the weight function $\rho_s$ reads
%
%{\small
\be    
\nonumber  
\rho_s(p^2)=-3\left(\frac{e}{4\pi}\right)^2
\left[rm\left(1-\frac{m^2}{p^2}\right)+ \int_{m^2}^{p^2}
d\alpha \SG_s(\alpha)\left(1-\frac{\alpha}{p^2}\right)\right]\Theta(p^2-m^2)\, .
\label{lowest}
\ee
%}
%
As we already saw in the previous Section  the peculiarity of the ladder approximation electron SDE in Landau gauge gives for the renormalization wave function  $A(\zeta;p^2)=1+a_R(\zeta;p^2)=1$ and renormalization group invariant mass function is 
\be
M(p^2)=B(p^2)=m(\zeta)+b_R(\zeta;p^2)\, .
\ee

The Unitary equations can be most easily derived from the following identity
\be\label{zakl}
S^{-1}S=1\, ,
\ee
where we use the SR (\ref{hobitka}) to express fermion propagator $S$ and SDE
in DR form for the inverse, which reads
\be
S^{-1}=\not\!p-m(\zeta)-\int ds \frac{[\not\!p
\rho_v(s)+\rho_s(s)]\frac{p^2-\zeta}{s-\zeta}}{p^2-s+\ep}\, .
\ee
For this purpose we introduce the following convenient notation:
\bea
\hat{g}=\Re S=\frac{r}{\not\! p-m}+P.\int ds \frac{\hat{\sigma}(s)}{p^2-s}\, &;& 
\hat{\sigma}(s)=\not\! p\sigma_v(s)+\sigma_s(s)\, ,
\nn \\
\nonumber \hat{d}=\Re S^{-1}=\not\! p-m(\zeta)-P.\int ds \frac{\hat{\rho}(s)}{p^2-s}\frac{p^2-\zeta}{s-\zeta}\, &;& 
\hat{\rho}(s)=\not\! p\rho_v(s)+\rho_s(s)\, ,\\
\label{ilona}
\eea
which allows us to rewrite the identity (\ref{zakl}) as
\be \label{kaf}
(\hat{g}-i\pi\hat{\sigma}(p^2))(\hat{d}+i\pi\hat{\rho}(p^2))=1\, ,
\ee
where we have used the well known functional identity for distributions (\ref{functional}).  Comparing the real and imaginary parts of (\ref{kaf}) we arrive at two coupled equations:
\bea
\hat{g}\hat{d}&=&1-\pi^2\hat{\sigma}(p^2)\pi\hat{\rho}(p^2)\, ,
\nn \\
\hat{\sigma}(p^2)\hat{d}&=&\hat{g}\hat{\rho}(p^2)\, .
\eea
Still in the matrix formulation introduced in Eq.~(\ref{ilona}) we can write
\be \label{usmev}
\hat{\sigma}(p^2)(\hat{d}^2+\pi^2\hat{\rho}^2(p^2))=\hat{\rho}(p^2)\, ,
\ee
i.e. $\hat{g}$ is eliminated.  Owing to this fact the principal value integrations over the unknown $\hat{\sigma}$ are not presented from now on. The Eq. (\ref{usmev}) represents matrix form of Unitary Equations. For the purpose of completeness we will write them down also in terms of $\sigma_{v,s}$ explicitly. To this end we define two scalar functions $c_v(p^2)$ and $c_s(p^2)$ such that
\be
\hat{d}^2+\pi^2\hat{\rho}^2(p^2)=\not\! p\, c_v+c_s\, ,
\ee
which implies
\be
c_s=p^2{\cal A}^2+{\cal B}+\pi^2(p^2\rho_v^2+\rho_s^2)\, \quad ; \quad
c_v=-2{\cal A}{\cal B}+2\pi^2\rho_v\rho_s\, ,
\ee
where we used shorthand notation for the real parts of the functions $A,B$
\bea
\cal{A}&=&\Re A(p^2)=1-P.\int ds \frac{\hat{\rho}_v(s)}{p^2-s}\frac{p^2-\zeta}{s-\zeta}\, ,
\nn \\
\cal{B}&=&\Re B(p^2)=m(\zeta)+P.\int ds \frac{\hat{\rho}_s(s)}{p^2-s}\frac{p^2-\zeta}{s-\zeta}\, .
\eea
Then ``projecting''  Eq. (\ref{usmev}) by $Tr $ and $Tr\!\not\!p $  we get, after a little algebra, the desired Unitary Equations:
\be  \label{Takk}
\SG_v= \frac{p^2c_v\rho_v-c_s\rho_s}{p^2\, c_v^2-c_s^2}\quad ;\quad
\SG_s= \frac{c_v\rho_s-c_s\rho_v}{p^2\, c_v^2-c_s^2}\quad ,
\ee
which are nonzero above the threshold, i.e. for $p^2>m^2$.  As in the previously  discussed scalar model, the form of the Unitary Equations (\ref{Takk}) does not depend on the details of interaction but on an analytical assumption -- a spectral  representation for the fermion propagator. Clearly, the Unitary Equations are formally equivalent to the Unitary Equations presented in the Ref.~\cite{SAULIJHEP}, but here with the main difference that all principal value integrals to be solved numerically are eliminated from now on.  In our approximations the function ${\cal A}=1$ and the real part of the mass function above the threshold is 
%
%{\small
\bea \label{gegek}
{\cal B}(p^2)&=&m(\zeta)+\frac{3\alpha_{QED}}{4\pi}
\left\{r\,m \left[1-
\left(1-\frac{m^2}{p^2}\right)\ln\left(\frac{p^2}{m^2}-1\right)\right]\right.
\nn \\
&+&\left.\int\limits_{m^2}^{\infty}
d\alpha \SG_s(\alpha)\left[1-
\left(1-\frac{\alpha}{p^2}\right)\ln\left|\frac{p^2}{\alpha}-1\right|\right]\right\}\, .
\eea
%}  
%
As we mentioned, the threshold singularity need not be real valued and the convential procedure of extracting a real number $r$ by on-shell differentiation of the self-energy function is inconsistent.  However, the solution of this puzzle is very simple, recall here that the fixed renormalization uniquely 
determine the whole propagator function and thus  the real pole part residuum $r$ in 
(\ref{hobitka}) as well.
Putting for instance $p=0$ in (\ref{zakl})  we get particularly simple 
formula for the residuum $r$
\be \label{rezid}
r=\frac{m}{m(\zeta)+\int_{m^2}^{\infty}\frac{\rho_s(x)\zeta}{x(x-\zeta)}dx}-
\int\limits_{m^2}^{\infty} \frac{\sigma_s(x)m}{x}dx \, .
\ee

The physical electron mass is defined by Eq. $\Re S^{-1}(m)=0$ or equivalently $M(m)=m$. Using the DRs (\ref{acko}),(\ref{becko}) the desired relation reads:
\be \label{masspole}
m=m(\zeta)+\int_{m^2}^{\infty} ds\frac{(m\rho_v(s)+\rho_s(s))(m^2-\zeta)}{(m^2-s)(s-\zeta)} \, .
\ee
However, here $\rho_{v,s}(m^2)=0$ is necessary in order to have the  correct definition of the physical mass. (This statement is clearly fulfilled in perturbation theory, see the first term in Eqs.~(\ref{lowest}),(\ref{gegek}) for this purpose.) 
 
The Unitary Equations (\ref{lowest}),(\ref{Takk}) have been solved by the method of iterations.  The residuum (\ref{rezid}), physical mass (\ref{masspole}) and the mass function (\ref{gegek}) have been evaluated at each step of the iteration procedure and substituted in the Unitary Equations until convergence is achieved.

\begin{figure}[t]
\centerline{\epsfig{figure=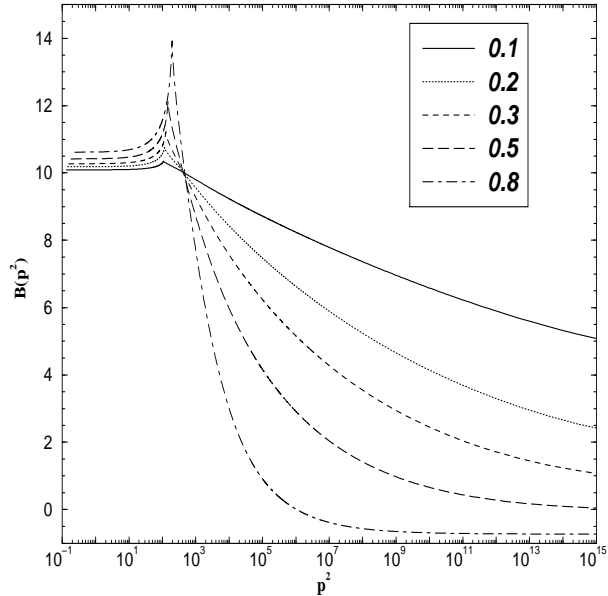,width=8truecm,height=8truecm,angle=270}}
\caption[caption]{Timelike mass functions of the  electron propagator as 
they have been obtained from  solution of the Unitary Equations.\label{Ftime10.eps}}
\end{figure}

To compare with the Euclidean space SDEs solutions we use the same renormalization condition in the both approaches, i.e. $B(-100)=10$.
Having all the numerical solutions stable and making comparison
between Euclidean and Minkowski results we have  found that they
perfectly agree  when the coupling is small enough.
The Unitary Equations resulting spacelike solutions for the mass function are compared
in the Fig.~\ref{Fspace.eps}.  In Fig.~\ref{Ftime10.eps} the Unitary Equations solutions are given for timelike momenta.
The meaning of the origin of small numerical difference for strong couplings is not 
satisfactory understood. We can speculate here that the analytical 
spectral representation Ansatz does not fully capture the structure
of the fermion propagator but we hope this point will be clarified in 
the future.

\subsection{Approximate solution of dynamical chiral symmetry breaking in Minkowski space}

In explicit chiral symmetry breaking  case, in which the
nonzero electron mass exists from the very beginning,  both
approaches-- Minkowski and Euclidean-- offer approximately the same
results  in the coupling constant domain
where  both solutions were obtained (in the spacelike momentum region, of course).
These solutions  were reviewed and compared in the previous Sections.

Until now, we have avoided the  phenomenon of spontaneous chiral symmetry breaking which  plays an important role in particle physics. 
The numerical methods used for rainbow Landau gauge approximation in the paper \cite{SAULIJHEP}  enable to find a non-trivial solution of Unitary Equations for zero fermion Lagrangian mass. The numerical solution tended to fail near  critical value $\alpha_c=\pi/3$ in rainbow approximation. In that case, approaching the coupling to its critical value the Unitary equations become less numerically stable. It was already observed that better stability can be achieved when nonzero mass is added into the gauge-boson propagator. The authors of the paper \cite{ADBESA} were looking for the Minkowski solution of the fermion SDE $\Sigma=\int K S$ with the effective kernel chosen as
\bea
K(q,p)_{ijkl}&=&\gamma^{\alpha}_{ij}\gamma^{\beta}_{kl}
\left[-g^{\alpha\beta}+\frac{q_{\alpha}q_{\beta}}{q^2}\right]\,
G(q^2) \, , 
\nn \\
G(q^2)&=&\frac{1}{q^2-m_B^2+i\epsilon}-\frac{1}{q^2-m_{\Lambda}^2+i\epsilon}
\eea
In the case when $m_B<<m_{\Lambda}$ the mass  $m_B$ can be understood as an infrared  regulator while $m_{\Lambda}$ represents ultraviolet (Pauli-Villars) one. For this simple model the gap equation has been solved in the Euclidean space and in Minkowski space as well. The nontrivial chiral symmetry breaking solutions have been compared for the first time. Within the spectral function Ansatz (\ref{hobitka})  the strategy of Unitary equations (\ref{Takk}) has been used for the later purpose.  In $A=1$ approximation the explicit evaluation of finite selfenergy  leads to the following result for the absorptive and dispersive part of the mass function
\begin{eqnarray} \label{pispunta}
\rho_s(\omega)&=&\frac{\alpha}{(4\pi)}\int d x \sigma_s(x) \left[X_0(\omega;m_B^2,x)-
X_0(\omega;m_{\Lambda}^2,x)\right]\, ,
\nn \\ 
{\cal M}(p^2)&=&\frac{3\alpha}{4\pi}
\int d x \sigma_s(x)\left[J(p^2,x,m_B^2)-J(p^2,x,m_{\Lambda}^2)\right]\, ,
\end{eqnarray}
where we have labeled
\begin{eqnarray}
J(p^2,y,z)&=&
-\frac{\Theta(-\lambda_p)\sqrt{-\lambda_p}}{p^2}
\left[\frac{\pi}{2}+arctg\frac{p^2-y-z}{\sqrt{-\lambda_p}}\right]+\frac{1}{2}\ln(16yz)
\nn \\
&-& \frac{\Theta(\lambda_p)\sqrt(\lambda_p)}{p^2}
\ln\left|\frac{p^2-y-z+\frac{\lambda_p}{T-p^2}}{p^2-y-z+\sqrt{\lambda_p}}\right|
+\frac{\Theta(\lambda_0)\sqrt{\lambda_0}}{p^2}
\ln\left|\frac{-y-z+\frac{\lambda_0}{T}}{-y-z+\sqrt{\lambda_0}}\right|\, ,
\nn \\
\end{eqnarray}
and  the following abbreviations is introduced:
\begin{eqnarray}
\lambda_p&=&\lambda(p^2,y,z) ,
\nn \\
\sqrt{\lambda_0}&=&|y-z|\, ,
\nn \\
T&=&(\sqrt{y}+\sqrt{z})^2\, .
\end{eqnarray}
The function $X$ is defined in (\ref{xnula}). 

These equations together with (\ref{Takk}) comprise the coupled set of integral equation that have been solve numerically in \cite{ADBESA}. Not strangely, the critical coupling $\alpha_c\sim \pi/3$ when $m_B<<m_{\Lambda}$ which is  expected from the study of rainbow QED with hard cutoff. More interestingly, the analytical continuation of (spectral) timelike solution to the spacelike regime is in rather good agreement with the Euclidean one, especially when the coupling is close enough to the critical value $\alpha_c$. Furthermore, the trivial solution was found bellow $\alpha_c$ in  both treatments.
However, there is apparent deviation when the coupling increases. This, together with the authour's observation of the  deviation from assumed analyticity  implies that spectral solution is not the exact and true solution of original SDE. It is suggested that deviation of 'Minkowski spectral' solutions from corresponding Euclidean ones reveals that the employed Ansatz, which fixes the analytical structure of propagator in Minkowski space, is not adequate for the case of the dynamical mass generation. It seems likely that with coupling increasing above its critical value, the pole of the propagator, located for subcritical case on  the real axis, moves gradually into complex plane (i.e. receives non-negligible imaginary part) and the standard Lehmann representation, employed for our calculations, is not valid anymore.   

However, checking  the analyticity in the whole Minkowski space and also comparing spectral solution with the Euclidean ones at spacelike, one can conclude that
the solution of Unitary Equations near the critical coupling is a rather good approximation. This approximative character become worse and worse when the coupling flows from its critical value, while the agreement is perfect at critical coupling (however were the result is trivial). At the end we can mention that within the Unitary Equations a given model neither provides the similar picture to the behaviour of light (say u,d,s) quark mass in QCD (here, the word 'similiar' means the  shape and slope of the mass function, or much simply the ratio of the value of mass function at two rather distinct scales). The analytical property of the solution in such a strong case remains unexplored at least by the presented method. For an different analytical guesses of the form of the quark propagator in modeled  QCD see for instance \cite{ADFM2003}.

%%%%%%%%%%%%%%%%%%%%%%%%%%%%%%%%%%%%%%%%%%%%%%%%%%%%%%%%%%%%%%%%%%%%%%%%%%
%%%%%%%%%%% CALCULATION OF UNQUENCHED QED - THE PAPER 2   %%%%%%%%%%%%%%%%
%%%%%%%%%%%%%%%%%%%%%%%%%%%%%%%%%%%%%%%%%%%%%%%%%%%%%%%%%%%%%%%%%%%%%%%%%%

\subsection{Unquenched QED - calculation of the running coupling}

In the previous Section we performed a numerical study of the electron mass function in the simple approximation  to the SDEs: the bare vertex and photon propagators were employed. In what follows we extend these calculations: we solve  the
corresponding SDEs in Landau gauge also for the photon polarization function.
 We  again compare  the solutions for propagators in
Minkowski and Euclidean space and we have found reasonable agreement
between solutions obtained in these two technically rather
different frameworks.

The 4D QED is trivial theory, in other words the presence of some regulator is required
for any nonzero value of the renormalized coupling constant. It is convenient to introduce nonperturbative invariant continuum regulator function which also render finite the ultraviolet divergences. Such  program, completed years ago, provides invariant regularization across all quantum field theories (for the review see the paper  \cite{HALPERN1993}). 
However in  this paper we simply use     
spectral cutoff regulator $\Lambda_S$ for the purpose of Minkowski calculation and the hard momentum cutoff $\Lambda_H$ for the purpose of Euclidean calculation. 
$\Lambda_H$ cuts the Euclidean momentum integrals while $ \Lambda_S $
represents the upper spectrum boundaries in GFs SR and DR.
They have been put to be equal, $\Lambda_S=\Lambda_H$ 
assuming no significant affect on the results.
By  construction the Minkowski regulator does not violate translation invariance of the theory
 while  Euclidean hard cutoff  $\Lambda_H$ do this. The same argument is also valid for the  gauge  invariance
of the theory (assuming gauge invariant truncation of SDEs). The observed discrepancy between the Minkowski and the Euclidean solutions, although not so large, is the consequence of the regularization scheme dependence in the trivial theory. 

There are many  attempts \cite{CUPE1990,BAPE1995,BURO1991,BAKIPE2000,BADE2004,BAPE1994,BASRAYII2004} to estimate the form and its  effect of the transverse part $\Gamma_T$ of the full vertex $\Gamma=\Gamma_{BC}+\Gamma_T$ on the behaviour of fermion propagator. The eight independent
component of  $\Gamma_T$ is the only unknown part of the fermion and photon  SDEs system when solely written. The longitudinal part, named Ball-Chiu vertex  $\Gamma_{BC}$ is the minimal Ansatz consistent with  Ward-Takahashi identity (\ref{WTI}) (and of course with perturbation theory  \cite{KRP1995}). To derive dispersion relation for fermion selfenergy and photon vacuum polarization 
is complicated task even when  $\Gamma_{BC}$ is only considered ( see \cite{SAULIPHD}),
therefore,  instead of making some complicated Ansatz for the gauge vertex we consider the skeleton expansions of the Green functions.  The skeleton expansion can be obtained from the sophisticated iteration of truncated SDEs system   corresponding thus with approximated effective action.  This so called $\Phi$ derivable approximation (or N-PI) of effective action  is  used in many branches of quantum field theory \cite{BLIARE1999,BLIAREii1999,BLIAREiii1999,AABBS,CODAMI2003,AJ2003,%
BRAEMM2002,BRAEMMii2002,ANDSTRI2004}. For an intimate relation between N-PI effective action and  the  full SDEs  see the discussion in \cite{BERGES}.  It is a systematically improvable method where the improvement is achieved by considering further and further terms in  the loop expansion of the effective action.  In the skeleton expansion there are only the classical vertices included in each skeleton diagram which  make the calculation of DR more tractable.   For a first attempt made in this paper we calculated  the one loop skeleton contribution, the resulting gap equation corresponds with the SDEs in the bare vertex approximation. The study of SDEs where self-energies are considered in two loop skeleton expansion is now under consideration.         

Further in order to  reduce the complicated hierarchy of equations we  explicitly put $A=1$. Within several percentage deviation such approximation can be justified in Landau gauge herein \cite{BLOCH2,KONNAK1992,BLOPEN1995}.

\subsection{Unquenched QED in Euclidean space}
The SDEs are  solved in Euclidean space after the Wick rotation $k_0 \rightarrow ik_{1E}$ is performed. Then the loop integrals should be free of kinematic singularities and the GFs are found for positive Euclidean momentum $k_E^2=k_1^2+k_2^2+k_3^2+k_4^2$.
 
Since we use the hard cutoff $\Lambda_H$ in our  Euclidean treatment the unpleasant quadratic divergence in the polarization function $\Pi$ is generated in this case. Due to this we follow usual strategy \cite{FISALK2003,BLOCH2,BLOPEN1995} and avoid the quadratic divergence by the use of Bloch-Pennington projector   
\bea
{\cal P}_{\mu\nu}^{(d)}(q)&=&\frac{1}{d}
\left[g_{\mu\nu}-(d+1)\frac{q_\mu q_\mu}{q^2}\right]\, ,
\nn \\
\Pi_E(q^2)&\equiv& \frac{\Pi^{\alpha\beta}(q)
{\cal P}_{\alpha\beta}^{(3)}(q)}{3q^2}\, .
\eea
Employing this projection in the SDE for the photon propagator we arrive at the following coupled equations:
%
%{\small
\bea  
\nonumber
\Pi_E(x)&=&\frac{2\alpha}{3x\pi^2}\int d y \frac{y}{y+M_E^2(y)}\int d\theta \sin^2\theta
\frac{2y-8y\cos^2\theta+6\sqrt{yx}\cos\theta}{z+M_E^2(z)}
, \\ 
\label{bloch1}\\
 M_E(x)&=&m_0+\frac{\alpha}{2\pi^2}\int d y \frac{y}{y+M_E^2(y)}
\int d\theta \sin^2\theta\frac{3M_E(y)}{z(1-\Pi_E(z))}\quad,\label{bloch2}
\eea%}
where the variables $x,y$ represent squares of Euclidean momenta, $z=x+y-2\sqrt{yx}\cos\theta$.

After the subtraction 
\bea
\Pi^R_E(\zeta';p^2)&=&\Pi_E(p^2)-\Pi_E(\zeta')\, ,
\nn \\
M^R_E(\zeta';p^2)&=&M_E(p^2)-M_E(\zeta')\, .
\eea
at  some arbitrary  $\zeta'$  and after renormalization, the Eqs.~(\ref{bloch1}),(\ref{bloch2}) have been solved numerically.  If needed the change of the renormalization scale choice is performed simply by the utilization of the identity
\bea
\Pi^R_E(\zeta;p^2)&=&\Pi^R_E(\zeta';p^2)-\Pi^R_E(\zeta';\zeta)\, .
\eea

\subsection{Unquenched QED  in  Minkowski space}

Likewise in the ladder approximation of QED the SDEs written in momentum space 
are converted to the coupled set of Unitary Equations.
The  Unitary Equations (\ref{Takk}),(\ref{rezid}),(\ref{masspole}) for fermion spectral functions is already derived in the previous Section.
They remain unchanged since they follow from general analytic structure of the fermion propagator function. The Unitary Equation for the photon spectral function is derived bellow.
Before doing this we review the DR for the fermion selfenergy in the presence of dressed photon propagator. 

 In $A=1$ approximation we can identify
 the electron mass function (\ref{defAB}) as
%
%{\small
\bea \label{bomba}
&&M(p^2)=m_o+\frac{Tr}{4}\Sigma(p)= m_o+
\nn \\
&&e^2 \frac{Tr}{4} \int da \int db \int \frac{d^4l}{(2\pi)^4}
 \gamma^{\nu}\frac{(\not\!p-\not l)
\bar{\SG}_v(a)+\bar{\SG}_s(a)}
{(p-l)^2-a}\gamma^{\mu}
\frac{-g_{\mu\nu}+\frac{l_{\mu}l_{\nu}}{l^2}}{l^2-b}\bar{\SG}_{\gamma}(b)
\, ,
\eea
%}
%
which after the renormalization 
(add  zero of the form $m(\zeta)-m(\zeta)$ into the Eq. (\ref{bomba}) and make the subtraction) 
leads to the DR 
%
%{\small
\bea   \label{brave}
M(p^2)&=&\int_{m^2}^{\infty} d\omega \frac{p^2-\zeta}{\omega-\zeta}
\frac{\rho_s(\omega)}{p^2-\omega+\ep} +m(\zeta)\, ,
\\
\rho_{s}(\omega)&=&-
3\left(\frac{e}{4\pi}\right)^2 \int da \, db \,\bar{\SG}_s(a)\bar{\SG}_{\gamma}(b)
\frac{\lambda^{1/2}(\omega,a,b)}{\omega}\Theta(\omega-(\sqrt{a}+\sqrt{b})^2) \, ,
\eea
%}
%
where $\sigma_{\gamma}$ is the photon spectral function defined 
by the SR for photon propagator, which in the Landau gauge reads
\bea \label{fotoni}
G^{\mu\nu}(q)&=&\int\limits_{0}^{\infty} db
\frac{\bar{\sigma}_{\gamma}(b)\left(-g^{\mu\nu}+\frac{q^{\mu}q^{\nu}}{q^2}\right)}
{q^2-b+\ep}\:,
\nn \\
\bar{\sigma}_{\gamma}(b)&=&r_{\gamma}\delta(b)+\sigma_{\gamma}(b)\, .
\eea
Since gauge invariance is correctly maintained the photon polarization 
function should posses at most logarithmically superficial divergence. 
Therefore only one subtraction is needed and the DR for renormalized polarization function then looks like
\bea
\Pi_R^{\mu\nu}(\zeta;q^2)&=&(q^2g^{\mu\nu}-q^{\mu}q^{\nu})\Pi_R(\zeta;q^2)\, ,
\nn \\
\label{momforp}
\Pi_R(\zeta;q^2)&=&\int\limits_{0}^{\infty}
d\omega \frac{\rho_{\gamma}(\omega)(q^2-\zeta)}{(q^2-\omega+\ep)(\omega-\zeta)} \, ,
\eea
where we use (not necessarily) the same renormalization scale as in the  case of electron propagator. 

 In order to arrive to the photon Unitary Equation, the  real and imaginary part of 
the identity $G_{\alpha\beta}^{-1}G^{\beta\gamma}=\delta_{\alpha\gamma}$ is evaluated. This, after using the definition of  photon propagator (\ref{photprop}) and  the SR assumption (\ref{fotoni}) leads to the following identity:
\be \label{husky}
[1-\Pi_R(\zeta;q^2)]\left[\frac{r_{\gamma}}{q^2}+\int\frac{ds\, \sigma_{\gamma}(s)}{q^2-s+\ep}\right]=1\, .
\ee
For the next purpose  we use shorthand notation
for the function
\be
b_{\gamma}(q^2)\equiv 1-\Re\Pi_R(\zeta;q^2)=1-P.\int\limits_{0}^{\infty}
ds \frac{\rho_{\gamma}(s)(q^2-\zeta)}{(q^2-s)(s-\zeta)}\, ,
\ee
noting  here that P. integration can be performed analytically at least at one (skeleton)
loop level. Evaluating the real  and the imaginary part of Eq. (\ref{husky}) gives
\be
b_{\gamma}(q^2)\left[r_{\gamma}+\, P.\int ds\frac{\sigma_{\gamma}(s)}{q^2-s}\right]=
1-\pi^2\, q^2\, \rho_{\gamma}(q^2)\sigma_{\gamma}(q^2)\, .
\ee
\be
q^2\, b_{\gamma}(q^2)\,  \sigma_{\gamma}(q^2)-\rho_{\gamma}(q^2)
\left[r_{\gamma} +q^2\, P.\int ds\frac{\sigma_{\gamma}(s)}{q^2-s}\right]
=0\, ,
\ee
Putting together it yields resulting {\it Unitary Equation } for the photon spectral function: 
\be  \label{eqforphot}
\sigma_{\gamma}(q^2)=\frac{\rho_{\gamma}(q^2)}{q^2[b^2_{\gamma}(q^2)+\pi^2\rho^2_{\gamma}(q^2)]}\, .
\ee
where the functions $\sigma_{\gamma}$ and $\rho_{\gamma}$ are zero under the threshold
$q^2=4m^2$.

The residuum $r_{\gamma}$ of the photon propagator can be calculated from
\be \label{rezidgam}
r_{\gamma}=\frac{1}{1-\Pi_R(\zeta;0)}\, .
\ee

In the next text we will sketched the derivation of DR (\ref{momforp}). First we briefly
review the method in its perturbative context.

In $4+\epsilon$ dimensions and for spacelike momentum $q^2<0$ the one loop
polarization function can be written   as \cite{COLMAC1974}
%
%{\small
\bea
\Pi(q^2)&=&\frac{4e^2}{3(4\pi)^2}\left\{\frac{2}{\epsilon}+\gamma_E-ln(4\pi)+
ln\left(\frac{m^2}{\mu_{t'H}^2}\right)
-\frac{4m^2}{q^2}-\frac{5}{3}
\right.
\nn \\
&+&\left.(1+2m^2/q^2)\sqrt{1-\frac{4m^2}{q^2}}
ln\left[\frac{1+\sqrt{1-\frac{4m^2}{q^2}}}{1-\sqrt{1-\frac{4m^2}{q^2}}}\right]
\right\}-\delta Z_3\mu_{t'H}^{-\epsilon}\, ,
\eea%}
where $\mu_{t'H}$ is t'Hooft  scale. The  mass-shell
subtraction scheme defines $Z_3$ so that $\Pi_R(0;0)=0$ which
implies that the photon propagator behaves as free one near
$q^2=0$. Choosing $\delta Z_3$ to cancel entire $O(e^2)$
correction we find
\be
\delta Z_3^{MASS}=\lim_{q^2\rightarrow 0}\Pi(q^2)=
\frac{e^2}{12\pi^2}\left[\frac{2}{\epsilon}+\gamma_E-ln(4\pi)+
ln\left(\frac{m^2}{\mu_{t'H}^2}\right)\right]
\ee
and renormalized polarization function in mass-shell renormalization
prescription satisfies well known DR:
\be
\label{drs}
\Pi_R(0;q^2)=\Pi(q^2)-\lim_{q^2\rightarrow0}\Pi(q^2)=\int\limits_{0}^{\infty}
d\omega \frac{q^2}{(q^2-\omega+\ep)\omega}\, \rho(\omega)
\ee
where the absorptive part reads:
\be
\label{oneloop}
\pi\rho(\omega)=\frac{\alpha_{QED}}{3}(1+2m^2/\omega)\sqrt{1-4m^2/\omega}
\, \Theta(\omega-4m^2) \, .
\ee
 Recall that the one loop 
$\Pi_R(0,q^2)$ represents also self-energy calculated in the  popular
$\overline{MS}$ scheme for the special choice of t'Hooft scale
$\mu_{t'H}=m$. Clearly in off-shell momentum subtraction renormalization
scheme:  $\delta Z_3=\Pi(\zeta)$ and 
the DR (\ref{drs}) is generalized to (\ref{momforp}).

In our approximation the fermion propagator entering the photon SDE  is dressed. 
Substituting SR for fermion propagator $S$ into the photon polarization (\ref{polar})
\be
\label{tensor}
\Pi^{\mu\nu}(q)=
ie^2 \int\frac {d^4l}{(2\pi)^4}
\, Tr \left[ \gamma^\nu \, S(l) \,
\gamma^{\mu} \, S(l-q) \right] \, ,
\ee
and performing the standard steps we can arrive at DR where the weight function is now:
%
%{\small
\bea \label{kutek}
\rho_{\gamma}(q^2)&=&\frac{e^2}{12\pi^2}
\int da \, db 
\frac{\lambda^{1/2}(q^2,a,b)}{q^2}
\left[1+\frac{a+b}{q^2}-
\frac{b-a}{q^2}\left(1+\frac{b-a}{q^2}\right)\right] 
\nn\\
&&
\bar{\SG}_v(a)\bar{\SG}_v(b)\Theta\left(q^2-(\sqrt{a}+\sqrt{b})^2\right)
\: .
\eea%}
Clearly form the expression (\ref{kutek}) follows that the functions 
$\rho_{\gamma},\sigma_{\gamma} $ are nonzero only  for $q^2>4m^2$.

\subsection{Numerical comparison}

Like in the ladder approximation considered previously  the resulting set of Unitary Equations have been solved by numerical iterations.
The   cutoffs $\Lambda_S^2=\Lambda_H^2=10^7M^2(0)$ have been used and  the renormalized mass is chosen such that $M(0)=1$.
     
\begin{figure}[t]
\vspace*{-4ex}

\centerline{\mbox{\psfig{figure=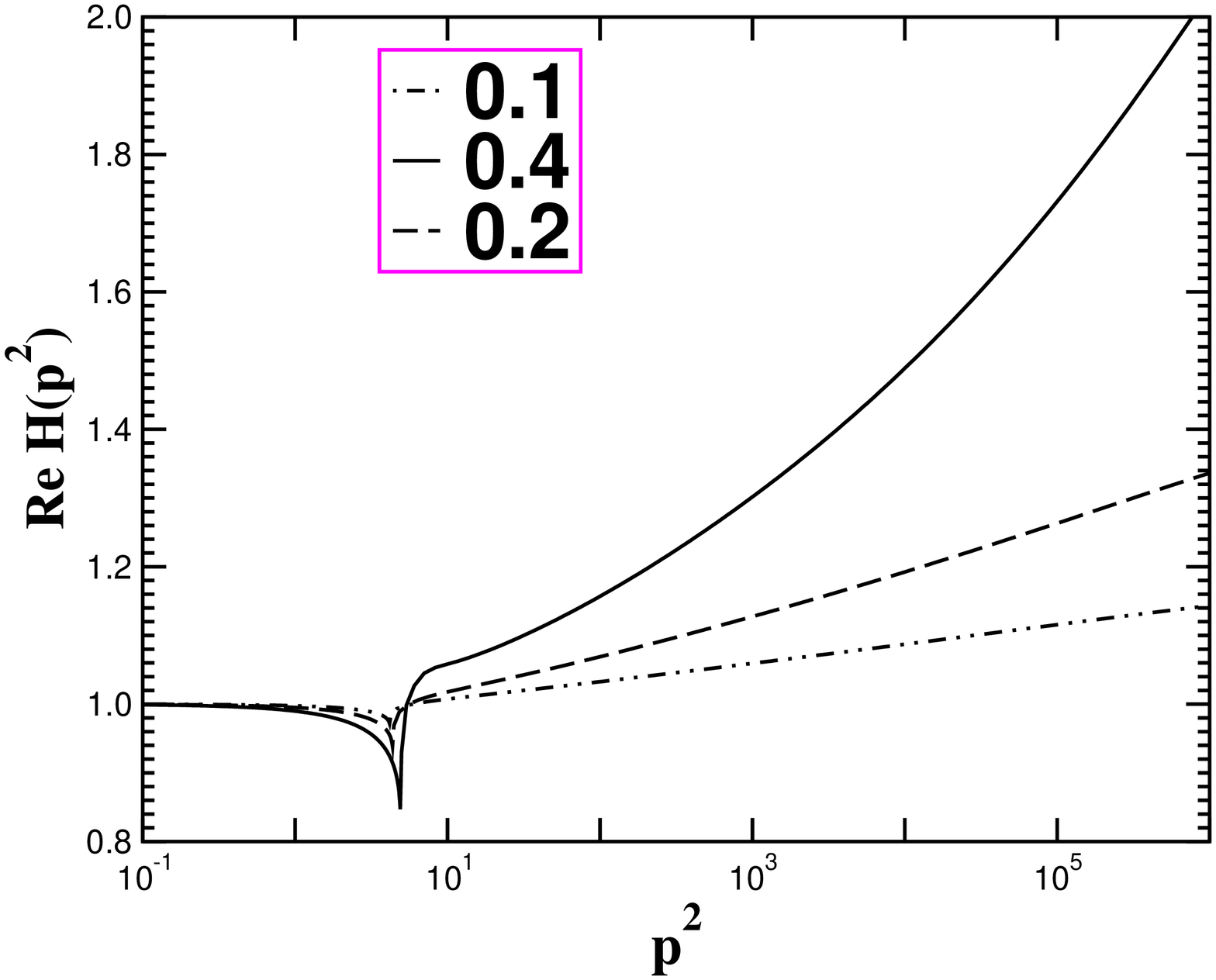,height=24em,angle=270}}}

\centerline{\mbox{\psfig{figure=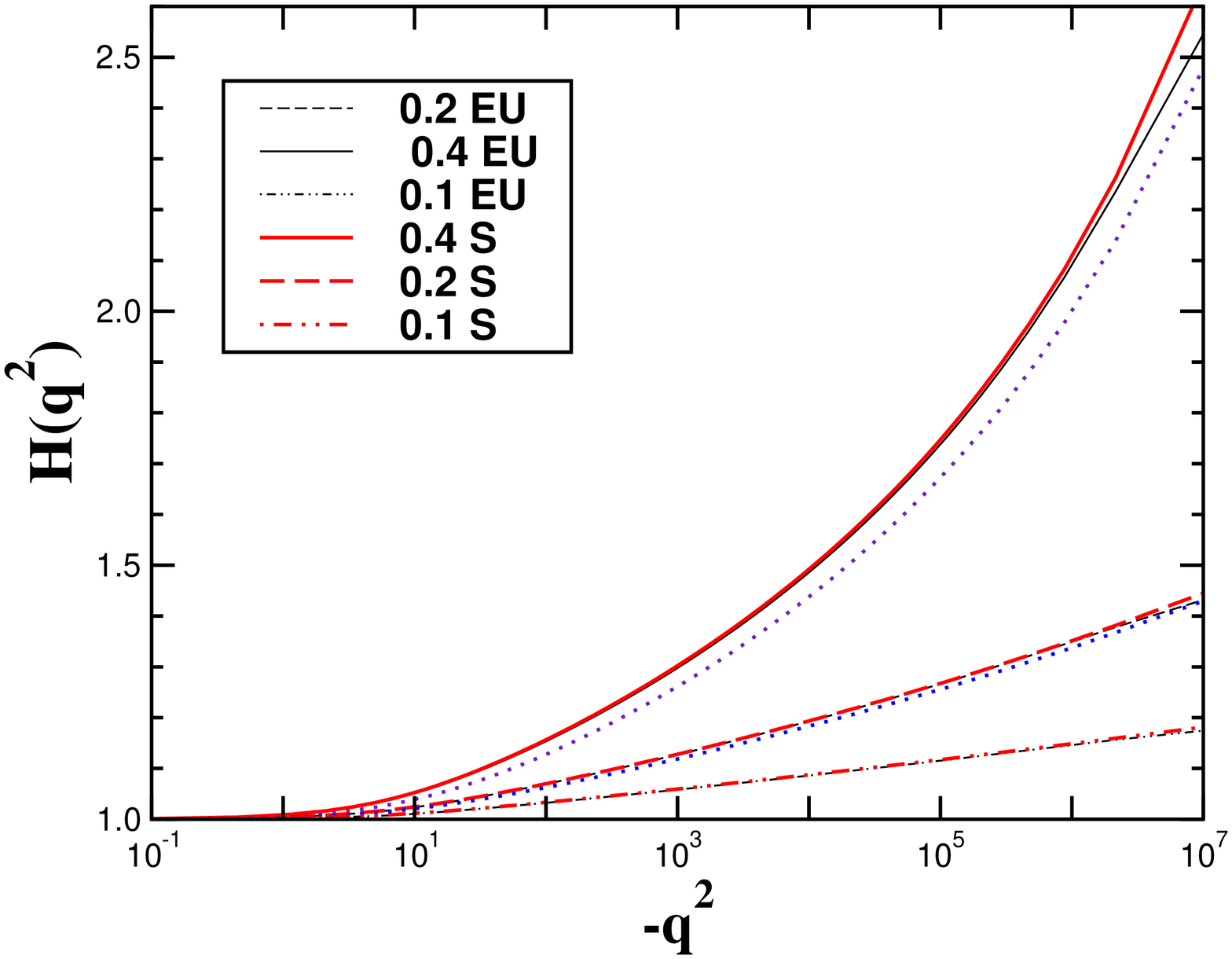,height=24em,angle=270}}}

\caption[caption]{a)~Top figure shows charge renormalization function $H=(1-\Pi_R)^{-1}$ for timelike momenta for the renormalized couplings: $\alpha(0)=0.1,0.2,0.4$. The down oriented peak corresponds to the threshold $4m^2$. \label{figU1} b)~Bottom figure \label{figU2} displays charge renormalization functions for spacelike momenta as obtained by the solution of the SDEs in Euclidean (EU) and Minkowski (S) space are shown in the right figure. Dotted lines stand for one loop perturbation theory.}
\end{figure}

\begin{figure}[t]
\vspace*{-4ex}

\centerline{\mbox{\epsfig{figure=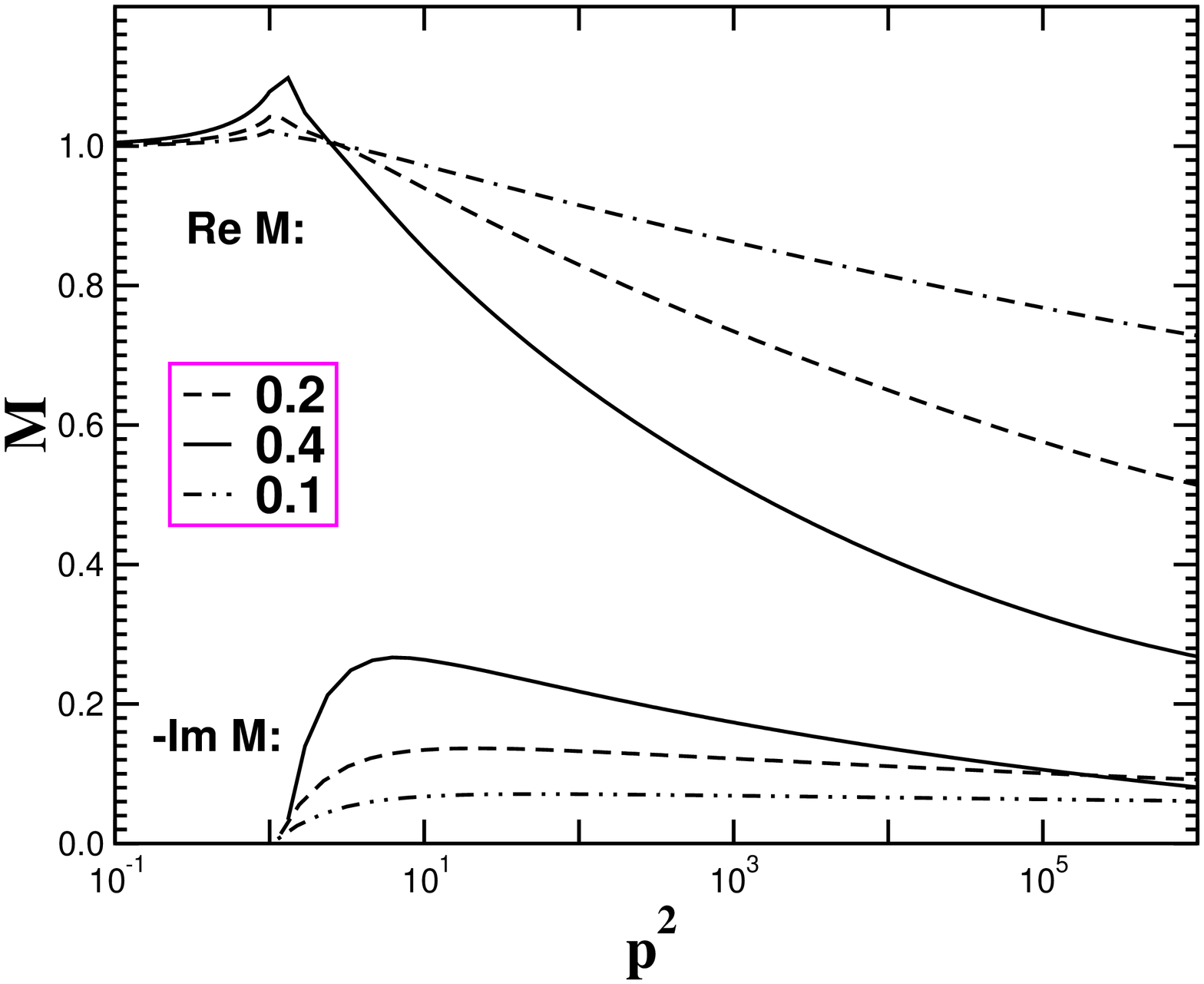,height=24em,angle=270}}}

\centerline{\mbox{\epsfig{figure=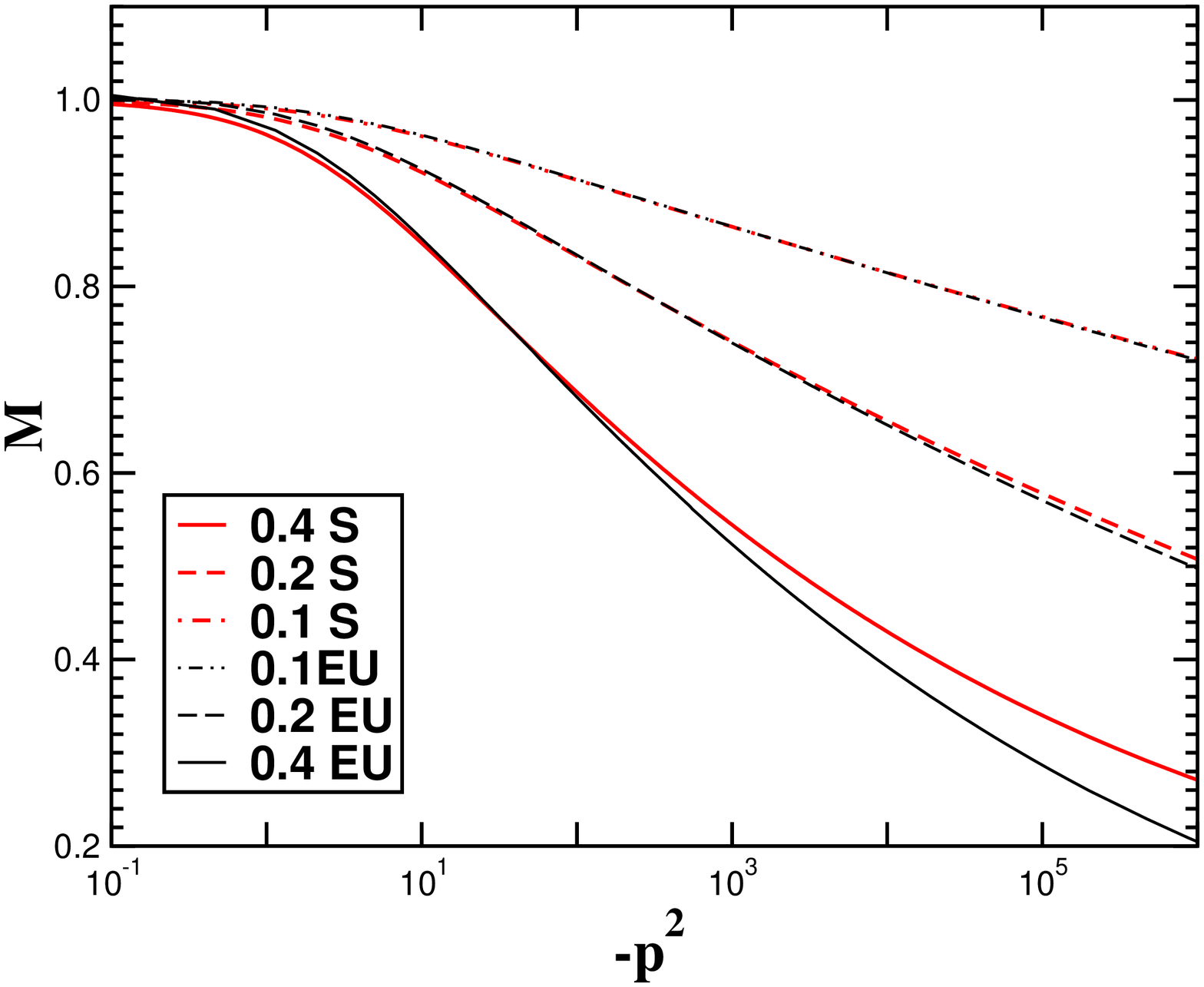,height=24em,angle=270}}}

\caption[caption]{a)\label{figU4} Upper panel displays the mass function $M(p^2)$ for timelike momenta. Note, the sign of the imaginary part,i.e. of ($\pi\rho_s(p^2)$). 
b)\label{figU3} The lower panel displays  comparison of  mass function $M(p^2)$
calculated in Euclidean (EU) and in Minkowski (S) formalism  for spacelike momenta now.}
\end{figure}

The solutions for the running couplings for timelike momenta as have been  obtained from the Unitary Equations  are  displayed in Fig.~\ref{figU1}.  For better visualization the so called photon renormalization function $H_R=[1-\Pi_R(\zeta;q^2)]^{-1}$ is actually plotted. The spectral and Euclidean solutions are compared in Fig.~\ref{figU2} too.  Fermion mass function  obtained from
SDEs is displayed for spacelike regime of momenta in Fig.~\ref{figU3}.  The observed
deviation of Euclidean results from the spectral ones can be explained as a consequence of different cutoffs concepts used in the Euclidean and Minkowski treatment.  The timelike solution for $M$ is shown in  Fig.~\ref{figU4}a) where the absorptive part is also added.  We compare, too, with one loop perturbation theory. In this case we use the pole mass obtained from Unitary Equations as an input for one loop  on-shell renormalized perturbation theory expression.

The large growth of the running coupling with its derivative close to the cutoff (see Fig.~\ref{figU2}) signals a good evidence for the Landau singularity \cite{LANDAU} and hence QED triviality. As the consequence of QED triviality the   collapse of the both  Spectral and Euclidean solution is observed. Contrary to quenched approximation, using the spectral approach here, we have always identified the physical pole mass which is probably  the consequence of the smallness of the coupling at low $q^2$. 

\section{Summary on the Unitary Equations}

%%%%%%%%%%%%%%%%%%%%%%%%%%%%%%%%%%%%%%%%%%%%%%%%%%%%%%%%%%%%%%%%%%%%%%%%%%%%%%%%%%%%%% 

\section{Analyticity of Green's functions in QCD}
\label{chapQCD}

\subsection{Motivation}

Quantum Chromodynamics (QCD) is the only experimentally studied strongly interacting relativistic quantum field theory.  This non-Abelian gauge theory with a gauge group $SU(3)$ has many interesting properties. The dynamical breaking of chiral
symmetry explains why the pions are light, identifying them with the
pseudo-Goldstone bosons associated with the symmetry breaking of the
group $SU(2)_L\times SU(2)_R$ to $SU(2)_V$ (in flavor space).
Asymptotic freedom \cite{GROWIL1973,POL1973} implies that the
coupling constant of the strong interaction decreases in the
ultraviolet region. For less than 33/2 quark flavors, QCD at high
energy becomes predictable by the perturbation theory. However, in the infrared region
the perturbation theory does not work and nonperturbative techniques have to be
applied.

One of the most straightforward nonperturbative approaches is a
solution of the SDEs for QCD. The extensive studies were undertaken
for a quark SDE, based on various model assumptions for a gluon
propagator. These approximate solutions, often accompanied by a
solution of the fermion-antifermion Bethe-Salpeter equation for meson states, have become
an efficient tool for studies of many nonperturbative problems,
e.g., the  chiral symmetry breaking, low energy electroweak hadron
form factors, strong form factors of exclusive processes, etc. (see
reviews \cite{ROBERTS,ALKSMEK2001,MARROB2003,ROBSMI2000,MARTAN2005} and also
recent papers \cite{FISALK2003,BENDER,BICEST2003,BIC2004,FISALKCON}).

\begin{figure}[t]
\begin{fmffile}{fig}
\begin{eqnarray}
\Pi^{\mu\nu}(q)&=&
\parbox{3.0\unitlength}{%
\begin{fmfgraph}(3.0,2.5)
\fmfpen{thick} \fmfleft{i} \fmfright{o} \fmf{zigzag}{v,v}
\fmf{zigzag}{i,v,v,o}
%\fmf{zigzag,left=90}{v,v}
%\fmfdot{v}
\end{fmfgraph}}
\, \, \,+ \, \, \,
\parbox{3.0\unitlength}{%
\begin{fmfgraph}(3.0,2.5)
\fmfpen{thick} \fmfleft{i} \fmfright{o} \fmf{zigzag}{i,v1}
\fmfblob{0.1w}{v1} \fmf{zigzag}{v2,o}
\fmf{zigzag,left,tension=0.25}{v1,v2}
\fmf{zigzag,right,tension=0.25}{v1,v2}
\end{fmfgraph}}% end of parbox
\nonumber \, \, \, + \, \, \,
\parbox{3.0\unitlength}{%
\begin{fmfgraph}(3.0,2.5)
\fmfpen{thick} \fmfleft{i} \fmfright{o} \fmf{zigzag}{i,v1}
\fmfblob{0.1w}{v1} \fmf{zigzag}{v2,o}
\fmf{dashes,left,tension=.2}{v1,v2}
\fmf{dashes,right,tension=.2}{v1,v2}
\end{fmfgraph}}% end of parbox
\\
\nonumber
\\
&+&\parbox{3.0\unitlength}{%
\begin{fmfgraph}(3.0,2.5)
\fmfpen{thick} \fmfleft{i} \fmfright{o} \fmf{zigzag}{i,v1}
\fmfblob{0.1w}{v1} \fmf{zigzag}{v2,o}
\fmf{zigzag,left,tension=0.2}{v1,v2} \fmf{zigzag,tension=0.2}{v1,v2}
\fmf{zigzag,right,tension=0.2}{v1,v2}
\end{fmfgraph}}
\, \, \, + \, \, \,
\parbox{3.0\unitlength}{%
\begin{fmfgraph}(3.0,2.5)
\fmfpen{thick} \fmfleft{i} \fmfright{o} \fmf{zigzag}{i,v1}
\fmfblob{0.1w}{v2} \fmfblob{0.1w}{v3} \fmf{zigzag}{v3,o}
\fmf{zigzag,left,tension=0.2}{v1,v3}
\fmf{zigzag,left,tension=0.1}{v1,v2}
\fmf{zigzag,right,tension=0.1}{v1,v2}
\fmf{zigzag,right,tension=0.2}{v2,v3}
\end{fmfgraph}}
\, \, \, + \, \, \,
\parbox{3\unitlength}{%
\begin{fmfgraph}(3.0,2.5)
\fmfpen{thick} \fmfleft{i} \fmfright{o} \fmf{zigzag}{i,v1}
\fmf{zigzag}{v2,o} \fmfblob{0.1w}{v1}
\fmf{fermion,left,tension=.2}{v1,v2,v1} \fmfv{dot}{v1}
\end{fmfgraph}}% end of parbox
\nonumber
\end{eqnarray}
\end{fmffile}
\caption{\label{SDEgluonfig}Diagrammatical representation of the
Schwinger-Dyson equation for gluon self-energy. All  internal
propagators are dressed, as well as  the vertices labeled by blobs.
Wavy lines represent  gluons, dashed lines represent  ghosts and the
solid lines stand for quarks.}
\end{figure}

However, to take gluons into account consistently  is much more
difficult than to solve the quark SDE alone. The SDE for the gluon
propagator (see Fig.~\ref{SDEgluonfig}) is more nonlinear than the
quark one. Moreover, the Fadeev-Popov ghosts have to be included
\cite{FADPOP1967} in a class of Lorentz gauges. 
In a class of covariant linear gauges the contributions to the  gluonic polarization function 
are sketched in Fig.~\ref{SDEgluonfig}.
Although it is not so difficult to derive and write down the formal dispersion relation for vacuum polarization function   (especially when one use bare vertices)
 we were not yet able to succeed 
with a numerical solution of Unitary Equations.
Therefore, after a  brief overview of selected representative  
SDEs solutions in Yang-Mills theory, 
instead of solving the SDEs we use the
generalized (nonpositive) spectral representation to fit the spectral function to
Euclidean solutions obtained in recent lattice simulations. 
This is a first step towards the understanding of analytical 
structure of infrared QCD.

%----------------------------------------------------------------------------

\subsection{Cornwall's approach}
 
In this section we mention a symmetry preserving  gauge invariant solution to gluon propagator obtained by Cornwall two decades ago \cite{COR1982}. To our knowledge this is the best published example, in which the behavior of the QCD Green function in the whole range of Minkowski formalism is addressed within the framework of the SDEs and spectral technique. To this end,  the gauge invariant GFs have been constructed by means of the Pinch Technique. To achieve this particular spectral Ansatz for GFs were employed.  The resulting vacuum polarization function is then by construction gauge-fixing independent. We leave aside the consistent nonperturbative proof of this statement, but we refer to the appropriate perturbative proof in ref.~\cite{BINPAP2004}. Let us now describe the basic steps of Cornwall's procedure.

As in the usual approach, the gauge fixing is imposed in the first
step: The light-cone axial gauge is employed. Avoiding thus SR for ghosts,
only the following  SR for the gluon propagator is assumed:
\bea
G_{\mu\nu}(q)&=& P_{\mu\nu}\, G(q^2) = P_{\mu\nu}\, \int d\omega
\frac{\sigma(\omega)}{q^2-\omega+i\epsilon} \, ,
\\
P_{\mu\nu}&=&-g_{\mu\nu}+
\frac{n_{\mu}q_{\nu}+n_{\nu}q_{\mu}}{(n\cdot q)} \, ,
\eea
where for $n_\mu$ it holds $n_\mu n^\mu=0, n\cdot A=0$. $G(q^2)$ is
the  gauge-fixing independent scalar part of the
 gauge-invariant propagator $G_{\mu\nu}(q)$. The  Ward identity
\be \label{gluonwti}
k_1^{\alpha}\Gamma_{\alpha\beta\gamma}(k_1,k_2,k_3)=
G_{\beta\gamma}^{-1}(k_2)-G_{\beta\gamma}^{-1}(k_3)  \, .
\ee
between the gluon propagator and the three-gluon  vertex
$\Gamma_{\alpha\beta\gamma}(k_1,k_2,k_3)$ has to be satisfied. To
enforce this identity the   Gauge Technique  was
used, which makes the following Ansatz for the longitudinal part of
the untruncated three-gluon proper GF:
\bea
\label{coans}
G_{\alpha\beta\gamma}&=&
G^{\beta\beta'}(k_2)\Gamma^{L}_{\alpha\beta'\gamma'}G^{\gamma\gamma'}(k_3)
\nn \\
&=&\int d\omega\,
\bar{\sigma}(\omega)\left[\frac{P^{\beta\beta'}(k_2)}{k_2^2-\omega+i\epsilon}\,
\Gamma_{\alpha\beta'\gamma'}^{(\omega)}(k_1,k_2,k_3)\,
\frac{P^{\gamma\gamma'}(k_3)}{k_3^2-\omega+i\epsilon}\right] \, .
\eea

Recall that the Gauge Technique was originally
developed by Salam \cite{SAL1963} as an attempt to solve a vector
electrodynamics and  later extended and applied to the scalar
\cite{SALDEL1964} and fermion \cite{STRAT} QED, for later studies based on the Gauge Technique see 
\cite{DELWES1977,WEST2,DEL1979,HOS2002}.
Substituting the Ansatz (\ref{coans}) into the approximate pinched
gluon SDE \cite{COR1982} yields for the inverse of $G(q^2)$:
\bea
\label{minkqcd}
G^{-1}(q^2)&=&q^2-\Pi(q^2) \nn\\
&=&Z_3\left\{q^2\left[1+\frac{ibg^2}{\pi^2}\int d^4k\int d\omega
\frac{\sigma(\omega)}{[(k+q)^2-\omega](k^2-\omega)}\right]\right.
\nn \\
&+&\frac{ibg^2}{11\pi^2}\int d^4k\int
d\omega \frac{\omega\sigma(\omega)}{[(k+q)^2-\omega](k^2-\omega)}
\,, \\
&-&\left.\frac{i4bg^2}{11\pi^2}\int d^4k\,  G(k^2)\right\} \, , \eea
where $b$ is the lowest-order coefficient in the $\beta$ function
\be \beta=-bg^3+ \dots\, , \quad \quad b= \frac{33}{48\pi^2} \, , \ee
and $Z_3$ is the coupling renormalization constant:
\be
\nn \\
Z_3= g^2/g_0^2 \, .
\ee

Equation (\ref{minkqcd}) was not solved directly in Minkowski space,
it was converted into the  Euclidean space. Defining
\be \label{tramin}
d(q^2)=-G(-q^2)\, ,
\ee
it was found  that the trial propagator
\bea
\label{corfr} d^{-1}(q^2)&=&[q^2+m^2(q^2)]\, b\, g^2 \,
\ln\left[\frac{q^2+4m^2(q^2)}{\Lambda^2}\right] \, ,
\nn \\
m^2(q^2)&=&m^2\left[\frac{\ln\left(\frac{q^2+4m^2}{\Lambda^2}\right)}
{\ln\frac{4m^2}{\Lambda^2}}\right]^{-12/11} \, ,
\eea
where $g, m$ and $\Lambda$ are fitted parameters, is an excellent fit
to the numerical solution of Eq.~(\ref{minkqcd}). The scale is fixed
by the condition $d^{-1}(q^2=\zeta)= \zeta$, where $\zeta$ is the
renormalization scale \cite{COR1982}.

Let us stress  the main physical results obtained from the
combination of the spectral Gauge and Pinch techniques.
First of all it seems  possible to reorganize  the SDEs of QCD
into a form  which allows to deal directly with a gauge independent
GFs. The second important feature is that the solution  has a mass gap.
There is no singularity corresponding with massless excitation since 
the gluon propagator is finite at zero momenta.
For low $q^2$  the mass $m(q^2)$ varies within $500\pm200$ MeV,  for
large $q^2$ it vanishes, in agreement with the expected
(perturbative) UV behavior. Note that  similar behavior was obtained
by the operator product expansion method several years later in
\cite{LAV1991}:
\be
m^2(Q^2)\simeq \frac{34 N_c}{\pi^2 9(N_c^2-1)}
\frac{<\frac{\alpha_s}{\pi} F_{\mu\nu} F^{\mu\nu}>}{Q^2} \, ,
\ee
where $Q$ is the gluon Euclidean momentum, $<\frac{\alpha_s}{\pi} F_{\mu\nu}
F^{\mu\nu}>$ is the  gluon condensate \cite{SVZ1979} and $N_c=3$.

Let us recall also some recent results supporting the idea (suggested
long time ago \cite{PARISI})  of the dynamical gluon mass. It should
be emphasized that the gluon ``mass" is not a directly measurable
quantity, but it can be related to other physical quantities. The
lattice results \cite{GUPTA1987,LAT4,LAT3,LAT2,LAT1,BOBOLEWIZA2001}
seem to support this concept. Parisi and Petronzio \cite{PARISI} were
first to introduce it into description of $J/\psi$ decays.

{\subsection{Gluons and ghost in linear covariant gauges}

 The solution of SDEs in strong coupling non-Abelian theory like infrared QCD is 
strongly dependent on the gauge fixing of the QCD action  and on the truncation of 
equations system as well. 
Since the coupling increases as one goes from the ultraviolet to the lower square 
of momenta, various approximations can  lead to the different analytical properties 
of Green's functions. A quite selfcontaining history of SDEs
approach and phenomenology in QCD can be find in the paper \cite{SMEALK2000}
where the present knowledge about 
the infrared behaviour of QCD Green's functions is 
summarized in the review (see eg. the Chapter 5).
Here we will  briefly mention the main results obtained in the Landau gauge, 
noting that the most of the 
recent results are obtained in this gauge.

In the initial study of SDEs \cite{MAND1979}
Mandelstam provisionally neglected 
the effect of ghosts. As the consequence of this first approximation it was shown that gluon 
propagator is more singular then simple pole in this case. Using the similar truncation this behaviour was confirmed in the studies 
\cite{BROPEN1989,BUTPEN1995}. It was first suggested in \cite{SMHAAL1997}
 that such behaviour is not an inherent 
property of nonpertubative dynamics of Landau gauged QCD but rather of deficiency of
truncation of SDEs system. It was suggested that this is not gluon, but rather ghost propagator 
that is highly singular in the zero momentum limit. The detailed analytical and numerical study of the coupled ghost and gluon equations in Landau gauge with bare vertices was made 
\cite{ATKBLO1998,ATKBLO1998II}. It was actually confirmed that 
the ghost loop  dominates 
the infrared behaviour of the gluon propagator.   
In recent papers
\cite{FISALK2003,BLOCH2003,LERSME2002,KONDO2003,AGUNAT2004} studies
of the coupled SDEs for gluon and ghost propagators in  the usual  Landau
gauge in Euclidean space were performed in various approximations.
An interesting result of the investigations in Landau gauge is the observation, 
that there is no qualitative difference of the solutions found with bare vertices or with vertices dressed by the use of Slavnov-Taylor identities. Note that the similar behaviour was observed in the wide class of ghost-antighost symmetric gauges \cite{AFRS2003}. 
The attempt to estimate the effect of 
two loop irreducible contributions was made in  \cite{BLOCH2003}. The numerical result 
indicates slight changes in the intermediate regime. Note that the inclination of 
this result towards the lattice data was shown already in the introduction of 
presented paper. 

\subsection{Analytical continuation of Euclidean solutions}
The  assumption of GF's analyticity and hence the assumption of
the propagator spectral representation 
was key point for  the formulation of the Unitary Equations and 
the solution of SDEs in Minkowski space. The requirement
of the propagator (or invariant charge) analyticity is 
supposed be a good assumption in QCD for many reasons 
\cite{KON2003,NES2003,NESPAP2004,ALEK2005} and the analytical 
structure of the GFs obtained form Euclidean SDEs in QCD  is  
under current investigation \cite{ADFM2003}.

In this Section we perform analytical continuation of  recent unquenched
lattice results \cite{lat2003}  for gluon propagator to the Minkowski space. 
Note that the similar fit has been already performed for the case of pure 
gluodynamics \cite{SAULIPHD} where the quenched
lattice date \cite{BOBOLEWIZA2001} has been used. In the paper
\cite{lat2003}
 the authors considered three flavor QCD.
For this purpose one massless and one massive quark have been taken into account.
The main difference that follows from unquenching of
lattice gluon propagator
is the partial suppression of infrared enhancement plateau. For some details see  the original 
paper \cite{lat2003} where the comparison is  made.
Of course, taking the effects of quark-antiquark pairs into account it
has no influence on the main characteristic feature of the Landau gauge gluon propagator:
it is obviously infrared vanishing function. 
Therefore, in contrast to the propagator of a
stable unconfined particle,  we assume that the spectral function is
a smooth real function, i.e., propagator has not simple pole structure.

In fact, what we actually do, it is the continuation of the data obtained for a real negative
momentum semiaxis to the border of analyticity domain. This is the standard situation when one 
make prediction from experimentally obtained data to the experimentally inaccessible regime (the extraction of nucleon spectral function from the electromagnetic form factor is a good example). This approach has been already extended to the case of gluon spectral function  of $SU(2)$ Yang-Mills theory ref.~\cite{LAREGA}. However, in our case, the smallness of errors of the lattice simulations  and the value of lattice spacing as well
allows to use the data directly for our purpose. No extrapolation to higher square of momenta is made and we neglect the computational errors as well. 

The gluon spectral function $\sigma(\omega)$ and the most probable analytical fit is obtained from
the minimization of
\bea
I_{fit}&=& \sum_i \left[ \tilde{H}(x_i)- H(x_i)
\right]^2  \, , \label{uzto}\\
\eea
where $H$ is the gluon form factor defined as $H(x)=xG(x)$ , $\tilde{H}(x)$ are the appropriate
lattice data evaluated for the square of momenta $x=q^2_E$  while $H(x)$ is obtained through the fit of the spectral function which enters the generic SR for the
gluon propagator in the Landau gauge which reads
\bea
G^{\alpha,\beta}_{AB}(q)&=&\delta_{AB}\,
\left[-g^{\alpha\beta}+\frac{q^{\alpha}q^{\beta}}{q^2}\right]\,
G(q^2) \, , \nn\\
G(q^2)&=&\int\limits_{0}^{\infty} d\omega\,
\frac{\sigma(\omega)}{q^2-\omega+i\epsilon} \, ,
\label{gluesp}
\eea
where $A,B$ are color indices. The Ansatz (\ref{gluesp}) should be considered as the
generalized spectral representation, since we do not assume (and do not obtain) the spectral function $\sigma(\omega)$ positive for all values of $\omega$. For completeness let us note that we use unrenormalized data, i.e. the suitable but for us  irrelevant constant prefactor should be added when considered for some application.

To achieve a reasonable accuracy we use the following iterative minimizations procedure:
The function $H_J$ evaluated at  some $J$-step of our iteration cycle is in each case 
obtained by the previously obtained fit for spectral function ,i.e. $\sigma_{J-1}$
such hat  the relation for $H_J$ reads:
\be
H_J(x_i)=x_i\int\limits_{0}^{\infty} d\omega\,
\frac{\sigma_{J-1}(\omega)}{x_i+\omega}
\sum_{l=0}^3 a_lP_l(z) \, , \quad \quad
z=\frac{\omega-c}{\omega+c} \, \, ,
\label{fit1}
\ee
where $a_i$ are dimensionless constant coefficients of Legendre polynomial which are
fitted during the minimization procedure, $c$ is positive massive parameter 
which is use for the fit as well. Some suitable zeroth iteration input 
is used for $\sigma_{0}$. For this purpose we use the convenient choice of the form:
$\sigma_0(\omega)=K \frac{log^{-\gamma-1}(e+\om/\Lambda^2)}{\om+\Lambda^2}$ with some 
suitable parameters $K,\gamma,\Lambda$.
Assuming that   the procedure is convergent then
the best fit is then obtained when $a_0=1$ and other coefficients of Legendre 
polynomial are zero, which should  be true irrespective of the value of the constant $c$.
We actually observe the relatively fast convergence of the described process 
and we stopped our minimization procedure when $a_0=1.0006$ and other $a$'s
are at most few of $10^{-3}$.
% At this stage $\chi=2$.
Such minimization procedures is much powerful then to perform  minimization 
at once time with some 'more suitable' function even with larger number of the fitting parameters.
As a bonus the computer time is reasonable short (seconds for office PC).

\begin{figure}[t]
\centerline{\epsfig{figure=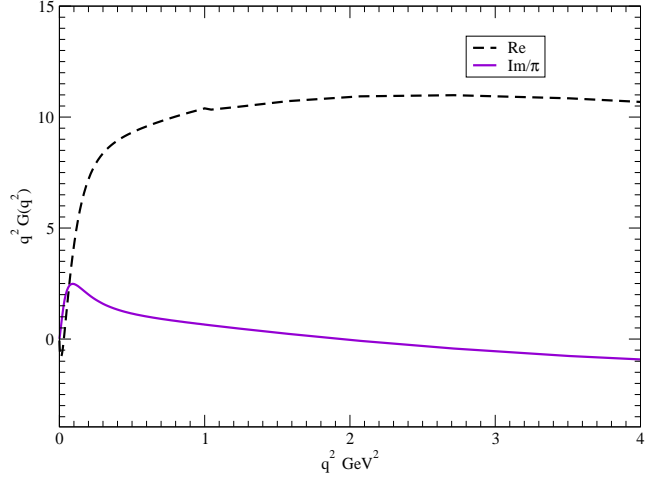,width=8truecm,height=10truecm,angle=270}}
\caption[caption]{\label{gluontime}
The extrapolated gluon form factor $H(x)= xG(x)$ at timelike momentum axis.}
\end{figure}

For a vanishing gluon propagator we have the following ``sum rule'' for $\sigma$
\be
G(0)=\int\limits_0^\infty \frac{\sigma(\omega)}{\omega} \simeq 0 \, ,
\ee
which is clearly impossible for the positive  $\sigma$.
Note that the violation of positivity was confirmed by some other lattice
simulations \cite{CUMETA}, where  it was  indicated 
by the study of Schwinger function.
The extrapolated  solution is plotted in Fig.~\ref{gluontime}. The spectral
function has a positive peak at the infrared  and
becomes negative for asymptotically large
$p^2$. The infrared behaviour describes the shortliving gluon excitation and is consistent 
with the absence of  gluon in the physical spectrum while 
the large momentum behaviour is the feature already expected   
from the perturbation theory. Clearly, the observed shape of the  
gluon propagator in Landau gauge is in
accord with our physical expectations, e.g. with the confinement of color
object and asymptotic freedom as well. 

\begin{figure}[t]
%\centerline{{\mbox{
\centerline{
\epsfig{figure=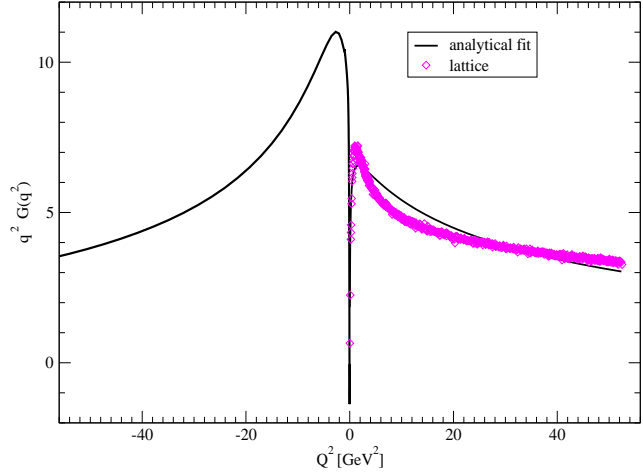,width=8truecm,height=10truecm,angle=270}}
\caption[caption]{\label{gluonspace} Gluon propagator for $N_f=3$ QCD, the lattice data
 \cite{lat2003} are represented by  diamonds. The fit (at spacelike) and the extrapolation (at timelike) is labeled by solid line.}
\end{figure}

In  Fig.~\ref{gluonspace} we also display our fit at the spacelike regime
and provide the comparison with the original lattice data \cite{lat2003}.
The real part of extrapolated  gluon propagator is added for comparison too
(negative $Q^2$ semiaxis now). From this we can see that we
were not able to fit the lattice data exactly. The origin of this small 
discrepancy can be a combination of several effects: it can be artifact of our 
iteration/minimization procedure, it can signal the the complex singularities 
however positioned rather close the real positive momentum square semiaxis, partially it can be 
a consequence of finiteness of the lattice spacing and finite lattice volume.
This complications could be clarified in future studies
but they have a little significance for the main statement:
The Landau gauge gluon propagator is the analytical function in the whole complex plane 
with the exception of (or small vicinity of) a real positive semiaxis of square of momenta and
it can be  parameterized (or at least reasonable approximated) 
by nonpositive spectral representation.

%%%%%%%%%%%%%%%%%%%%%%%%%%%%%%%%%%%%%%%%%%%%%%%%%%%%%%%%%%%%%%%%%%%%%%%%%%%%%%%%%%%%%%%%%%%%%%%%%%%%%%%%%%%%%%%%%%%%%%%%%%%%%%%%%%%%%%%%%%%%%%%%%%%%%%%%%%%%%%%%%%%%%%%%%%%%%%%%%%%%%%%%%%%%%%%%%%%%%%%%%%%%%%%%%%%%%%%%%%%%%%%%%%%%%%%%%%%%%%%%%%%%%%%%%%%%%%%%%%%%%%%%%%%%%%%%%%%%%%%%%

\section{Summary, conclusion and outlook}

An important question, in field theory in general, and in the SDE approach particular, is the 
connection between Euclidean-space and physical Minkowski space. As we explicitly demonstrated
it is clear that to perform SDEs studies directly in Minkowski space is possible. 

We began with an introduction to the analytical concept in the formalism of Schwinger-Dyson equations with specific emphasizes to the derivation and solution of the Unitary Equations in scalar models and QED. For the later case we presented
 studies of strong-coupling QED in  different approximations. 
One of them was unquenched QED, which represents the study where the running of the coupling has been correctly taken into account. In this case the QED triviality plays its own crucial 
role in the   asymptoticly large spacelike momenta and we explain how to deal with this trivial theory correctly within the formalism of spectral representations and dispersion relations.
The Unitary Equations presented  are systematically improvable by considering further and further skeleton contributions to the effective action. 
If the analyticity is correctly maintained then the all momentum integrations can be performed analytically. For instance, let us consider the two-loop improved approximation of our scalar model considered in the Section 3.  In this case it is sufficient to add the contribution from the two-loop graphs to the weight function (\ref{PP}), i.e.
\be
\rho(p^2)\rightarrow \rho(p^2)+\rho_{sun}(p^2)+... \,,
\ee   
where $\rho_{sun}(p^2)$ is given by rhs. of Eq.(\ref{enda}) and the dots stand for the terms stemming from the two other known skeletons not explicitly considered here.
This is a clear technical advantage of the presented method: 
we can immediately apply the known original results for absorptive parts of  Feynman diagrams when they were already evaluated in perturbation theory. 

For an explicitly massive theory and up to the certain values of the coupling constant the  accurate comparison of SDEs solutions at spacelike momentum axis with the solutions performed by more conventional strategy in Euclidean space gives us strong belief in the relevance of presented methods.  
In this cases the only limitation of spectral  method is given by the increasing number  
of spectral integrals in the expression for DR that is directly dictated by the number of internal propagators. Nevertheless, within contemporary power of computer facilities many of exciting calculations can be made and remain to be done.

The strong coupling QED is often regarded as an reasonable pedagogical tool  for SDE studies
and their application to QCD. We also attempted to extend the discussion to QCD where
the direct Minkowski space formulation and solution is presently prevented by the partial lack of information on the timelike axis. While many important steps have been accomplished, 
it is apparent that much more needs to be done. Furthermore, at present time and up to a few exeptions, the vast amount of   studies 
dealing with the deep nonperturbative phenomena like chiral symmetry breaking and dynamical mas generation  were carried out in Euclidean space. Regarding the real QCD, it is not clear when or if it will be possible to perform such symmetry breaking SDEs studies directly in Minkowski space.      

Another interesting prospective is to further explore the timelike infrared behavior of Green's functions by analytical continuation of presented lattice studies. We offered the solution for
Landau gauge gluon propagator, but the information about the timelike structure of quark propagator is currently unavailable. The  present day great improvements in the 
QCD lattice studies could provide significant benefits at this area in the recent and in the near future.

We also collected the necessary information and illustrated how it can be used in relativistic bound state calculations using the spectral representation  for solving the Bethe-Salpeter equation in (3+1) Minkowski space. To further develop the method, it would be interesting to extend it to more complicated  kernel: trying to include the cross boxed contributions, $s$ and $u$ channel interactions etc. It is already known that the ``spectral approach'' used here is  suitable even for more complicated systems. The same or very similar tricks and technology when successfully performed in the Minkowski calculation could become great goals in hadronic physics in the not too distant future. As it is already known the obstacles due to fermionic degrees of freedom can be overcome. The solution of the spinless bound state (pion) composed from two quarks interacting via dressed gluon exchange is under auspicious consideration.

\begin{center}{\bf Acknowledgments}\end{center}
I am grateful to Ji\v{r}\'{i} Adam, jr. for the stimulating discussions and useful comments.
The author thank to P. Bowmann for sending me the recent lattice data.
The author  also thank to  W. Guttinger, M.B. Halpern and B. Mihaila for the useful correspondence. This work is supported in part by the grant GA\v{C}R 202/03/0210 and by the ASCR project K1048102.}

\appendix
\section{}
%%
%\normalsize{
The standard trick is to re-write the integral $J(p^2)$ in
terms of  the Feynman parameterization:
\bea
\label{tycka}
J(p^2)&=&\int\frac{d^dk_1}{(2\pi)^d}\int \frac{d^dk_2}{(2\pi)^d}
\int_0^1 dx\, \frac {D(k_2+p,\al_3)}
{\left[k_1^2+k_2^2\, x(1-x)-\al_1x-\al_2(1-x)
+\ep \right]^2}
\nn \\
&=&\int \frac{d^dk_2}{(2\pi)^d}\int _0^1 dx\,
\frac{i\,D(k_2+p,\al_3)\Gamma(2-d/2)}{(4\pi)^{d/2}[x
(1-x)]^{2-d/2}\left[k_2^2-\frac{\al_1}{1-x}-\frac{\al_2}{x}+\ep\right]^{2-d/2}}
\nn \\
&=&\int_0^1 dx dy \int \frac{d^dk_2}{(2\pi)^d}
\frac{i(4\pi)^{-d/2}\Gamma(3-d/2) y^{1-d/2}}{[x(1-x)]^{2-d/2}}
\nn \\
&&\frac{1}{\left[k_2^2+p^2 y(1-y)-[\frac{\al_1}
{1-x}-\frac{\al_2}{x}]y-\al_3(1-y)+\ep \right]^{3-d/2}}
\nn \\
&=&\int_0^1 dx dy \frac{(-)^{2-d/2}\Gamma(3-d) y^{1-d/2}}
{(4\pi)^{d}[x(1-x)]^{2-d/2}[y(1-y)]^{3-d}\left(p^2-\Omega+\ep\right)^{3-d}}
\nn \\
&&\Omega=\frac{S(x)}{1-y}+\frac{\al_3}{y}
\quad ; \quad
S(x)=\frac{\al_1}{1-x}+\frac{\al_2}{x} \, .
\eea
To make the subtractions (\ref{subtrakce}) it is convenient to use the
following Feynman formula:
\be A^{-q}-B^{-q}=\int_0^1 dw \frac{q(B-A)}{[(A-B)w+B]^{q+1}} \, .
\ee
Using variable $z$ for the mass subtraction and the variable $u$ for
the subtraction of the term with derivative, we can immediately write
down a formula for renormalized $J$:
\bea  \label{rex}
&&J_R(p^2)= J(p^2)- J(\zeta) -
\left. \frac{dJ(p^2)}{d p^2} \right|_{p^2=\zeta}(p^2-\zeta)= \nn\\
&&
\int_0^1 
\frac{dx\, dy\, dz\, du\, \Gamma(5-d)\, z(-1)^{2-d/2}\, y^{1-d/2}\,(p^2-\zeta)^2}
{(4\pi)^d\, [x(1-x)]^{2-d/2}\, [y(1-y)]^{3-d}
\left[(p^2-\zeta)zu+\zeta-\Omega+\ep \right]^{5-d}}\, .
\eea
In four dimensional spacetime the integral (\ref{rex}) is already
finite and the limit $d \rightarrow 4$ can be simply taken.

In addition we make the following substitution:
$\omega=\zeta+\frac{\Omega-\zeta}{zu}$. After changing  ordering of
the integrations $d\omega \leftrightarrow dz $ and  integrating over
 $z$ gives:
\be
\label{rexato}
J_R(p^2)=\int_0^1 dx dy\, \int_{\Omega}^{\infty} d\omega\,
\frac{(p^2-\zeta)^2(\omega-\Omega)}
{(4\pi)^4(\omega-\zeta)^2(p^2-\omega+\ep)}\, .
\ee
Notice that all $x,y$ dependence is  now hidden in the lower bound
$\Omega$, which is defined in last line  of (\ref{tycka}). It is
still possible to evaluate one more integral analytically, we have
chosen to take the integral over $y$.  Changing again the order of
integrations we get for boundaries:
\be
\int_0^1 dy \int_{\Omega}^{\infty} d\omega
\rightarrow
\int_{(\sqrt{S(x)}+\sqrt{\al_3})^2}^{\infty} d\omega  \int_{y_-}^{y_+} dy
\quad,
\ee
where $y_-, y_+$ are the roots of the equation $\omega-\Omega=0$:
\be y_{\pm}=\frac{\omega-s+\al_3\pm\sqrt{\lambda(S(x),\omega,\al_3)}}
{2\omega} \, ,
\ee
where $\lambda(x,y,z)$ is the triangle function defined in (\ref{xnula}).  After the explicit integration over $y$ and changing order of the last two integrals we arrive at the following result:
\bea \label{result}
J_R(p^2)&=& \int_{(\sqrt{\al_1}+\sqrt{\al_2}+\sqrt{\al_3})^2}^{\infty}
d\omega\, \frac{(p^2-\zeta)^2 \rho_J(\omega)}{(\omega-\zeta)^2(p^2-\omega+\ep)} \,
\\
\rho_J(\omega)&=&\frac{1}{(4\pi)^4}\int_{x_{-}}^{x_{+}}dx
\left[\frac{(\omega-S(x)+\al_3)\sqrt{\lambda(S(x),\omega,\al_3)}}{\omega}\right.
\nn \\
&-&\left.\al_3\ln\left(\frac{\omega-S(x)+\al_3+\sqrt{\lambda(S(x),\omega,\al_3)}}
{\omega-S(x)+\al_3-\sqrt{\lambda(S(x),\omega,\al_3)}}\right)\right] \, ,
\label{result2}
\\
x_{\pm}&=&\frac{(\sqrt{\omega}-\sqrt{\al_3})^2+\al_1-\al_2\pm
\sqrt{\lambda(\al_2,\al_1,(\sqrt{\omega}-\sqrt{\al_3})^2)}}
{2(\sqrt{\omega}-\sqrt{\al_3})^2} \, .
\nn
\eea
If  the omitted prefactor is restored Eq.~(\ref{result}) we can simply 
identify the one dimensional integral representation for imaginary part of
$\Pi_R/\pi$.

This result can be rewritten into the form already available in
literature \cite{BBBB},\cite{BDR}.  To this end we have to make
the following substitution $x\rightarrow s$ such that
\bea
s&=&S(x)\quad; \quad  \Longrightarrow
\nn \\
x&=&\frac{s+\al_1-\al_2+\sqrt{\lambda(\al_2,\al_1,s)}}{2s}
\quad  \mbox{for}\quad  x>x(s=(\sqrt{\al_1}+\sqrt{\al_2})^2)
\nn \\
x&=&\frac{s+\al_1-\al_2-\sqrt{\lambda(\al_2,\al_1,s)}}{2s}
\quad  \mbox{for}\quad x<x(s=(\sqrt{\al_1}+\sqrt{\al_2})^2)
\eea
which leads to the relation given in \cite{BBBB},\cite{BDR}:
\be
\rho_J(\omega)=\frac{1}{(4\pi)^4}
\int\limits_{(\sqrt{\al_1}+\sqrt{\al_2})^2}^{(\sqrt{\omega}-\sqrt{\al_3})}ds
\frac{ \sqrt{\lambda(s,\omega,\al_3)} \sqrt{\lambda(\al_2,s,\al_1)}}{s\rm\omega}
\, \Theta\left(\omega-(\sum_{i=1}^{3}\al_i^{1/2})^2\right) \, .
\ee

%%%%%%%%%%%%%%%%%%%%%%%%%%%%%%%%%%%%%%%%%%%%%%%%%%%%%%%%%%%%%%%%%%%%%%%%%%%%%%%%%%%%%%%%%%%%%%%%%%%%%%%%%%%%%%%%%%%%%%%%%%%%%%%%%%%%%%%%%%%%%%%%%%%%%%%%%%%%%%%%%%%%%%%%%%%%%%%%%%%%%%%%%%%


\begin{thebibliography}{99}

%
%
\bibitem{lat2003}
P.O. Bowman, U.M. Heller, D.B. Leinweber, M.B. Parapilly, A.G. Wiliams,
 Phys.Rev. {\bf D70}, 034509 (2004). 
% 
\bibitem{BHLPW2004}
P.O. Bowman, U.M. Heller, D.B. Leinweber, M.B. Parappilly, A.G. Williams,
{ Phys.Rev.} {\bf D70},034509 (2004).
%
\bibitem{BHLPW2005}
P.O. Bowman, U.M. Heller, D.B. Leinweber, M.B. Parappilly, A.G. Williams, Jianbo Zhang,  Phys.Rev. {\bf D71}, 054507 (2005) 
%
\bibitem{ROBERTS}
C.D. Roberts, A.G. Williams, { Prog. Part. Nuc. Phys.} {\bf 33} (Pergamon Press, Oxford,) 1994.
%
\bibitem{ALKSMEK2001}
R. Alkofer, L. von Smekal,
{ Phys. Rep.} {\bf 353}, 281 (2001).
%
\bibitem{MARROB2003}
P. Maris, C.D. Roberts, 
{  Int. J. Mod. Phys.} {\bf E12}, 297 (2003).
%
\bibitem{FISALK2003}
C.S. Fischer R. Alkofer, {
 Phys. Rev.} {\bf D67}, 094020  (2003).
%
\bibitem{MAPA2003}
N.E. Mavromatos, J. Papavassiliou,
, cond-mat/0311421, Invited review to appear in 'Recent Research Developments in Physics' (TRN)
, and references therein
%
\bibitem{BLOCH2003}
J.C.R. Bloch,
{\it Few Body Syst.} {\bf33}, 111 (2003).
%
\bibitem{BOBOLEWIZA2001}
F.D.R. Bonnet, P.O. Bowman, D.B. Leinweber, A.G. Williams, J.M. Zanotti,
{Phys. Rev.}  {\bf D64}, 034501  (2001).
%
\bibitem{BCLM2002}
J.C.R. Bloch, A. Cucchieri, K. Langfeld, T. Mendes,
hep-lat/0209040.
%
\bibitem{FIES2004}
C. S. Fischer, F. Llanes-Estrada, R. Alkofer, 
hep-ph/0407294, 
Summary of a talk given at the international conference QCD DOWN UNDER 2004, Adelaide.
%
\bibitem{SAULIJHEP}
V.\v{S}auli,  {\it JHEP} 0302, 001 (2003).
%
\bibitem{SAULIRUN}
V.\v{S}auli,  {\it J. Phys.} {\bf G}30, 739 (2004).
%
%the end of section K2
\bibitem{NAMBU1955}
Y. Nambu, { Phys. Rev.}
{\bf 100}, 394 (1955).
%
\bibitem{NAMBU1956}
Y. Nambu, { Phys. Rev.}
{\bf 101}, 459 (1956).
%
\bibitem{DEWITT1960}
C. De Witt, R. Omnes (editors), {\it Dispersion relations and elementary particles},
Hermann, Paris and John Wiley \& Sons Inc. New York (1960).
%
\bibitem{MAN1959}
S. Mandelstam, 
{ Phys. Rev.} {\bf 115}, 1741 (1959).
%
\bibitem{ELOP1966}
R.J. Eden, P.V. Landshoff, D.I. Olive, J.C. Polkinghorn,
{ The Analytic S-matrix}, Cambridge (1966).
%
\bibitem{UMEKAM1951}
H. Umezawa, S. Kamefuchi,
{ Prog. Theor. Phys.} 6, 543 (1951).
%
\bibitem{KAL1952}
 G. K\"allen, { Helv. Phys. Acta.} {\bf 25}, 417 (1952).
%
\bibitem{KALL1958}
G.Kallen, {Quantenelektrodynamik},in "Handbuch der Physik," Vol.5, Part I.
J.Springer Verlag,Berlin ,1958; also Kgl. Danke Viedenskab. Selskab,
Mat. fys. Medd. 27, No.12 (1953);
%
\bibitem{LEH1954}
H. Lehmann, 
{ Nuovo Cim.} {\bf 11}, 342 (1954).
%
\bibitem{BBBB}
S.Bauberger, F.A.Berends, M. Bohm, M.Buza,
{ Nucl. Phys.} {\bf B434}, 383 (1995).
%
\bibitem{BDR}
A. Bashir, R. Delbourgo, M.L. Roberts,
{ J. Math. Phys.} {\bf 42},  5553 (2001).
%
\bibitem{WEISCHW}
S. Weinberg, 
{ Phys.Rev.} {\bf D8}, 3497 (1973).
%
\bibitem{NAK1971}
N. Nakanishi, { Graph Theory and Feynman Integrals},
(eds. Gordon and Breach, New York,1971).
%
\bibitem{DGR1959}
S. Deser, W. Gilbert, E.C.G. Sudarshan,  {\it Phys. Rev.} {\bf 115}, 731 (1959).
%
\bibitem{IDA1960}
M. Ida, { Prog. Theor. Phys.} {\bf 23},  1151 (1960).
%
\bibitem{KUSIWI1997}
K. Kusaka, K. Simpson, A.G. Williams,
{ Phys. Rev.} {\bf D 56}, 5071 (1997).
%
\bibitem{SAUADA2003}
V.\v{S}auli, J. Adam, 
{ Phys. Rev.}, {\bf D}67, 085007 (2003).
%
\bibitem{FEL1960}
D. Feldman,
{ Phys. Rev.} {\bf 117},279  (1960).
%
\bibitem{ITZYKSON}
C. Itzykson, J.B. Zuber,  {Quantum Field Theory}, McGraw-Hill Inc. (1980).
%
\bibitem{RIVERS}
R.J. Rivers, {\it Path integral methods in quantum field theory},\\
Cambridge University Press, 1987.
%
\bibitem{SAULIPHD}
V.\v{S}auli, {\it PHD Thesis}, hep-ph/0108160.
%
\bibitem{BETSAP1951}
H. Bethe, E. Salpeter,
{Phys. Rev.} {\bf 84},1232 (1951).
%
\bibitem{WICK}
G.C. Wick, 
{ Phys. Rev.} { \bf 96}, 1124 (1954).
%
\bibitem{NIETJO1996}
T. Nieuwenhuis, J.A. Tjon,
 { Few Body Syst.} {\bf 21},167 (1996).
%
\bibitem{AHLALK1999}
S. Ahlig, R. Alkofer,
{ Ann. Phys.} {\bf 275}, 113 (1999).
%
\bibitem{KUWI1995}
K. Kusaka, A.G. Williams,
{ Phys. Rev.} D {\bf51},  7026 (1995).
% 40:
\bibitem{SAGRTJ2004}
C. Savkli, F. Gross, J. Tjon, {\it Phys. Atom. Nucl.} {\bf 68}, 842 (2005); {\it Yad.Fiz.}
{\bf 68}, 874 (2005).
% 
\bibitem{MAN1955}
S. Mandelstam, 
{ Proc. Roy. Soc.} A {\bf 233}, 248 (1955).
%
\bibitem{HSW1995}
F.T.Hawes, T.Sizer, A.G.Williams,
{ Phys. Rev.} D {\bf51}, 3081 (1995).
%
\bibitem{HSW1997}
F.T.Hawes, T.Sizer, A.G.Williams,
{ Phys. Rev.} D {\bf55}, 3866 (1997).
%
\bibitem{KONNAK1992}
K. Kondo, H. Nakatani,
{ Prog. Theor. Phys.} {\bf 88}, 737 (1992).
%
\bibitem{PAGELS}
H. Pagels,  Phys. Rev. {\bf D21}
,2336 (1980).
%
\bibitem{WEI1979}
S. Weinberg, 
{ Phys. Rev.} D{\bf19}, 1277 (1979).
%
\bibitem{HOL1985}
B. Holdom,  { Phys. Lett.}, B{\bf 150}, 301
(1985).
%
\bibitem{CHIV2001}
 R.S. Chivukula,hep-ph/0011264
%
 \bibitem{APP2003}
T. Appelquist, R. Shrock, {  Phys. Rev. Lett.} {\bf 90},
201801 (2003).
%
\bibitem{CHRSHR2005}
Neil D. Christensen, Robert Shrock, hep-ph/0501294.
%
\bibitem{FUKKUG1976}
R.Fukuda, T.Kugo,  { Nucl. Phys.} B {\bf 117}, 250 (1976).
%
\bibitem{GOHOLIRA1998}
M.Gockeler, R. Horsely, V. Linke, P. Rakow, G.Schierholz, H. Stuben, { Phys. Rev. Lett.}
{\bf 80}, 4119(1998).
%
\bibitem{FAPO1967}
L.D. Fadeev, V.N. Popov, 
{ Phys. Lett. } {\bf B25}, 29 (1967).
%
\bibitem{BINPAP2002}
D. Binosi, J. Papavassiliou,  { Phys. Rev.} D {\bf 65}, 085003 (2002).
%
\bibitem{BINPAP2004}
D. Binosi, J. Papavassiliou,  J. Phys. {\bf G30},
203 (2004).
%
\bibitem{NIEL}
N. K. Nielsen, Nucl.Phys. {\bf B101} 173 (1975). 
%
\bibitem{LKT1}
L.D. Landau, I.M. Khalatnikov , Zh. Eksp. Teor. Fiz {\bf 29} 89 (1956).
%
\bibitem{LKT2}
L.D. Landau, I.M. Khalatnikov , Sov. Phys. JETP {\bf} 69 (1959).
%
\bibitem{PROGRES}
M.Namiki, K.Okano,
{Prog. Theor. Phys. Suppl.}{\bf 111} , (1993).
%
\bibitem{COR1982}111
John M. Cornwall,
{ Phys. Rev.} D {\bf 26}, 1453 (1982).
%
\bibitem{CORPAP1989}112
J.M. Cornwall, J. Papavassiliou,
 { Phys. Rev.} D {\bf 40}, 3474 (1989).
%
\bibitem{ZWAN2004}
D. Zwanziger, { Phys. Rev.} {\bf D69},016002 (2004).
%
\bibitem{SALDEL1964}
A. Salam, R. Delbourgo,  { Phys. Rev.} {\bf 135}, 1398 (1964).
%
\bibitem{DELZHA1984}
R.Delbourgo, R.B.Zhang,
{ J. Phys.} {\bf 17A}, 3593 (1984).
%
\bibitem{DELB1999}
R. Delbourgo, {\it Austral. J. Phys.} {\bf 52}, 681 (1999).
%
\bibitem{BALCHI1980}
J.S. Ball, T-W. Chiu, { Phys. Rev.} D {\bf 22}, 2542 (1980).
%
\bibitem{BASRAY}
A. Bashir, A. Raya,  Phys. Rev. {\bf D66} 105005 (2002).
%
\bibitem{MONTANO}
Alfredo Raya Montano, PhD Thesis, hep-th/0404138.
%
\bibitem{BAPERA2006}
A. Bashir, M.R. Pennington, A. Raya, hep-ph/0511291 
%
\bibitem{BARA2005} 
A. Bashir and A. Raya, Nucl.Phys. {\bf B709} 307 (2005).
%
\bibitem{ARI}
 A. Arrizabalaga, J. Smit,  Phys.Rev.  D{\bf66},065014 (2002).
%
\bibitem{MIRANSKY}
V. Miransky,  Nuovo Cim. {\bf A90}, 149 (1985).
%
\bibitem{VOLODA}
V. Miransky,  Sov. Phys. ,JETP {\bf 61}, 905 (1985).
%
%\bibitem{COLWEI1973}
%S. Coleman, E. Weinberg,
%{ Phys. Rev.} D{\bf7}, 1888 (1973).
\bibitem{ADBESA}
J. Adam, P. Bicudo, V. \v{S}auli, prepared for publication.
%
\bibitem{ADFM2003}
R. Alkofer, W. Detmold, C.S. Fischer, P.Maris, Phys. Rev., {\bf D70}, 014014 (2004).
%
\bibitem{HALPERN1993}
M.B. Halpern,  Prog. Theor. Phys. Suppl. 
{\bf 111}, 163 (1993).
%
\bibitem{KRP1995}
 A. Kizilersu, M. Reenders, M.R. Pennington,
 { Phys. Rev.} D{\bf 52}, 1242 (1995).
%
\bibitem{CUPE1990}
D.C. Curtis, M.R. Pennington, {\it Phys. Rev} {\bf D42},4165 (1990).
%
\bibitem{BAPE1995}
A. Bashir, M.R. Pennington, Phys. Rev. {\bf D53}, 4694 (1996).
%
\bibitem{BURO1991}
C.J. Burden, C.D. Roberts, { Phys. Rev} {\bf D44},540 (1991). 
%
\bibitem{BAKIPE2000}
A. Bashir, A. Kizilersu, M.R. Pennington, 
 Phys. Rev.  {\bf D62},085000 (2000).                                                            
%
\bibitem{BADE2004}
A. Bashir, R. Delbourgo,  J. Phys. {\bf A37},6587 (2004).
%
\bibitem{BAPE1994}
A. Bashir, M.R. Pennington, 
Phys. Rev. {\bf D50}, 7679 (1994).
%
\bibitem{BASRAYII2004}
A. Bashir, A. Raya, Nucl.Phys. {\bf B709}, 307 (2005).
%
\bibitem{BLIARE1999}
J.-P. Blaizot, E. Iancu, A. Rebhan, 
 Phys. Rev. Lett. {\bf 83}2906 (1999).
%
\bibitem{BLIAREii1999}
J.-P. Blaizot, E. Iancu,  Phys. Rev.
{\bf D63} (2001), 065003.
%     
\bibitem{BLIAREiii1999}
J.-P. Blaizot, E. Iancu, A. Rebhan,  Phys. Lett.
{\bf B470}, 181 (1999).
% 
\bibitem{AABBS}
G. Aarts, D. Ahrensmeier, R. Baier, J. Berges, J. Serreau, 
{ Phys. Rev.} {\bf D66}, 045008 (2002).
%
\bibitem{CODAMI2003}
F. Cooper, J. Dawson, B. Mihaila, { Phys.Rev.} {\bf D67},056003 (2003), and references therein.
%
\bibitem{AJ2003}
G. Aarts,  Jose M. Martinez Resco,  Phys. Rev. {\bf D68}, 085009 (2003).
%
\bibitem{BRAEMM2002}
E. Braaten, E. Petitgirard,  Phys. Rev.  {\bf D65}, 085039 (2002).
%
\bibitem{BRAEMMii2002}
E. Braaten, E. Petitgirard,  Phys. Rev.  {\bf D65},041701  (2002).
%          
\bibitem{ANDSTRI2004}
J.O. Andersen, M. Strickland,  Phys.Rev. {\bf D71}, 025011 (2005).
%
\bibitem{BERGES}
J. Berges,  hep-ph/0409233.
%
\bibitem {BLOCH2}
J.C.R. Bloch,  Ph.D. thesis - University of Durham (1995),
hep-ph/0208074.
% 
\bibitem{BLOPEN1995}
J.C.R. Bloch, M.R. Pennington, { Mod. Phys. Lett.} A {\bf
10}, 1225, (1995).
%
\bibitem{COLMAC1974}
J.C. Collins, A.J.MacFarlane, { Phys. Rev.} D {\bf10}, 1201
(1974).
%
\bibitem{LANDAU}
L.D. Landau, {\it On the Quantum Field Theory, in Niels Bohr and
the Development of Physics,} ed. W. Pauli Pergamon, London (1955).
%
\bibitem{HESKNO}
H.van Hees, J. Knoll, Phys.Rev. {\bf D65},105005 (2002).
%
\bibitem{COMIDA2004}
F. Cooper, B. Mihaila, J.F. Dawson,   Phys.Rev. {\bf D70},105008 (2004).
%
\bibitem{MAR1995}
P. Maris, Phys. Rev. D {\bf52}, 6087 (1995).
%
\bibitem{MAR1992}
P. Maris,  Int. J. Mod. Phys. A {\bf7}, 5369 (1992).
%
\bibitem{ALEK2005}
A. I. Alekseev,  hep-ph/0503242.
%
\bibitem{GROWIL1973}
D.J. Gross, F. Wilczek,  Phys. Pev. Lett. {\bf 30}, 1343 (1973).
%
\bibitem{POL1973} 
H.D. Politzer,  Phys. Pev. Lett.
{\bf 30}, 1346 (1973).
%
\bibitem{ROBSMI2000}
C.D. Roberts, S.M. Schmidt, {Prog. Part. Nucl. Phys.} {\bf 45S1}, 103 (2000).
%
\bibitem{MARTAN2005}
P. Maris and P.C. Tandy,
nucl-th/0511017, contribution to the proceedings of LC05, Cairns, Australia, July 2005.
%
\bibitem{BENDER}
 A. Bender, W. Detmold, C.D. Roberts, A.W. Thomas,
 {Phys. Rev.} C{\bf65},  065203 (2002).
%
\bibitem{BICEST2003}
F.J. Llanes-Estrada, P. Bicudo, Phys. Rev., {\bf D68}, 094014 (2003). 
%
\bibitem{BIC2004}
P. Bicudo, Phys. Rev. {\bf D69}, 074003 (2004).
%
\bibitem{FISALKCON}
C.S. Fischer,  R. Alkofer,
 hep-ph/0411347. 
%
\bibitem{FADPOP1967}
L.D. Fadeev, V.N. Popov,  
Phys. Lett{ \bf 25B}, 29 (1967).
%
\bibitem{SAL1963}
A. Salam, { Phys. Rev.} {\bf 130}, 1287 (1963).
%
\bibitem{STRAT}
J. Strathdee, 
{ Phys. Rev.} {\bf 135}, 1428 (1964).
%
\bibitem{DELWES1977}
R. Delbourgo, P.C. West,   {\it J. Phys.} {\bf A 10}, 1049 (1977).
%
\bibitem{WEST2}
R. Delbourgo, P.C. West,
 {  Phys. Lett.} {\bf B 72}, 96 (1977).
%
\bibitem{DEL1979}
R.Delbourgo, Nuovo Cim. {\bf 49}, 484 (1979).
%
\bibitem{HOS2002}
Y. Hoshino,  hep-th/0202020.
%
\bibitem{LAV1991}
M. Lavelle, Phys. Rev. {\bf D44} R26 (1991).
%
\bibitem{SVZ1979}
M.A. Shifman, A.I. Vainshtein, V.I. Zakharov, { Nucl. Phys.} {\it B147}, 385 (1979).
%
\bibitem{PARISI}
G. Parisi, R. Petronzio, 
{ Phys. Lett.} {\bf 94B}, 51 (1980).
%
\bibitem{GUPTA1987}
R. Gupta et all., 
  { Phys. Rev. } {\bf D36}, 2813 (1987).
%
\bibitem{LAT4}
P. Marenzoni, G. Martinelli, N. Stella, M. Testa,
{ Phys. Lett.} {\bf B318}, 511 (1993)
%
\bibitem{LAT3}
C. Bernard, C. Parrinello, A. Soni, { Phys. Rev. } {\bf D49}, 1585 (1994).
%
\bibitem{LAT2}
J.I. Skullerud et al. (UKQCD Collaboration), Nucl. Phys. Proc. Suppl. {\bf 73}. 626 (1999).
%
\bibitem{LAT1}
A. Cucchieri, 
{ Phys. Rev.} {\bf D 60},034508 (1999).
%
%\bibitem{MIHARA}
%A. Mihara, A.A. Natale, 
% { Phys. Lett.} {\bf B482}, 378 (2000).
%
%\bibitem{AGUILAR}
%A.C. Aguilar, A. Mihara, A.A. Natale,
% { Phys. Rev. } {\bf D65}, 054011 (2000).
%
%\bibitem{AGUILAR2}
% A.C. Aguilar, A.A. Natale, P.S. Rodrigues da Silva,
%{  Phys. Rev. Lett.} {\bf 90}, 152001 (2003).
%
\bibitem{SMEALK2000}
 R. Alkofer, L. von Smekal, {\it Phys.Rept. }, {\bf 353}, 281 (2001).
%
\bibitem{MAND1979}
S. Mandelstam, {\it Phys. Rev. }{\ bf D}20, 3223 (1979).
%
\bibitem{BROPEN1989}
N. Brown, M.R. Pennington ,{\it Phys.Rev.} {\bf D}39, 2723 (1989).
%
\bibitem{BUTPEN1995}
K. Buttner, M.R. Pennington, {\it  Phys.Rev.} {\bf D}52, 5220 (1995).
%
\bibitem{SMHAAL1997}
L. von Smekal, R. Alkofer, A. Hauck, {\it Phys. Rev. Lett.} {\bf 79}, 3591 (1997).
%
\bibitem{ATKBLO1998}
 D. Atkinson, J.C.R. Bloch, {\it Phys.Rev.} {\bf D}58, 094036n (1998)
%
\bibitem{ATKBLO1998II}
D. Atkinson,  J.C.R. Bloch, 
Mod. Phys. Lett. {\bf A13}, 1055 (1998). 
%
\bibitem{HAROWI1994}
F.T. Hawes, C.D. Roberts, A.G. Williams, {\it Phys. Rev.}{\bf D}49, 4683 (1994).
%
\bibitem{LERSME2002}
C. Lerche, L. Smekal, {\it Phys.Rev.} {\bf D}65 ,125006 (2002).
%
\bibitem{KONDO2003}
K. Kondo,  Phys.Lett. {\bf B551},324 (2003).
%
\bibitem{AGUNAT2004}
A. C. Aguilar, A. A. Natale, JHEP 0408 (2004) 057.
%
\bibitem{AFRS2003} 
 R. Alkofer, C.S. Fischer, H. Reinhardt, L. Smekal,
{\it Phys.Rev.}{\bf D}68, 045003 (2003).
%
\bibitem{KON2003}
K. Kondo, hep-th/0303251.
%
\bibitem{NES2003}
V.A. Nesterenko,  Int. J. Mod. Phys. {\bf A18},5475 (2003).
%
\bibitem{NESPAP2004}
A.V. Nesterenko,  J. Papavassiliou, Phys.Rev. {\bf D71},016009 (2005).
%
%\bibitem{JONIVAR1991}
%D.B.Leinweber, J.I.Skullerud, A.G.Williams, C.Parrinello,
% Phys.Rev. {\bf D60},094507 (1999) ; Erratum-ibid. {\bf D61}, 079901 (2000).
%
\bibitem{LAREGA}
K. Langfeld, H. Reinhardt, J. Gattnar,
 Nucl.Phys. {\bf B621}131 (2002).
%
\bibitem{CUMETA}
A. Cucchieri, T. Mendes, A. R. Taurines, Phys.Rev. {\bf D71}, 051902 (2005). 


\end{thebibliography}
\end{document}